\documentclass[
showkeys,11pt,
preprint,preprintnumbers,nofootinbib,
groupedaddress,superscriptaddress,amsmath,amssymb]{revtex4}
\usepackage{graphicx}
\usepackage{amsmath,amssymb}
\usepackage{bm}
\usepackage{color}
\def\Babar{{\mbox{\slshape B\kern-0.1em{\smaller A}\kern-0.1em B\kern-0.1em{\smaller A\kern-0.2em R}}}}

\usepackage{hyperref}
\hypersetup{ 
 setpagesize=fake,
 bookmarksnumbered=true,%
 bookmarksopen=true,%
 colorlinks=true,%
 linkcolor=blue,
 citecolor=red,
}

\newcommand{\bp}{\begin{pmatrix}}
\newcommand{\ep}{\end{pmatrix}}
\newcommand{\bb}{\begin{bmatrix}}
\newcommand{\eb}{\end{bmatrix}}

\usepackage[normalem]{ulem}

\allowdisplaybreaks[3]
\begin{document}

\title{LHC constraints and prospects for $S_1$ scalar leptoquark \\ explaining the $\bar B \to D^{(*)} \tau \bar\nu$ anomaly}

\author{B\'eranger Dumont}
\affiliation{Center for Theoretical Physics of the Universe, Institute for Basic Science (IBS), Daejeon 305-811, Republic of Korea}

\author{Kenji Nishiwaki}
\email{nishiken@kias.re.kr}
\affiliation{School of Physics, Korea Institute for Advanced Study, Seoul 02455, Republic of Korea}
\preprint{KIAS-P16023}

\author{Ryoutaro Watanabe}
\email{wryou1985@ibs.re.kr,watanabe@lps.umontreal.ca}
\affiliation{Center for Theoretical Physics of the Universe, Institute for Basic Science (IBS), Daejeon 305-811, Republic of Korea}
\preprint{CTPU-16-07}

\date{\today}


\begin{abstract}
Recently, deviations in flavor observables of $\bar B \to D^{(*)} \tau \bar\nu$ have been shown between the predictions in the Standard Model and the experimental results reported by BaBar, Belle, and LHCb collaborations. 
One of the solutions to this anomaly is obtained in a class of leptoquark model with a scalar leptoquark boson $S_1$, 
which is a $SU(3)_c$ triplet and $SU(2)_L$ singlet particle with $-1/3$ hypercharge interacting with a quark-lepton pair.   
With well-adjusted couplings, this model can explain the anomaly and be compatible with all flavor constraints. 
In such a case, the $S_1$ boson can be pair-produced at CERN's Large Hadron Collider (LHC) and subsequently decay as $S_1^* \to t \tau$, $b\nu_\tau$, and $c \tau$. 
This paper explores the current 8 and 13~TeV constraints, as well as the detailed prospects at 14~TeV, of this flavor-motivated $S_1$ model.
From the current available 8 and 13~TeV LHC searches, we obtain constraints on the $S_1$ boson mass for $M_{S_1} < 400\,\text{GeV}$ - $640\,\text{GeV}$ depending on values of the leptoquark couplings to fermions.  
Then we study future prospects for this scenario at the 14~TeV LHC using detailed cut analyses 
and evaluate exclusion/discovery potentials for the flavor-motivated $S_1$ leptoquark model from searches for the $(b\nu)(\bar{b}\bar{\nu})$ and $(c\tau)(\bar{c}\bar{\tau})$ final states. 
In the latter case, we consider several scenarios for the identification of charm jets. 
As a result, we find that the $S_{1}$ leptoquark origin of the $\bar B \to D^{(*)} \tau \bar\nu$ anomaly can be probed with $M_{S_1} \lesssim 600/800\,\text{GeV}$ at the 14~TeV LHC 
with $\mathcal{L} = 300/3000\,\text{fb}^{{-1}}$ of accumulated data. 
One can also see that the 14~TeV LHC run~II with $\mathcal{L} = 300\,\text{fb}^{-1}$ can exclude the $S_{1}$ leptoquark boson up to $M_{S_1} \sim 0.8\,\text{TeV}$ at $95\%$ confidence level, 
whereas a future 14~TeV LHC with $\mathcal{L} = 3000\,\text{fb}^{-1}$ data has a potential to discover the $S_1$ leptoquark boson with its mass up to $M_{S_1} \sim 1.1\,\text{TeV}$ with over $5\sigma$ significance, 
from the $(b\nu)(\bar{b}\bar{\nu})$ and/or $(c\tau)(\bar{c}\bar{\tau})$ searches. 
\end{abstract}

\maketitle

\section{Introduction}
\begin{figure}[t]
\begin{center}
\includegraphics[viewport=0 0 466 203, width=32em]{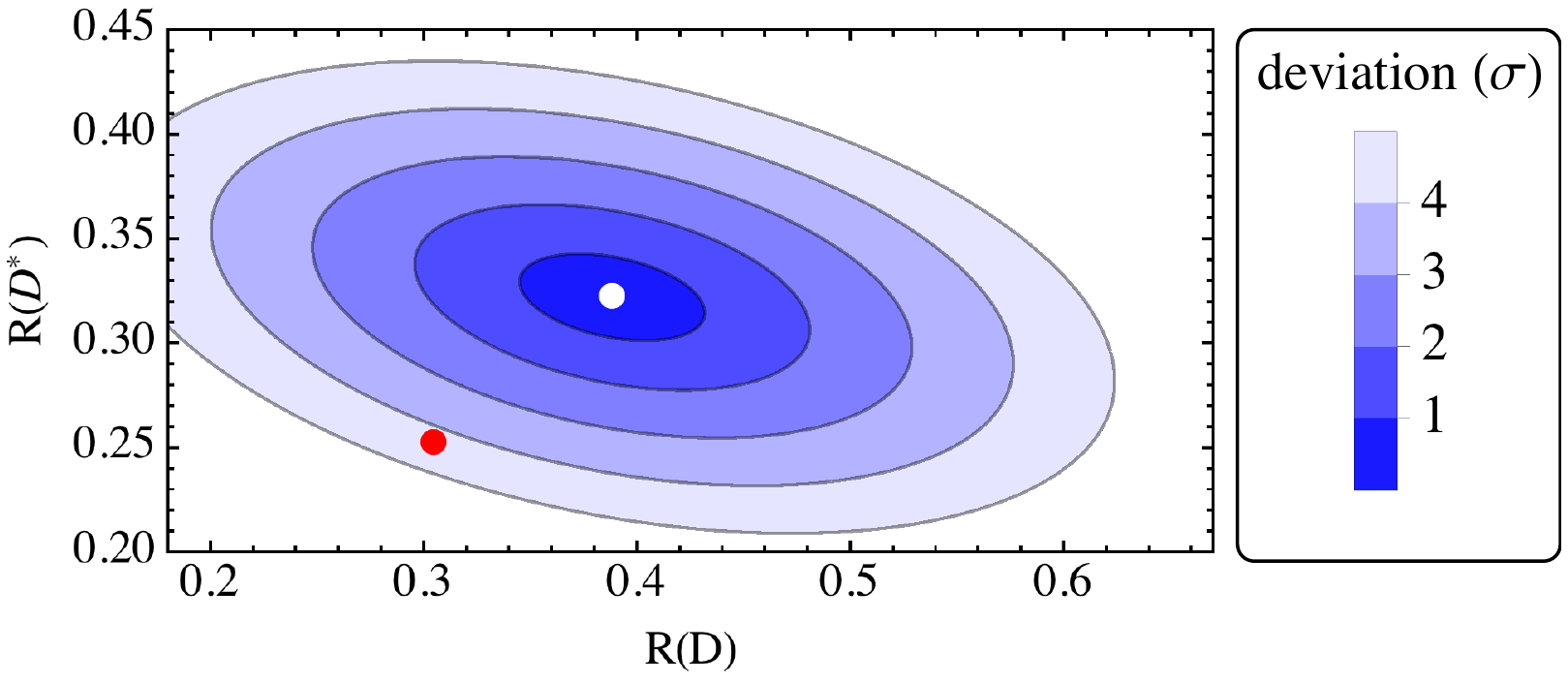}
\caption{
Correlation between combined measurements of $R(D)$ and $R(D^{*})$~\cite{Lees:2012xj,Huschle:2015rga,Aaij:2015yra} and comparison with the SM prediction. 
The red and white dots indicate the central values of the SM predictions and the combined experimental results, respectively. 
Both the theoretical and experimental uncertainties are taken into account when calculating the deviation contours. 
}
\label{Fig:Comparison}
\end{center}
\end{figure}

An excess in the search for $\bar B \to D^{(*)} \tau \bar\nu$ reported by the BaBar and Belle collaborations in Refs.~\cite{Lees:2012xj,Lees:2013uzd,Matyja:2007kt,Adachi:2009qg,Bozek:2010xy} 
has provided hints of an indirect evidence of new physics, even though the full data sample was not yet used in the Belle results~\cite{Matyja:2007kt,Adachi:2009qg,Bozek:2010xy}. 
The observables, defined as 
\begin{align}
R(D) \equiv \frac{\mathcal{B}(\bar B\to D\tau^-\bar\nu_\tau)}{\mathcal{B}(\bar B\to D\ell^-\bar\nu_\ell)} \,, \quad
R(D^{*}) \equiv \frac{\mathcal{B}(\bar B\to D^{*}\tau^-\bar\nu_\tau)}{\mathcal{B}(\bar B\to D^{*}\ell^-\bar\nu_\ell)} \,, 
\label{Eq:RDdefinition}
\end{align}
where $\ell=e$ or $\mu$, are introduced for these processes in order to reduce theoretical uncertainties and separate the issue of the determination of $|V_{cb}|$ from new physics study. 
The standard model (SM) predicts precise values of $R(D^{(*)})$ with the help of the heavy quark effective theory~\cite{Caprini:1997mu,Amhis:2012bh}.  
In May 2015, the latest results from the BaBar~\cite{Lees:2012xj,Lees:2013uzd}, Belle~\cite{Huschle:2015rga} and LHCb~\cite{Aaij:2015yra} collaborations have finally appeared all together. 
As a result, we can see the significant deviations between the combined experimental results~\cite{Lees:2012xj,Lees:2013uzd,Huschle:2015rga,Aaij:2015yra} and the SM predictions~\cite{Sakaki:2013bfa}, 
which reads 
\begin{align}
 & \label{Eq:deviationRD} R(D)^\text{exp.} - R(D)^\text{SM} = 0.089 \pm 0.051 \,, \\
 & \label{Eq:deviationRDst} R(D^*)^\text{exp.} - R(D^*)^\text{SM} = 0.070 \pm 0.022  \,,
\end{align}
where the combined experimental results are privately evaluated assuming Gaussian distributions and the experimental and theoretical uncertainties are taken into account in the errors. 
The standard deviation with a correlation is also shown in Fig.~\ref{Fig:Comparison} and we can see that the discrepancy reaches $\sim 4 \sigma$.
It is interesting that both of the deviations are ``excesses'' of the experimental results from the SM predictions despite negative correlations ($\sim -0.3$) in the experiments. 
We put individual and combined values of the experimental results in Appendix~\ref{App:DetailedExperiment}.

In recent years, several new physics scenarios have been investigated with respect to the excesses.   
In particular, as the two-Higgs-doublet model (2HDM) can give a large contribution to the tauonic $B$ meson decays~\cite{Hou:1992sy,Tanaka:1994ay,Kamenik:2008tj,Nierste:2008qe,Tanaka:2010se}, 
it is studied in Refs.~\cite{Tanaka:2012nw,Sakaki:2012ft,Bailey:2012jg,Fajfer:2012jt,Celis:2012dk,Crivellin:2012ye,Bhattacharya:2015ida,Ko:2012sv} to explain the large deviation in $\bar B \to D^{(*)} \tau \bar\nu$. 
Their results imply that it is hard to accommodate the excesses in $R(D)$ and $R(D^{*})$  simultaneously for the type-I, II, X, and Y 2HDMs, whereas there is still allowed parameter space for the general 2HDM. 
The $R$-parity violating minimal supersymmetric standard model is considered in Refs.~\cite{Tanaka:2012nw,Erler:1996ww,Chemtob:2004xr,Deshpande:2012rr}. 
It turns out that this scenario is not likely to explain the excesses at the same time with satisfying the constraint from $\bar B \to X_s \nu\bar\nu$. 
The extra gauge boson is also studied in the context of $\bar B \to D^{(*)} \tau \bar\nu$ in reaction to the recent update~\cite{Freytsis:2015qca,Hati:2015awg}. 

The other feasible and interesting scenario is given in the leptoquark model~\cite{Buchmuller:1986zs} on which we focus in this paper. 
Its potential for explaining the $\bar B \to D^{(*)} \tau \bar\nu$ anomaly is studied in Refs.~\cite{Tanaka:2012nw,Biancofiore:2013ki,Dorsner:2013tla,Sakaki:2013bfa}. 
As a consequence of the recent study in Ref.~\cite{Sakaki:2013bfa}, three types of the leptoquark bosons can explain the excess without any inconsistency with the constraint from $\bar B \to X_s \nu\bar\nu$. 
By limiting the flavor structure of leptoquark couplings, correlations to other processes, especially to the $R_K$ anomaly, are also discussed in Refs.~\cite{Freytsis:2015qca,Bhattacharya:2014wla,Calibbi:2015kma,Bauer:2015knc,Fajfer:2015ycq,Barbieri:2015yvd}. 
Note that scalar leptoquarks are also useful for explaining the $h \to \mu \tau$ anomaly in CMS (and ATLAS)~\cite{Dorsner:2015mja,Cheung:2015yga,Baek:2015mea,Bauer:2015knc}.

To explain the central combined experimental values of $R(D^{(*)})$ in any case, somewhat large couplings of the leptoquark boson to the third (and second, in part) generation quarks and leptons are required. 
Hence, the leptoquark search for the third generation at the Large Hadron Collider (LHC) can be significant. 
Since the color $SU(3)$ charge is assigned, the leptoquark bosons are dominantly pair-produced at the hadron collider and its cross section is independent on the couplings to fermions. 
Thus, the direct search of the leptoquark boson gives a constraint on a branching ratio of its decay into fermions. 
In this paper, we study the leptoquark search at the LHC, including the second and third generation quarks and leptons in the final state, where it is motivated by the flavor anomaly in $\bar B \to D^{(*)} \tau \bar\nu$.

This paper is organized as follows. 
At first, after briefly reviewing the leptoquark model, we show a current status of explaining the $\bar B \to D^{(*)} \tau \bar\nu$ anomaly and constraints from a related flavor process on the model in Sec.~\ref{Sec:LQflavor}. 
Then, we summarize present collider studies at the LHC and apply them to the model in Sec.~\ref{Sec:LHCstudy}. 
In Sec.~\ref{Sec:Analysis}, we provide detailed analysis cuts, which are performed for 14 TeV LHC searches. 
In turn, we show our result and discuss future prospects for exclusion and discovery potentials of the leptoquark boson in Sec.~\ref{Sec:Prospect}. 
Finally, a summary is  {provided} in Sec.~\ref{Sec:Summary}.

\section{Leptoquark model and flavor observables}
\label{Sec:LQflavor}
Here, we give a brief review on the possible types of leptoquarks and their lepto-quark interactions. 
Then we summarize the contribution to the process in $b \to c \tau\bar\nu$, which leads to $\bar B \to D^{(*)} \tau \bar\nu$ at hadron level, for all possible cases.

\subsection{Classification}
Some of new physics scenarios, especially for grand unifications of the fundamental interactions, contain new scalar and vector bosons which interact with quarks and leptons. 
This kind of boson is called as leptoquark and carries both the baryon and lepton numbers together with color and electric charges. 
It is known~\cite{Buchmuller:1986zs} that there are ten types of leptoquarks with the general dimensionless $SU(3)_c\times SU(2)_L\times U(1)_Y$ invariant and {\it flavor non-diagonal} couplings.\footnote
{In this paper, we do not consider possible ``di-quark'' interactions even though they are allowed by the SM gauge invariance in general.
As widely known, if leptoquark and di-quark interactions coexist, both the baryon and lepton numbers are violated so that the proton becomes unstable.
Note that among the three scalar leptoquarks shown in Table~\ref{tab:LQ_numbers}, $R_{2}$ can avoid such an unstable proton (within renormalizable interactions) 
since no renormalizable di-quark interaction is written down~\cite{Arnold:2013cva}.
}
Among them, six leptoquark (LQ) bosons are relevant for the process $b \to c \ell \bar\nu$. 
The Lagrangian for the term interacting with SM fermions is given by 
\begin{align}
      & \mathcal L^{\rm LQ} = \mathcal L_{F=0}^{\rm LQ} + \mathcal L_{F=-2}^{\rm LQ} \,, \label{EQ:LagLQ1} \\[1em]
      & \mathcal L_{F=0}^{\rm LQ}
      = \left( {h_{1L}^{ij}}\,\bar Q_L^i \gamma_\mu L_{L}^j + {h_{1R}^{ij}} \,\bar d_R^i \gamma_\mu \ell_{R}^j \right)U_{1}^\mu + {h_{3L}^{ij}} \,\bar Q_{L}^i {\bm\sigma}\gamma_\mu L_{L}^j {\bm U}_3^\mu \notag \\
      &\hspace{3em} + \left( { h_{2L}^{ij}} \,\bar u_{R}^i L_{L}^j + {h_{2R}^{ij}} \,\bar Q_{L}^i i\sigma_2 \ell_{R}^j \right)R_2 +\text{h.c.} \,, \\[1em]
      &\mathcal L_{F=-2}^{\rm LQ}
      =\left( {g_{1L}^{ij}} \,\bar Q_{L}^{c,j} i\sigma_2 L_{L}^j + {g_{1R}^{ij}}\,\bar u_{R}^{c,i} \ell_{R}^j \right)S_1 + {g_{3L}^{ij}} \,\bar Q_{L}^{c,i} i\sigma_2{\bm\sigma} L_{L}^j {\bm S}_3 \notag \\
      &\hspace{3em}+ \left( {g_{2L}^{ij}} \,\bar d_{R}^{c,i} \gamma_\mu L_{L}^j + {g_{2R}^{ij}} \,\bar Q_{L}^{c,i} \gamma_\mu \ell_{R}^j \right)V_{2}^\mu +\text{h.c.}  \,,  \label{EQ:LagLQ2}
\end{align}
where $h^{ij}$ and $g^{ij}$ are the dimensionless couplings; 
$S_1$, ${\bm S}_3$, and $R_2$ are scalar leptoquark bosons;  
$U_{1}^\mu$, ${\bm U}_3^\mu$, and $V_2^\mu$ are vector leptoquark bosons; 
index $i$ ($j$) indicates the generation of quarks (leptons); 
$\psi^c = C\bar \psi^T=C\gamma^0\psi^*$ is  {the} charge-conjugated fermion field of $\psi$. 
These six leptoquark bosons ($S_1$, ${\bm S}_3$, $R_2$, $U_{1}$, ${\bm U}_3$, and $V_2$) can contribute to $\bar B \to D^{(*)} \tau\bar\nu$. 
In Table~\ref{tab:LQ_numbers}, we summarize the quantum numbers of the leptoquark bosons.
{Here we define the fermions in the gauge eigenbasis and} follow the treatment in Ref.~\cite{Sakaki:2013bfa} such that Yukawa couplings of the up-type quarks and the charged leptons are diagonal, 
while the down-type quark fields are rotated into the mass eigenstate basis by the Cabibbo-Kobayashi-Maskawa~(CKM) matrix. 
\begin{table}[t]
 \begin{center}\begin{tabular}{cccccc}
 \hline\hline
    		& \,\,spin\,\, 	& \,\,$F=3B+L$\,\, 	& \,\,$SU(3)_c$\,\, 	& \,\,$SU(2)_L$\,\, 	& \,\,$U(1)_{Y=Q-T_3}$\,\, \\ 
 \hline
 $S_1$ 	& $0$	 	& $-2$			& $3^*$ 		& $1$			& $1/3$        \\
 \hline
 ${\bm S}_3$ 	& $0$		& $-2$ 			& $3^*$		& $3$			& $1/3$        \\     
 \hline
 $R_2$ 	& $0$		& $0$			& $3$		& $2$			& $7/6$        \\
 \hline
 $V_2$ 	& $1$		& $-2$			& $3^*$		& $2$		 	& $5/6$        \\
 \hline
 $U_1$ 	& $1$		& $0$ 			& $3$		& $1$			& $2/3$        \\
 \hline
 ${\bm U}_3$ 	& $1$		& $0$			& $3$		& $3$			& $2/3$        \\
 \hline\hline
 \end{tabular}
 \caption{Quantum numbers of scalar and vector leptoquarks.}
 \label{tab:LQ_numbers}
 \end{center}
\end{table}

\subsection{Contribution to $\bar B \to D^{(*)} \tau\bar\nu$}
The leptoquark bosons which have interactions in Eqs.~(\ref{EQ:LagLQ1})-(\ref{EQ:LagLQ2}) can contribute to $\bar B \to D^{(*)} \tau\bar\nu$ at the tree level. 
The effective Lagrangian for $b \to c \tau \bar\nu_l$ is written~\cite{Sakaki:2013bfa} as 
\begin{equation}
   - \mathcal L_\text{eff} =  
   ( C_\text{SM} \delta_{l\tau} + C_{\mathcal V_1}^l) \mathcal O_{\mathcal V_1}^l + C_{\mathcal V_2}^l \mathcal O_{\mathcal V_2}^l 
   + C_{\mathcal S_1}^l \mathcal O_{\mathcal S_1}^l + C_{\mathcal S_2}^l \mathcal O_{\mathcal S_2}^l + C_{\mathcal T}^l \mathcal O_{\mathcal T}^l  \,,
\end{equation}
where the effective operators are defined as
\begin{align}
     & \mathcal O_{\mathcal V_1}^l = (\bar c_L \gamma^\mu b_L)(\bar \tau_L \gamma_\mu \nu_{lL}) \,, \\
     &\mathcal O_{\mathcal V_2}^l = (\bar c_R \gamma^\mu b_R)(\bar \tau_L \gamma_\mu \nu_{lL}) \,, \\
     & \mathcal O_{\mathcal S_1}^l = (\bar c_L b_R)(\bar \tau_R \nu_{lL}) \,, \\
     & \mathcal O_{\mathcal S_2}^l = (\bar c_R b_L)(\bar \tau_R \nu_{lL}) \,, \\
     & \mathcal O_{\mathcal T}^l = (\bar c_R \sigma^{\mu\nu} b_L)(\bar \tau_R \sigma_{\mu\nu} \nu_{lL}) \,, 
\end{align}
and the Wilson coefficients in the leptoquark model are given by 
\begin{align}
      & C_\text{SM} = 2 \sqrt 2 G_F V_{cb} \,, \label{EQ:CVsm} \\
      & C_{\mathcal V_1}^l =  \sum_{k=1}^3 V_{k3} 
      \left[ 
      {g_{1L}^{kl}g_{1L}^{23*} \over 2M_{S_1}^2} - {g_{3L}^{kl}g_{3L}^{23*} \over 2M_{{\bm S}_3}^2} + {h_{1L}^{2l}h_{1L}^{k3*} \over M_{U_1}^2} - {h_{3L}^{2l}h_{3L}^{k3*} \over M_{{\bm U}_3}^2}
      \right] \,,  \label{EQ:CV1} \\
      & C_{\mathcal V_2}^l = 0 \,, \\
      & C_{\mathcal S_1}^l = \sum_{k=1}^3 V_{k3} 
      \left[ 
      -{2g_{2L}^{kl}g_{2R}^{23*} \over M_{V_2}^2} - {2h_{1L}^{2l}h_{1R}^{k3*} \over M_{U_1}^2} 
      \right] \,, \\
      & C_{\mathcal S_2}^l = \sum_{k=1}^3 V_{k3} 
      \left[ 
      -{g_{1L}^{kl}g_{1R}^{23*} \over 2M_{S_1}^2} - {h_{2L}^{2l}h_{2R}^{k3*} \over 2M_{R_2}^2} 
      \right] \,, \label{EQ:CS2} \\
      & C_{\mathcal T}^l = \sum_{k=1}^3 V_{k3} 
      \left[ 
      {g_{1L}^{kl}g_{1R}^{23*} \over 8M_{S_1}^2} - {h_{2L}^{2l}h_{2R}^{k3*} \over 8M_{R_2}^2} 
      \right] \,, \label{EQ:CT}
\end{align}
at the energy scale $\mu = M_X$, where $X$ represents a leptoquark. 
The SM contribution is given by $C_\text{SM}$. 
The index $l$ denotes the generation of the neutrino which, in general, needs not be the third one in this case. 
The CKM matrix element is denoted as $V_{ij} \equiv V_{u_i d_j}$. 
We note that we take the correct mass eigenstate basis for the fermions and thus the CKM matrix elements appear in the Wilson coefficients.

As can be seen in Eqs.~(\ref{EQ:CV1})-(\ref{EQ:CT}), several leptoquark bosons with several combinations of the couplings can contribute to $b \to c \tau \bar\nu_l$. 
Those contributions can be classified as 
\begin{itemize} 
 \item $C_{\mathcal S_2}^l =-4C_{\mathcal T}^l$ mediated by $S_1$ boson with nonzero value of $(g_{1L}g_{1R}^*)$, 
 \item $C_{\mathcal S_2}^l =4C_{\mathcal T}^l$ by $R_2$ boson with $(h_{2L}h_{2R}^*)$, 
 \item $C_{\mathcal V_1}^l$ by $S_1$, ${\bm S}_3$, $U_1$, or ${\bm U}_3$ bosons with $(g_{1L}g_{1L}^*)$, $(g_{3L}g_{3L}^*)$, $(h_{1L}h_{1L}^*)$, or $(h_{3L}h_{3L}^*)$, 
 \item $C_{\mathcal S_1}^l$ by $U_1$ or $V_2$ bosons with $(h_{1L}h_{1R}^*)$ or $(g_{2L}g_{2R}^*)$.
\end{itemize}
It is interesting that the tensor type operator appears in the $S_1$ and $R_2$ type leptoquark models~\cite{Lee:2001nw}. 
To evaluate those effects on the observables $R(D)$ and $R(D^*)$, the running effect of  $C_Y^l(\mu)$ ($Y$ showing types of the effective operators) from $\mu=M_X$ to $\mu=\mu_b$, 
where $\mu_{b}$ is the mass scale of the bottom quark, must be taken into account. 
Due to the fact that the vector and axial-vector currents are not renormalized and their anomalous dimensions vanish, $\mathcal V_{1,2}$ do not receive the running effect. 
On the other hand, a scale dependence in the scalar $\mathcal S_{1,2}$ and tensor $\mathcal T$ currents exist and is approximately evaluated as 
\begin{align}
  C_{\mathcal S_{1,2}}(\mu_b) &= \left[ \alpha_s(m_t) \over \alpha_s(\mu_b) \right]^{-\frac{12}{23}} \left[ \alpha_s(m_{\rm LQ}) \over \alpha_s(m_t) \right]^{-\frac{4}{7}}\, C_{\mathcal S_{1,2}} (m_{\rm LQ}) \,, \\ 
  C_{\mathcal T}(\mu_b) &=\left[ \alpha_s(m_t) \over \alpha_s(\mu_b) \right]^{\frac{4}{23}} \left[ \alpha_s(m_{\rm LQ}) \over \alpha_s(m_t) \right]^{\frac{4}{21}}\, C_{\mathcal T} (m_{\rm LQ}) \,, 
\end{align}
where $\alpha_s (\mu)$ is a running QCD coupling at a scale $\mu$. 
In the following study, we take $\mu_b=4.2$~GeV and the flavor observables are evaluated at this scale.

The branching ratios of $\bar B \to D^{(*)} \tau\bar\nu$ can be calculated, given hadronic form factors that are precisely estimated with use of the heavy quark effective theory. 
The formulae in terms of the helicity amplitudes are found, {\it e.g.}, in Refs.~\cite{Sakaki:2013bfa,Tanaka:2012nw}.

\subsection{Present bound from $\bar B \to D^{(*)} \tau\bar\nu$ and $\bar B \to X_s \nu\bar\nu$}
In Ref.~\cite{Sakaki:2013bfa}, a precise study has been done for the present constraints on the leptoquark bosons from $\bar B \to D^{(*)} \tau\bar\nu$ together with $\bar B \to X_s \nu\bar\nu$, 
which is also affected by $S_1$, ${\bm S}_3$, $V_2$, and ${\bm U}_3$ leptoquark bosons~\cite{Grossman:1995gt} with partly same combinations of the couplings~\cite{Sakaki:2013bfa}.  
The experimental upper limit on the inclusive branching ratio of $\bar B \to X_s \nu\bar\nu$ is given as 
\begin{align}
  \mathcal B (\bar B \to X_s \nu\bar\nu) < 6.4 \times 10^{-4} \,,
\end{align}
at the $90\%$ confidence level (CL) by the ALEPH collaboration~\cite{Barate:2000rc}. 
As an illustration for the bound from $\bar B \to D^{(*)} \tau\bar\nu$ and $\bar B \to X_s \nu\bar\nu$, we show the allowed range of the product of the couplings in Table~\ref{Tab:LQ_limit}. 
In this table, we assume that only one specific combination of the product, having a real or pure imaginary value,\footnote
{When the product of the couplings can be real and pure imaginary, we show only the real case. 
}
and one type of leptoquark bosons exist with its mass to be 1~TeV. 
We also neglect the couplings with $k\neq 3$ due to double Cabibbo suppressions. 
Namely, we keep only the leading terms proportional to $V_{33} = V_{tb}$ in Eqs.~(\ref{EQ:CV1})-(\ref{EQ:CT}). 
We can see that the ${\bm S}_3$ and ${\bm U}_3$ leptoquarks cannot satisfy both constraints from $\bar B \to D^{(*)} \tau\bar\nu$ and $\bar B \to X_s \nu\bar\nu$ at the same time.
The $V_2$ leptoquark has no way to explain the anomaly in $\bar B \to D^{(*)} \tau\bar\nu$. 
As for the $R_2$ and $U_1$ leptoquarks, the condition from $\bar B \to D^{(*)} \tau\bar\nu$ is fulfilled, whereas no constraint comes from $\bar B \to X_s \nu\bar\nu$. 
\begin{table}[t!]
\begin{center}
\begin{tabular}{ccc}
 \hline\hline       
   Leptoquark     &  $\bar B \to D^{(*)} \tau\bar\nu$		&  $\bar B \to X_s \nu\bar\nu$ \\
 \hline       
   $S_1$  	& $\begin{matrix} -0.87 < g_{1L}^{33} g_{1R}^{23*} < -0.54 \\ 
                    1.64 < |g_{1L}^{3i} g_{1R}^{23*}| < 1.81 \quad (i = 1, 2) \\
                    0.19 < g_{1L}^{33}g_{1L}^{23*} < 0.48,\,\,\, -5.59 < g_{1L}^{33}g_{1L}^{23*} < -5.87  \\
                    1.04 < |g_{1L}^{3i} g_{1L}^{23*}| < 1.67 \quad (i = 1, 2) \end{matrix}$ &  $|g_{1L}^{3i}g_{1L}^{2j*}| \lesssim 0.15$ \\
 \hline
   ${\bm S}_3$  	& $\begin{matrix} 0.19 < g_{3L}^{33}g_{3L}^{23*} < 0.48,\,\,\, -5.59 < g_{3L}^{33}g_{3L}^{23*} < -5.87  \\
                    1.04 < |g_{3L}^{3i} g_{1L}^{23*}| < 1.67 \quad (i = 1, 2) \end{matrix}$ 		& $|g_{3L}^{3i}g_{3L}^{2j*}| \lesssim 0.15$  \\
 \hline
   $R_2$  		& $1.64 < \left | \text{Im}( h_{2L}^{2i}h_{2R}^{33*}) \right | < 1.81$ 				& - \\
 \hline
   $V_2$  		& $g_{2L}^{3i}g_{2R}^{23*}$: no region within $2\sigma$ 				& $|g_{2L}^{3i}g_{2L}^{2j*}| \lesssim 0.07$ \\
 \hline
   $U_1$  		& $\begin{matrix} 0.10 < h_{1L}^{23}h_{1L}^{33*} < 0.24,\,\,\, -2.94 < h_{1L}^{23}h_{1L}^{33*} < -2.80 \\ 
                             0.52 < |h_{1L}^{2i}h_{1L}^{33*}| < 0.84 \quad (i = 1, 2) \\
                             h_{1L}^{2i}h_{1R}^{33*} \text{: no region within } 2\sigma \end{matrix}$ & -  \\
 \hline
   ${\bm U}_3$ 	& \hspace{2em} $\begin{matrix} 0.10 < h_{3L}^{23}h_{3L}^{33*} < 0.24,\,\,\, -2.94 < h_{3L}^{23}h_{3L}^{33*} < -2.80 \\ 
                             0.52 < |h_{3L}^{2i}h_{3L}^{33*}| < 0.84 \quad (i = 1, 2)  \end{matrix}$ \hspace{2em} & \hspace{2em}$|h_{3L}^{2i}h_{3L}^{3j*}| \lesssim 0.04$\hspace{2em} \\
 \hline\hline
\end{tabular}
\caption{
Allowed ranges for the products of leptoquark couplings assuming nonzero value in only one specific product of the couplings and zero in the others, at the leptoquark mass to be 1~TeV. 
The values are $2\sigma$ boundaries of the allowed region for the $\bar B \to D^{(*)} \tau\bar\nu$ case. 
The constraints from $\bar B \to X_s \nu\bar\nu$ are presented at $90\%$ CL which can be applied for each possible combinations of fermion generation $(i,j)$.
Here, we assume that the product of the couplings is real or pure imaginary.
When the value can be real and pure imaginary, we show only the real case.
}
\label{Tab:LQ_limit}
\end{center}
\end{table}

A further more interesting result is obtained in the $S_1$ leptoquark case as follows. 
The allowed region for $g_{1L}^{3i}g_{1L}^{23*}$ from $\bar B \to D^{(*)} \tau\bar\nu$ is inconsistent with that for $|g_{1L}^{3i}g_{1L}^{2j*}|$ from $\bar B \to X_s \nu\bar\nu$. 
On the other hand, the $S_1$ leptoquark boson can satisfy both of the constraints, 
in the case that $g_{1L}^{2j}$ is sufficiently small and the product $g_{1L}^{3i} g_{1R}^{23*}$ has $O(1)$ magnitude (for $M_{S_1}= O(1)\,\text{TeV}$). 
In particular,  when $g_{1L}^{3i} g_{1R}^{23*}$ is real, the best fit value to explain the anomaly is given as 
\begin{align}
 \label{EQ:condition}
 \frac{g_{1L}^{3i} g_{1R}^{23*} }{ 2M_{S_1}^2}  \,\simeq\,
 \begin{cases}
  - 0.26\, C_\text{SM} & \text{for $i=3$} \\
  \pm 0.64\, C_\text{SM} & \text{for $i \neq 3$} \,\, (i = 1 \text{ or } 2)
 \end{cases} \,,
\end{align}
where $C_\text{SM}$ is defined in Eq.~(\ref{EQ:CVsm}) and the other couplings are assumed to be zero.  
This means that $25\%$ of the SM contribution is required for the case of $i=3$.
In the case of $i = 1 \text{ or } 2$, the sign of the right-hand side of Eq.~(\ref{EQ:condition}) is not determined.
Also, this sign does not affect the physics discussed in this paper since no interference term appear in the decay sequence of $S_{1}$ in collider. 
Such a large effect, {\it motivated by the flavor anomaly}, can be significant at the collider search and thus will be studied below. 
In the following, we focus on the $S_1$ leptoquark boson and study the collider phenomenology at the LHC with keeping the condition to explain the anomaly in $\bar B \to D^{(*)} \tau\bar\nu$.

\section{Collider study}
\label{Sec:LHCstudy}
In general, the leptoquark model contains a lot of interaction terms to quarks and leptons and thus there are many possible signals for a collider search. 
Given the condition in Eq.~(\ref{EQ:condition}) motivated by the anomaly in $\bar B \to D^{(*)} \tau\bar\nu$, the minimal setup is 
\begin{align}
 \label{EQ:minimal}
 g_{1L}^{3i} \neq 0, \quad g_{1R}^{23} \neq 0, \quad \text{others} = 0  \,, 
\end{align}
namely, nonzero couplings only in the terms $\bar Q_{L}^{c,3} i\sigma_2 L_{L}^i S_1$ and $\bar c_{R}^{c} \tau_{R} S_1$  {(and their Hermitian conjugates)}. 
In our study, we obey this setup and thus consider the phenomenology for the decays $S_1^* \to t \ell$, $b \nu_\ell$ and $c \tau$ at the LHC. 
As is the case in the previous section, we ignore the doubly-Cabibbo-suppressed terms from the CKM matrix elements and consider only the $V_{33} = V_{tb}$ terms of Eqs.~(\ref{EQ:CS2}) and (\ref{EQ:CT}) in the following paper.

\subsection{Production process}

\begin{figure}[t]
\begin{center}
\includegraphics[viewport=0 0 400 461, width=18em]{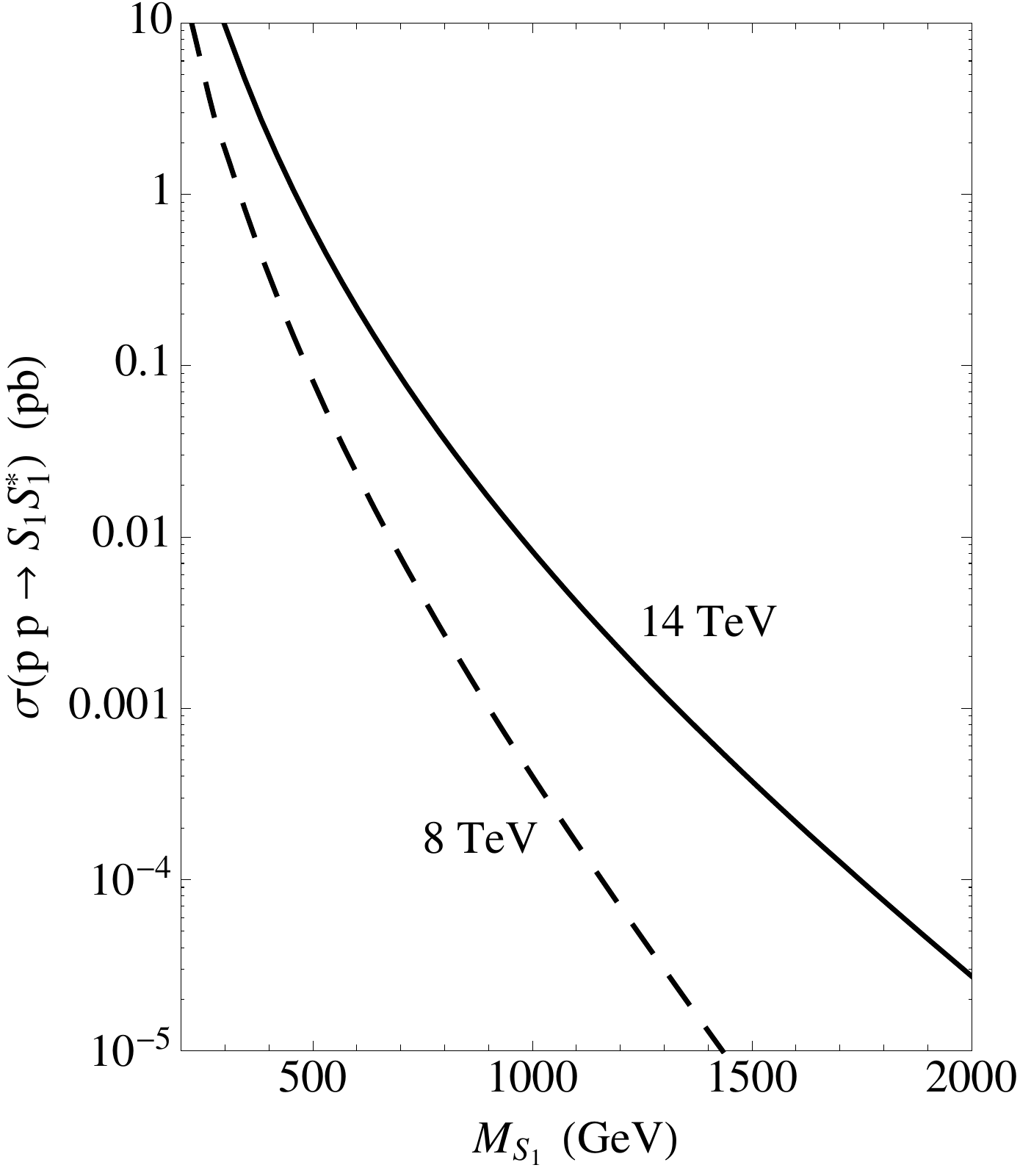} \quad
\includegraphics[viewport=0 0 400 478, width=18em]{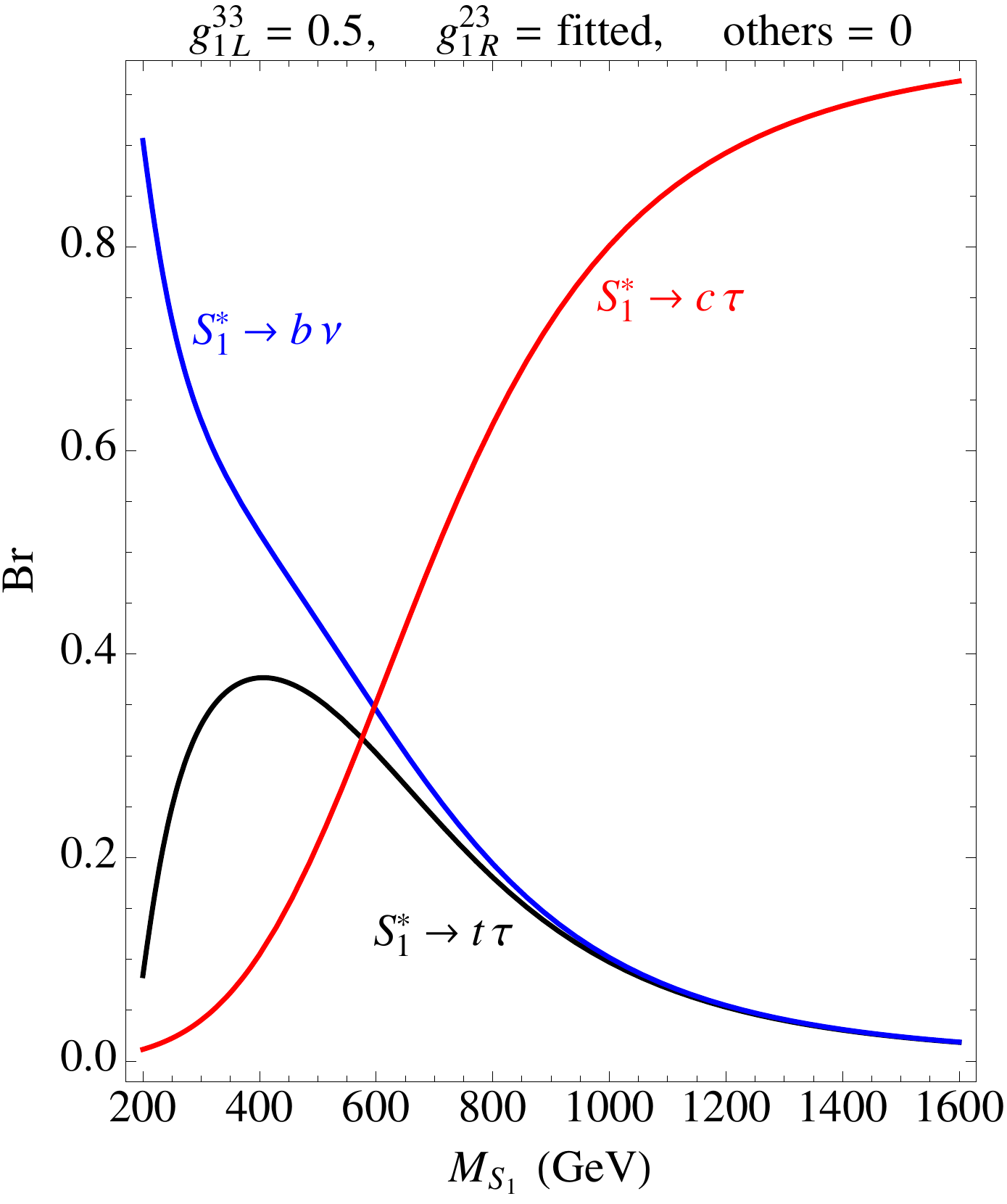}
\caption{
Pair production cross sections (left) and decay branching ratios (right) of the  {$S_{1}$} leptoquark boson as a function of its mass. 
The NLO cross sections at 8 and 14 TeV are shown as indicated by the legend in the left figure. 
The branching ratios for $S_1^* \to t \tau$, $S_1^* \to b \nu_{\tau}$, and $S_1^* \to c \tau$ are denoted by black, blue, and red curves in the right figure, respectively. 
We take $g_{1L}^{33}=0.5$ and $g_{1R}^{23}$ is fixed by following Eq.~(\ref{EQ:condition}) so as to explain $R(D)$ and $R(D^*)$ simultaneously. 
}
\label{Fig:CsNLO_Br}
\end{center}
\end{figure}
Since a leptoquark boson has $SU(3)$ color charge, it is expected that a pair production of leptoquark bosons by the QCD interaction is significant. 
We note that the QCD pair production does not depend on the couplings defined in Eqs.~(\ref{EQ:LagLQ1})-(\ref{EQ:LagLQ2}).  
In this paper, we investigate the pair-produced leptoquark bosons by QCD at the LHC.\footnote
{A $t$-channel exchange of a lepton can also produce a pair of leptoquark bosons by the couplings in Eqs.~(\ref{EQ:LagLQ1})-(\ref{EQ:LagLQ2}). 
This contribution is however much suppressed unless the couplings are very large such as $g_{1L}^{11} \sim 2$, {\it e.g.}, see Ref.~\cite{Dorsner:2014axa}. 
When the leptoquark couplings are much larger, single production in association with a lepton becomes important as well~\cite{Hammett:2015sea,Mandal:2015vfa}. 
On the other hand, in our configuration, only the charm, bottom and top quarks appear through the leptoquark interactions, 
which are highly parton distribution function (PDF) suppressed or do not exist as a parton when $\sqrt{s} = 8\,\text{or}\,14\,\text{TeV}$.
Thereby, only the QCD pair production is relevant in our setup even when the couplings are $g_{1L}^{3i}, g_{1R}^{23} \sim 2$. 
}

Thus, our target signal at the LHC is produced through $p p \to S_1 S_1^*$, where $p$ indicates a proton.  
The production cross section in the leptoquark model has been evaluated at the next-to-leading order (NLO)~\cite{Beenakker:1996ed,Kramer:2004df,Hepdata}.
With the use of {\tt Prospino2.1}~\cite{Beenakker:1996ed,Prospino2}, we show the plot for $\sigma (p p \to S_1 S_1^*)$ as a function of $M_{S_1}$ at $\sqrt s =8$ and $14$~TeV in Fig.~\ref{Fig:CsNLO_Br}.

\subsection{Decay process}
In the minimal setup for our study, the possible decay processes are $S_1^* \to t \ell^i$, $b \nu_{\ell^i}$ for $g_{1L}^{3i}\neq 0$ and $S_1^* \to c \tau$ for $g_{1R}^{23} \neq 0$, 
where we define $\ell^1=e$, $\ell^2=\mu$, and $\ell^3=\tau$. 
To see the feature, we show the branching ratios for these three decay modes for $g_{1L}^{33} = 0.5$ in Fig.~\ref{Fig:CsNLO_Br} as an example. 
Here, the coupling $g_{1R}^{23}$ is automatically fixed as the relation in Eq.~(\ref{EQ:condition}), namely, $g_{1R}^{23}= - 0.52 \, C_\text{SM} M_{S_1}^2 \big/ g_{1L}^{33}$. 
The decay branch $S_{1}^{\ast} \to c \tau$ becomes the dominant one for $S_{1}$ with a large mass.

Therefore there are six final states of the signal event from the pair production for each lepton generation $\ell^i$. 
The final states can be categorized by two part (here we omit the particle/anti-particle assignment):  
\begin{itemize}
	\item independent on the flavor of $\ell$: \quad $(b \nu_\ell) (b \nu_\ell)$, \quad $(c \tau) (c \tau)$, \quad $(b \nu_\ell) (c \tau)$.  	
	\item dependent on the flavor of $\ell$: \quad $(t \ell) (t \ell)$, \quad $(t \ell) (b \nu_\ell)$, \quad $(t \ell) (c \tau)$.  
\end{itemize}
The final states in the former category are independent on the choice of $\ell$, and thus can be analyzed without specifying $\ell$. 
As for the latter category, on the other hand, it is required to investigate every lepton flavor due to differences in the efficiency, acceptance, and tagging methods.

\subsection{Current status}
\label{Sec:curentbounds}

\subsubsection{$(b \nu_\ell) (\bar{b}\bar{\nu}_\ell)$ and $(t \tau) (\bar{t}\bar{\tau})$}
Up to the present, there exist two CMS and ATLAS searches which can be applied to the final states of $(b \nu_\ell) (\bar{b}\bar{\nu}_\ell)$ for the LHC run~I. 
In Refs.~\cite{Aad:2013ija,Khachatryan:2015wza}, the ATLAS and CMS collaborations have searched for the third-generation squarks 
and obtained exclusion limit in terms of the lightest bottom squark ($\tilde{b}_{1}$) and lightest neutralino ($\tilde{\chi}^0_1$) masses, 
where the final state is $(b  \tilde{\chi}^0_1) (\bar{b} \tilde{\chi}^{0}_1)$ with zero or more jets. 
Results obtained for $M_{\tilde{\chi}^0_1}=0$ can be directly translated into results for $(b \nu_\ell) (\bar{b}\bar{\nu}_\ell)$ in the scalar leptoquark model. 
The CMS analysis in Ref.~\cite{Khachatryan:2015wza} gives the observed limit on the branching ratio for $\text{LQ} \to b \nu_{\ell}$. 
On the other hand, a direct bound on third generation leptoquarks through the $(b \nu_\ell) (\bar{b}\bar{\nu}_\ell)$ channel was provided by ATLAS~\cite{Aad:2015caa}. 
In addition, results of the bottom squark search at the 13~TeV LHC have been recently reported by the ATLAS collaboration~\cite{ATLAS-CONF-2015-066}. 
However, since this report lacks information for the observed limit on the cross section, we only obtain a rough bound for the leptoquark case as shown below. 
In Ref.~\cite{Khachatryan:2015bsa}, the CMS collaboration has also analyzed the pair production of third-generation scalar leptoquarks decaying into $(t \tau) (\bar{t}\bar{\tau})$.

\begin{figure}[t]
\begin{center}
\includegraphics[viewport=0 0 360 370, width=18em]{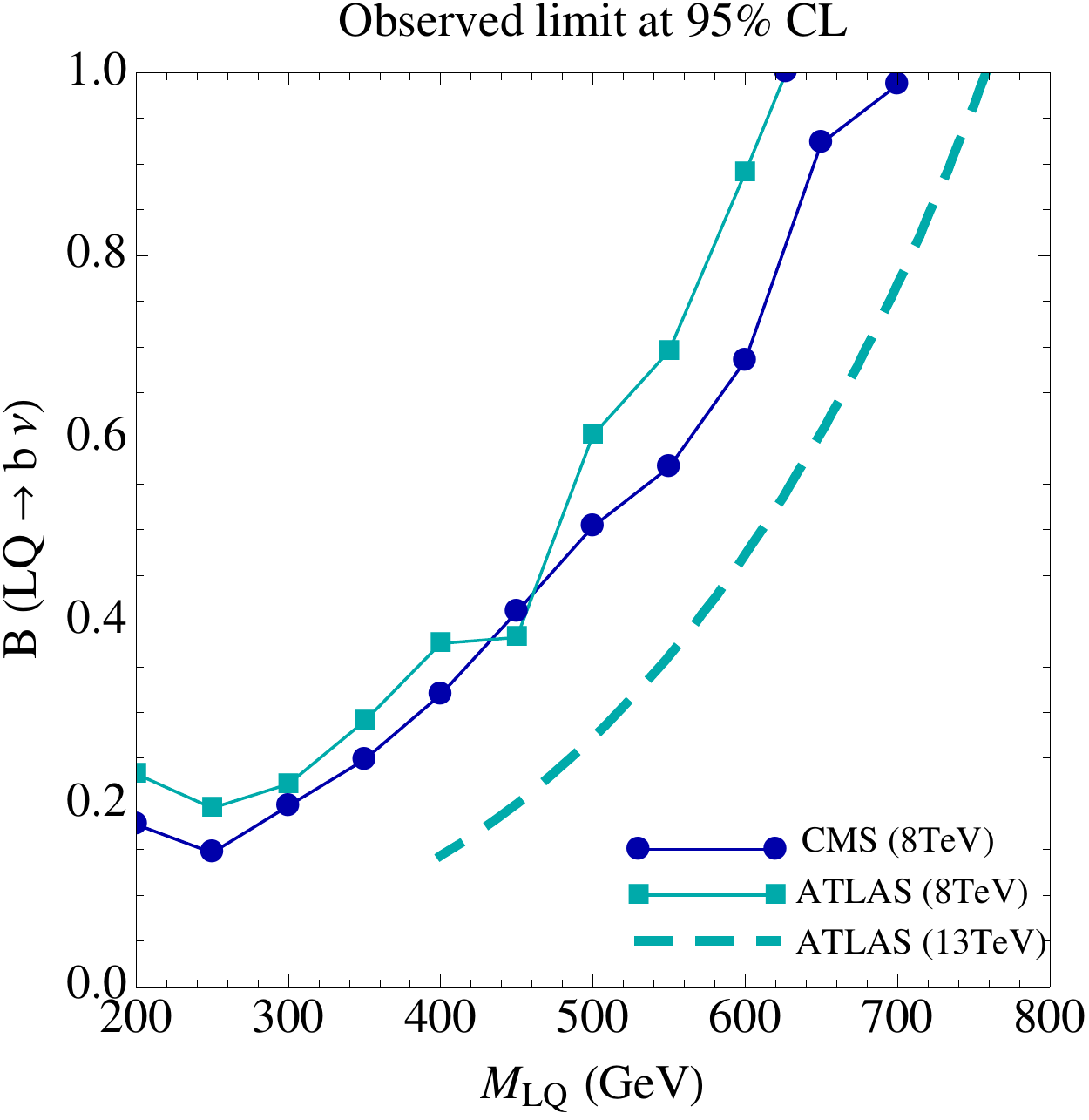} \quad\quad
\includegraphics[viewport=0 0 360 370, width=18em]{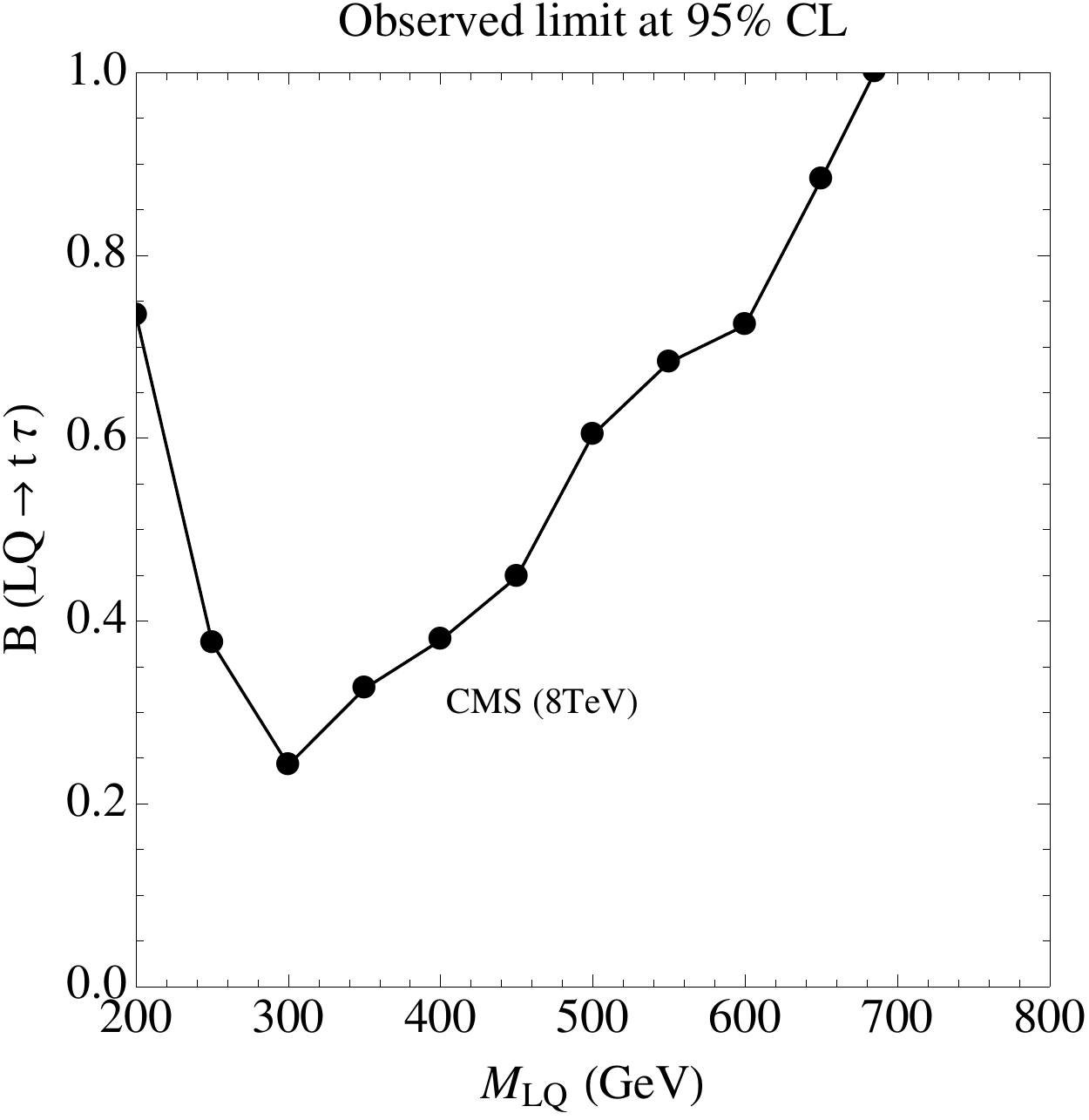}\\[-0.7em]
\hspace{1.3em} (a) \hspace{18.5em} (b)
\caption{
Observed upper limits on the branching ratio at 95\% CL for (a) $\text{LQ} \to b \nu_\ell$ from the CMS (blue) and ATLAS (cyan) analyses, and (b) $\text{LQ} \to t \tau$ obtained from the CMS analysis. 
}
\label{Fig:LimitLQ}
\end{center}
\end{figure}
In Fig.~\ref{Fig:LimitLQ}, we show the exclusion plot for $\mathcal B (\text{LQ} \to b \nu_\ell)$ and $\mathcal B (\text{LQ} \to t \tau)$ as a function of the LQ mass, where LQ indicates an arbitrary scalar leptoquark boson.  
The result from the ATLAS search is translated from the one in Ref.~\cite{Aad:2013ija}, 
by taking into account the NLO cross section of LQ pair production~\cite{Hepdata} and by assuming the narrow width approximation for the total decay width of LQ.
We confirmed that our interpretation from the ATLAS bottom squark search is close to the ATLAS official bound in Ref.~\cite{Aad:2013ija}. 
Note that the 13~TeV recast shown in the figure is estimated 
by obtaining the observed limit on the cross section as $\sigma(pp \to \tilde{b}_{1}\, \tilde{b}_{1}) \simeq 22.8\,\text{fb}$ at the $95\%$ CL exclusion point~\cite{ATLAS-CONF-2015-066} 
and then applying it to the leptoquark case. 
In this rough estimation, the mass dependence on the observed limit is neglected since such information is not available in this report.
Hence, this estimation should not be applied to the small LQ mass region less than around $400\,\text{GeV}$ because the acceptance times efficiency can be drastically changed in this region.

\subsubsection{$(c \tau) (\bar{c}\bar{\tau})$}
\label{Sec:current_ctauctau}

\begin{figure}[t]
\begin{center}
\includegraphics[viewport=0 0 360 370, width=18em]{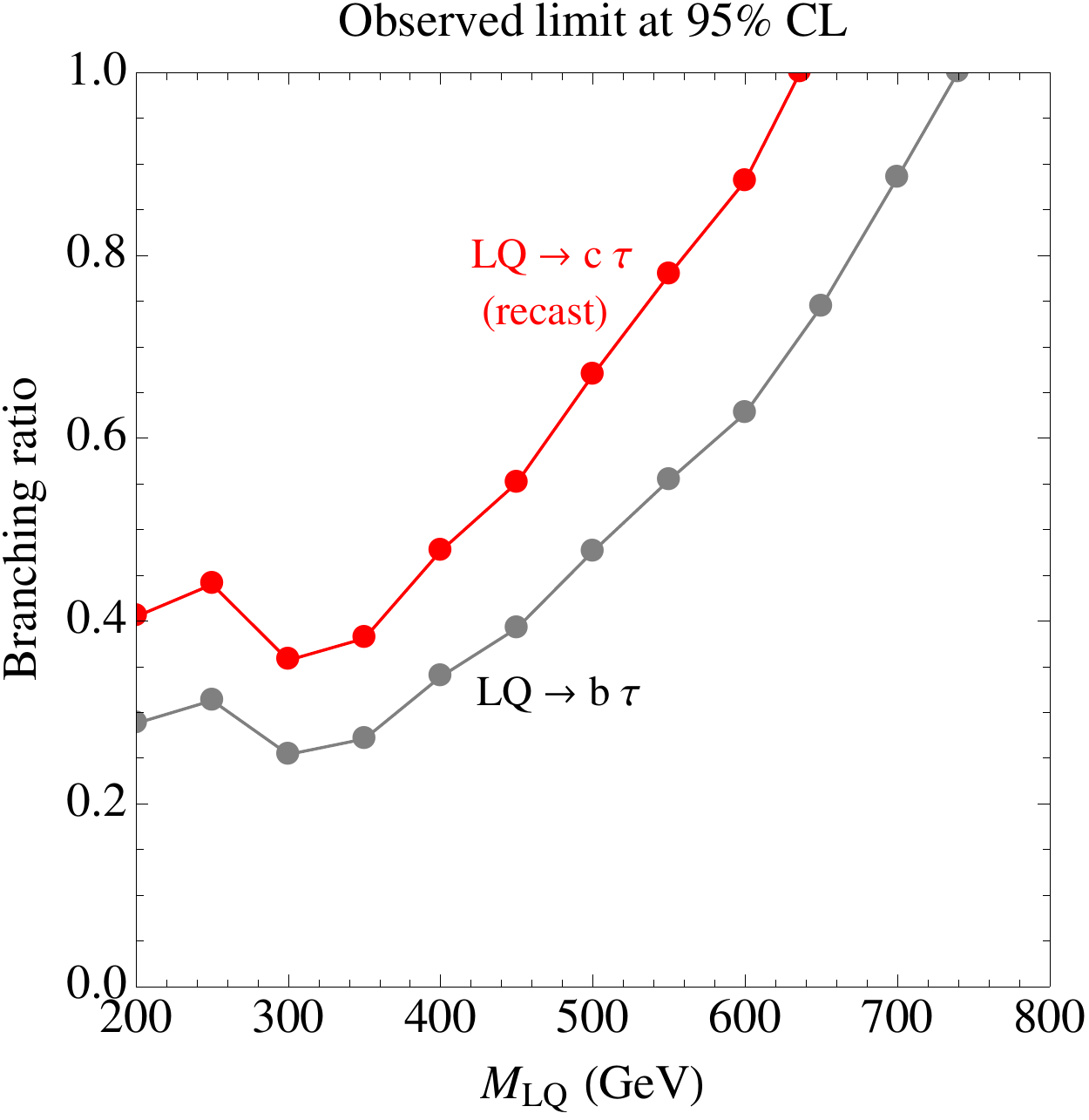}
\caption{
Observed upper limits on the branching ratio at 95\% CL for $\text{LQ} \to b \tau$ (gray) and $\text{LQ} \to c \tau$ (red) as a function of the leptoquark mass. 
}
\label{Fig:LimitLQctau}
\end{center}
\end{figure}
There is a CMS search for the pair-produced scalar leptoquarks decaying to $(b \tau) (\bar{b}\bar{\tau})$~\cite{Khachatryan:2014ura}. 
It is possible to reinterpret this result to put a constraint on the leptoquark boson decays into $(c \tau) (\bar{c}\bar{\tau})$, 
since $c$-jets are close cousins of $b$-jets, and the $b$-tagging algorithms actually have a reasonably high probability of tagging a $c$-jet as a $b$-jet (mis-tagging).\footnote
{Note that similar discussions are found in how to measure the charm Yukawa coupling to the Higgs boson in Refs.~\cite{Delaunay:2013pja,Perez:2015aoa,Perez:2015lra}.
}
For this, however, it is necessary to quantify the probability of mis-identifying $c$-jets as being $b$-jets.

In this analysis, jets are $b$-tagged using the combined secondary vertex (CSV) algorithm with the loose operating point (CSVL). 
Furthermore, only one jet is required to be $b$-tagged, while the second one is selected whether or not it is $b$-tagged. 
The latest preliminary note on $b$-tagging at $\sqrt{s} = 8$~TeV is obtained in Ref.~\cite{CMS:2013vea} but does not contain the information we need. 
However, tagging and mis-tagging efficiencies for the CSVL can be found in the $\sqrt{s} = 7$~TeV $b$-tagging paper~\cite{Chatrchyan:2012jua}. 
There, we find
\begin{align}
 \varepsilon_\text{CSVL}^\text{$b$-jet} = 85\% \,, \quad \varepsilon_\text{CSVL}^\text{$c$-jet} = 45\% \,.
\end{align}
The CMS analysis has two relevant signal regions: $e\tau_\text{h}$ and $\mu\tau_\text{h}$, targeting final states with two $\tau$ leptons, one decaying hadronically and the other leptonically. 
In each of these two signal regions the number of expected events per integrated luminosity $\mathcal{L}$ for a scalar LQ boson decaying into $c\tau$ is given by
\begin{equation}
 n_{\text{LQ} \to c\tau} \big/ \mathcal{L} = 
 \sigma_{p p \to \text{LQ}\, \text{LQ}^*} \times (A \times \varepsilon)_{\text{LQ} \to b\tau} 
 \times 
 \frac{\varepsilon_\text{CSVL}^\text{$c$-jet}}{\varepsilon_\text{CSVL}^\text{$b$-jet}} 
 \approx 0.53\, \sigma_{p p \to \text{LQ} \text{LQ}^*} \times (A \times \varepsilon)_{\text{LQ} \to b\tau} \,,
\end{equation}
where $(A \times \varepsilon)_{\text{LQ} \to b\tau}$ is the acceptance times efficiency of the selection criteria.
As the nature of the jet has very little influence on the acceptance times efficiency, apart from the tagging requirement, 
the factor $\varepsilon_\text{CSVL}^\text{$c$-jet} \Big/ \varepsilon_\text{CSVL}^\text{$b$-jet}$ can be considered as a rescaling factor for the cross section. 
Therefore, it is straightforward to recast the results in Ref.~\cite{Khachatryan:2014ura} for $(c \tau) (\bar{c} \bar{\tau})$. 
In Fig.~\ref{Fig:LimitLQctau}, we show the exclusion plot for $\mathcal B (\text{LQ} \to b \tau)$ and $\mathcal B (\text{LQ} \to c \tau)$.

\subsubsection{Constraint on $S_1$ leptoquark model}

\begin{figure}[t!]
\begin{center}
\includegraphics[viewport=0 0 360 367, width=18em]{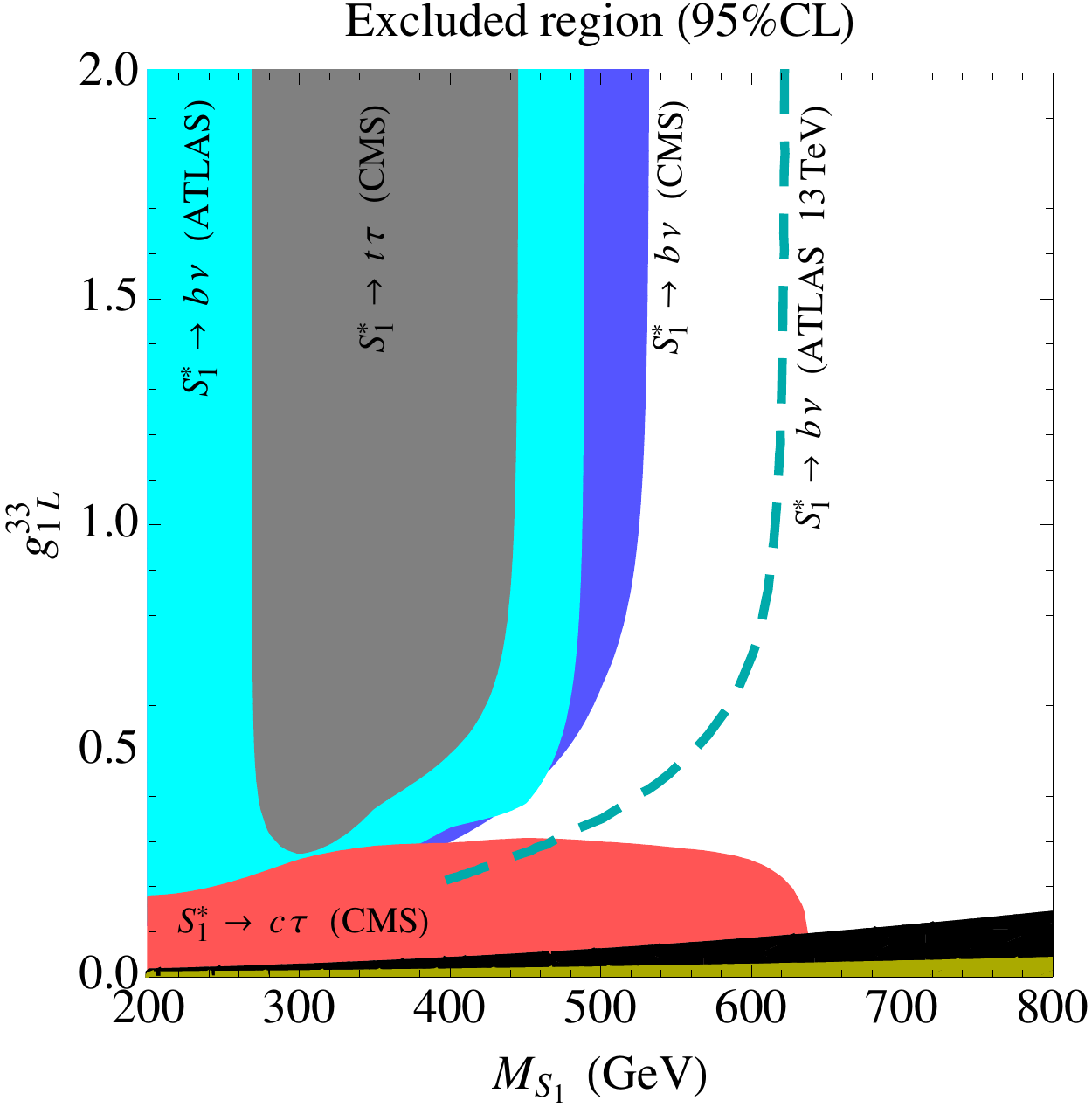} \quad
\includegraphics[viewport=0 0 360 367, width=18em]{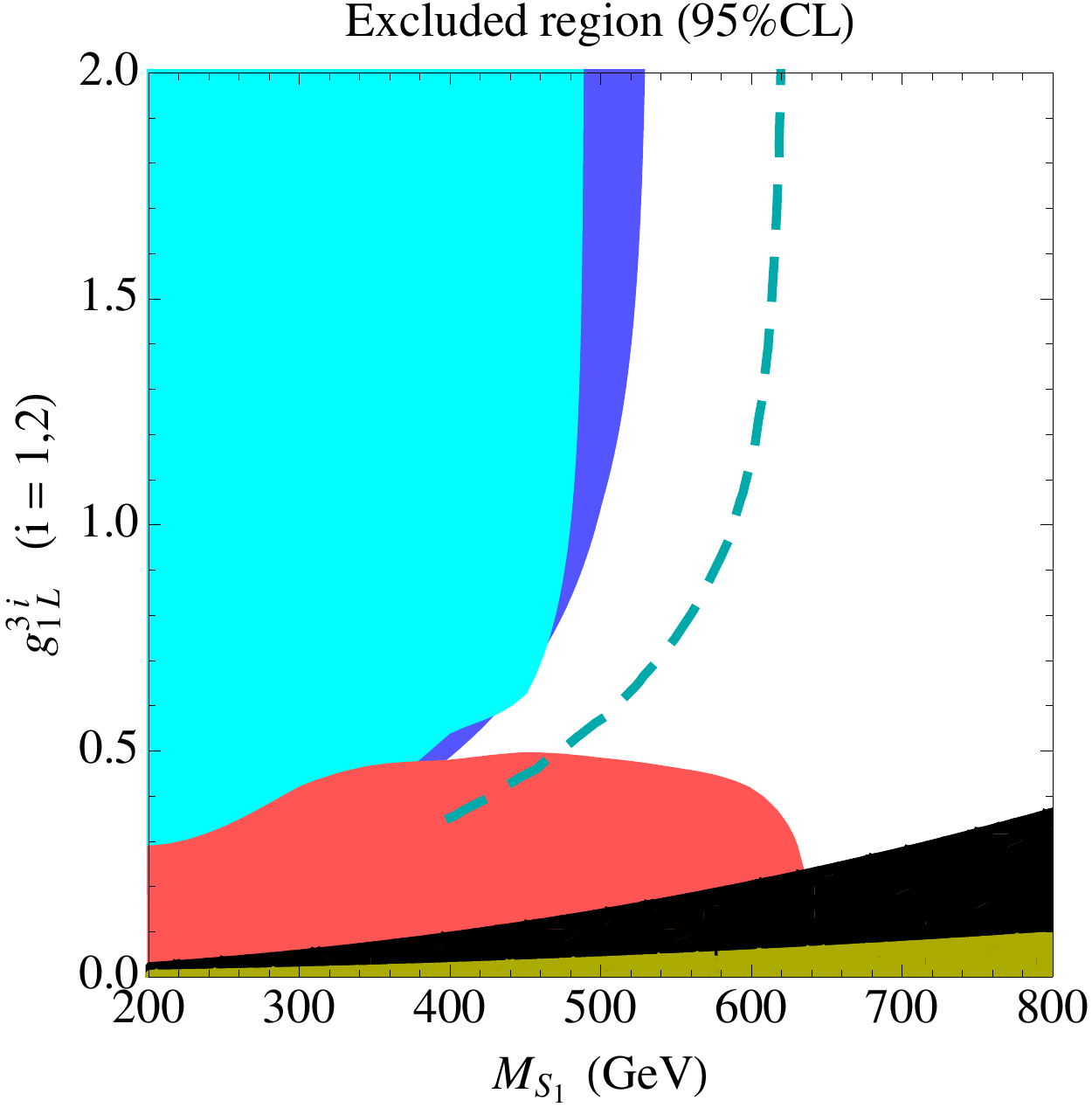} 
\caption{
(left) An excluded region plot in the $(M_{S_1}, g_{1L}^{33})$ plane for the $S_1$ leptoquark model, 
obtained by assuming that $g_{1R}^{23} = - 0.52 \, C_\text{SM} M_{S_1}^2 \big/ g_{1L}^{33}$ and the other couplings are zero; 
(right) a plot in the $(M_{S_1}, g_{1L}^{3i})$ plane for $i=1$ or $2$ by assuming that $g_{1R}^{23} = - 1.28 \, C_\text{SM} M_{S_1}^2 \big/ g_{1L}^{3i}$ and others are zero.  
Each colored region is excluded from the ATLAS or CMS analyses for the decay modes as exhibited in the legend. 
In the {black} region, the ratio of the width to mass of $S_1$ boson becomes larger than $0.2$, where the narrow width approximation does not work correctly. 
The dark yellow color shows the region for $g_{1R}^{23} > 4\pi$. 
}
\label{Fig:LimitS1}
\end{center}
\end{figure}
We can apply the present limits on the branching ratios shown above to the specific model. 
For the $S_1$ leptoquark with the minimal setup of Eq.~(\ref{EQ:minimal}), the branching ratios for $S_1^* \to t \ell^i$, $b \nu_{\ell^i}$, and $c \tau$ are controlled by $g_{1L}^{3i}$, $g_{1R}^{23}$, and $M_{S_1}$. 
If we take $g_{1L}^{3i}=0$ for $i =1,2$ and keep the condition in Eq.~(\ref{EQ:condition}), two of $g_{1L}^{33}$, $g_{1R}^{23}$, and $M_{S_1}$ remain free parameters. 
The excluded region in the $(M_{S_1}, g_{1L}^{33})$ plane for this case is given in Fig.~\ref{Fig:LimitS1}, where the coupling $g_{1R}^{23}$ is fixed as $g_{1R}^{23}= - 0.52 \, C_\text{SM} M_{S_1}^2 \big/ g_{1L}^{33}$. 
The colored regions are excluded from the corresponding searches at ATLAS or CMS as denoted in the figure. 
We can see that $M_{S_1} < 530\,\text{GeV}$ and $M_{S_1} < 640\,\text{GeV}$ are ruled out for $g_{1L}^{33} \gtrsim 0.5$ and $g_{1L}^{33} \lesssim 0.2$, respectively from the 8~TeV LHC searches. 
The rough estimate for $S_1^* \to b \nu_{\ell^i}$ from the 13~TeV analysis is also shown with the dashed line. 
In this setup, for a small $g_{1L}^{33}$ and a large $M_{S_1}$, the coupling $g_{1R}^{23}$ and the total decay width $\Gamma_{S_1}$ can be large. 
Thus we show the regions for $g_{1R}^{23} > 4\pi$ and $\Gamma_{S_1}/M_{S_1} > 0.2$ with dark yellow and black colors, respectively. 
The right panel in Fig.~\ref{Fig:LimitS1} shows the exclusion in the $(M_{S_1}, g_{1L}^{3i})$ plane for $i = 1$ or $2$ 
with the condition $g_{1R}^{23} = - 1.28 \, C_\text{SM} M_{S_1}^2 \big/ g_{1L}^{3i}$ assuming the other couplings to be zero. 
In this case, the search for $S_1^* \to t \tau$ is irrelevant. 
To conclude, the white regions in the figure are totally allowed by both the 8~TeV LHC searches and the flavor observables in $\bar B \to D^{(*)} \tau\bar\nu$ and $\bar B \to X_s \nu\bar\nu$.

\section{Analysis at 14\,TeV LHC}
\label{Sec:Analysis}
Recently, the LHC run II successfully started at an energy of $13\,\text{TeV}$.
The updated LHC experiments at $13$ and $14\,\text{TeV}$ will greatly improve the discovery potential for the leptoquark models as well as many other new physics candidates.
In this section, we provide the detailed procedure of our analyses to obtain our numerical results at the 14~TeV LHC. 
Based on the analyses given in this section, prospects and results by simulations for our leptoquark model are shown in the next section. 
Our target signals for the analyses are $(b \nu_\ell)(\bar{b}\bar{\nu}_\ell)$ and $(c \tau)(\bar{c}\bar{\tau})$ from the $S_1^{(\ast)}$ pair production. 
Signal and background events are simulated in the cluster system provided at CTPU-IBS.

\subsection{$S_1^\ast \to b \nu$}
As already mentioned in the previous section, the event topology of the final state from $p p \to S_1^* S_1 \to (b \nu_\ell)\, (\bar{b} \bar{\nu}_\ell)$ is very similar to 
that from $p p \to \tilde{b}_1^* \tilde{b}_1 \to (b \tilde{\chi}_1^0)\, (\bar{b}  \tilde{\chi}_1^{0})$ in a supersymmetric (SUSY) model, where $\tilde{b}_1$ is the lightest bottom squark and $\tilde{\chi}_1^0$ is the lightest neutralino. 
Therefore, we can straightforwardly adopt the way of such kind of SUSY searches at the LHC in this category.
The ATLAS official prospects for this SUSY search at $14\,\text{TeV}$ were communicated in Ref.~\cite{ATL-PHYS-PUB-2014-010} 
assuming that each $\tilde{b}_1$ decays into $b \tilde{\chi}_1^0$ with a $100\%$ branching ratio. 
Details of analysis cuts are almost the same with the $8\,\text{TeV}$ analysis which gave the lower mass bound $\sim 650\,\text{GeV}$ for a massless $\tilde{\chi}_1^0$~\cite{Aad:2013ija}.
In our analysis for the $(b \nu_\ell)\, (\bar{b} \bar{\nu}_\ell)$ final states, we follow the method in Refs.~\cite{ATL-PHYS-PUB-2014-010,Aad:2013ija}.
Before proceeding with the leptoquark case, we reproduce the 14~TeV prospects for the {bottom squark} search reported in Ref.~\cite{ATL-PHYS-PUB-2014-010}, 
in order to verify our methodology and confirm our result to be robust.

\subsubsection{Procedure of our analysis}
\label{SubSec:procedure}

At first, we describe procedure of our event simulation and cut analysis. 
Later, we apply this procedure to the SUSY and $S_1$ leptoquark cases.

The final state of our targeting process is categorized as ``two $b$-jets with missing particles''. 
Trigger cuts for reconstructed objects are required to be $p_\text{T} > 20\,\text{GeV},\ |\eta| < 2.8$ for jets; $p_\text{T} > 7\,\text{GeV},\ |\eta| < 2.47$ for electrons; 
and $p_\text{T} > 6\,\text{GeV},\ |\eta| < 2.4$ for muons~\cite{Aad:2013ija}, where $p_\text{T}$ and $\eta$ are transverse momentum and pseudorapidity, respectively. 
After that, an isolation cut based on the distance between two objects, defined as $\Delta R = \sqrt{(\Delta \eta)^2 + (\Delta \phi)^2}$, is imposed on each pair of objects. 
The isolation $\Delta R > 0.2$ is required between jet and light lepton candidates to remove jet candidates, and then $\Delta R > 0.4$ is required afterward to remove light lepton candidates~\cite{Aad:2013ija}. 
Finally we also require a lepton veto.

\begin{table}[t]
\centering
\begin{tabular}{cc}
\hline
\hline
Category 											& Cut condition (in SRA) \\ 
\hline 
Lepton veto 										& no $e/\mu$ after the isolation \\ 
\hline
$E^\text{miss}_\text{T}$ 								& $> 150\,\text{GeV}$ \\ 
\hline
Leading jet $p_\text{T}\,(j_1)$ 							& $> 130\,\text{GeV}$ \\ 
\hline
Second jet $p_\text{T}\,(j_2)$ 							& $> 50\,\text{GeV}$ \\ 
\hline
Third jet $p_\text{T}\,(j_3)$ 							& veto if $> 50\,\text{GeV}$ \\ 
\hline
$b$-tagging 										& \begin{tabular}{c} for leading two jets, $n_{b\text{-jets}} = 2$ \\ ($p_\text{T} > 20\,\text{GeV},\ |\eta| < 2.5$) \end{tabular} \\
\hline
$\Delta \phi_\text{min}$ 								& $> 0.4$ \\ 
\hline
$E_\text{T}^\text{miss}/m_\text{eff}(k)$ 					& $> 0.25$ for $k=2$ \\ 
\hline
$m_{bb}$ 											& $> 200\,\text{GeV}$ \\ 
\hline
$m_\text{CT}$ 										& $> 300,\ 350,\ 450,\ 550,\ 650,\ 750\,\text{GeV}$ \\ 
\hline
\hline
\end{tabular}
\caption{Summary of the event selection cuts (in SRA) after the physics object reconstruction (trigger cuts and isolation), {based on Refs.}~\cite{ATL-PHYS-PUB-2014-010,Aad:2013ija}.}
\label{tab:SRA_cuts}
\end{table}

The above step is followed by event selection cuts for our analysis. 
We summarize it in Table~\ref{tab:SRA_cuts}. 
We require $E_\text{T}^\text{miss} > 150\,\text{GeV}$ for the missing transverse energy and $p_\text{T}\,(j_{1(2)}) > 150 \, (130)\,\text{GeV}$ for the leading (second) jet transverse momentum. 
The two leading jets are then required to be $b$-tagged. 
Events are discarded if any other additional jets are hard enough ($p_\text{T} > 50\,\text{GeV}$). 
For rejecting QCD multi-jet backgrounds, we use the two variables $\Delta \phi_\text{min}$ and $m_\text{eff}(k)$ which are defined as
\begin{align}
\Delta \phi_\text{min} &= \text{min}
\left(
|\phi_1 - \phi_{{\bf p}_\text{T}^\text{miss}}|,\
|\phi_2 - \phi_{{\bf p}_\text{T}^\text{miss}}|,\
|\phi_3 - \phi_{{\bf p}_\text{T}^\text{miss}}|
\right), \\
m_\text{eff}(k) &= \sum_{i=1}^{k} (p_\text{T}^\text{jet})_i + E_\text{T}^\text{miss}.
\end{align}
The variable $\Delta \phi_\text{min}$ describes the minimal azimuthal distance ($\Delta \phi$) between any of the three leading jets and the ${\bf p}_\text{T}^\text{miss}$ vector.  
The variable $m_\text{eff}(k)$ indicates the scalar sum of the $p_\text{T}$ up to the $k$-th leading jet and $E_\text{T}^\text{miss}$. 
They are required to satisfy the condition $\Delta \phi_\text{min}>0.4$ and $E_\text{T}^\text{miss}/m_\text{eff}(2) > 0.25$. 
The invariant mass of the two $b$-tagged jets $m_{bb}$ is used for suppressing backgrounds with two $b$-jets, 
(from single/double top productions and Z-bosons in association with heavy-flavor jets), required as $m_{bb}> 200\,\text{GeV}$.

As the final step, we adopt contransverse mass cuts for the signal region A (SRA)\footnote
{In the previous analysis~\cite{Aad:2013ija} at $8\,\text{TeV}$ by ATLAS, another signal region, SRB, targets scenarios with small mass splitting between the parent (bottom squark) and invisible-daughter (neutralino) particles. 
This is not the case for the $S_1$ leptoquark since the counterpart of the neutralino is the neutrino and the mass splitting is always large.
}
in Refs.~\cite{ATL-PHYS-PUB-2014-010,Aad:2013ija}, which is effective for the case of large mass splitting between parent and invisible-daughter particles in the decays, 
(corresponding to $\tilde{b}_1$ and $\tilde{\chi}_1^0$ for the SUSY case; $S_1$ and $\nu$ for the leptoquark case).  
The boost-corrected contransverse mass $m_\text{CT}$ is designed to measure the masses of pair-produced semi-invisibly decaying heavy particles~\cite{Tovey:2008ui,Polesello:2009rn}, and defined as 
\begin{align}
m_\text{CT}^2 
= \left[ E_\text{T}(v_1) + E_\text{T}(v_2) \right]^2 - \left[ {\bf p}_\text{T}(v_1) - {\bf p}_\text{T}(v_2) \right]^2,
\end{align}
for the case of two identical decays of heavy particles ($v_1$ and $v_2$) into two visible and invisible particles. 
As for the choice of $m_\text{CT}$ thresholds, the six subdivisions of SRA, such as $m_\text{CT} > 300,\ 350,\ 450,\ 550,\ 650,\ 750\,\text{GeV}$ as in Ref.~\cite{ATL-PHYS-PUB-2014-010}, are prepared in advance. 
Among them, an appropriate threshold is selected so that a signal significance is maximized for each model parameter point $(M_{\tilde{b}_1},\, M_{\tilde{\chi}_1^0})$.

\subsubsection{SUSY case}

\subsubsection*{\bf Computation method for signal event:}

To reproduce the result of $14\,\text{TeV}$ prospects in the MSSM, we utilize the default MSSM model file provided by {\tt FeynRules}~\cite{Christensen:2008py,Alloul:2013bka} to generate signal events. 
Since the production process $p p \to {\tilde{b}_1}^*\, \tilde{b}_1$ is produced by QCD interactions and $\mathcal B(\tilde{b}_1 \to b \tilde{\chi}^0_1)=100\%$ is assumed, 
relevant model parameters for the process $p p \to \tilde{b}_1 \tilde{b}_1^\ast \to b \bar{b} \tilde{\chi}_1^0 \tilde{\chi}_1^{0*}$ are the masses of bottom squark $M_{\tilde{b}_1}$ and neutralino $M_{\tilde{\chi}^0_1}$. 
Thus, we investigate (reproduce) discovery potentials and exclusion limits on the plane of $(M_{\tilde{b}_1}, M_{\tilde{\chi}^0_1})$ at the $14\,\text{TeV}$ LHC, 
setting all the other mass parameters as $10^6\,\text{GeV}$ to be decoupled.

For parton-level event generations, we use the event generator {\tt MadGraph5\_aMC@NLO} version 2.2.2~\cite{Alwall:2011uj,Alwall:2014hca} with the PDF set {\tt CTEQ6L}~\cite{Pumplin:2002vw}. 
At the $14\,\text{TeV}$ LHC, jets become harder and considering jet merging becomes more important.
In our setup, we examine merged events with one and two additional hard jet(s) in the $k_\text{T}$ MLM matching scheme~\cite{Hoche:2006ph,Mangano:2006rw,Alwall:2007fs,Alwall:2008qv} with $x_q^\text{cut} = M_{\tilde{b}_1}/4$.

The effects of parton-showering, hadronization, and jet merging are simulated by the {\tt pythia-pgs} package~\cite{Sjostrand:2006za} implemented in {\tt MadGraph5\_aMC@NLO}, 
and the resultant events are recorded in the {\tt StdHep} format. 
Detector simulations are performed using {\tt DelphesMA5tune}~\cite{Dumont:2014tja}, 
a modified version of {\tt Delphes\,3}~\cite{deFavereau:2013fsa} provided in the {\it expert mode} of {\tt MadAnalysis5}~\cite{Conte:2012fm,Conte:2014zja} version 1.1.11. 
In {\tt DelphesMA5tune}, jets are found with the help of the package {\tt Fastjet}~\cite{Cacciari:2005hq,Cacciari:2011ma}. 
We use the default configuration for jet finding written in the modified Delphes card ``{\tt delphesMA5tune\_card\_ATLAS\_05.tcl}'' obtained in Ref.~\cite{MA5PAD} 
(for the anti-$k_\text{T}$ algorithm~\cite{Cacciari:2008gp} with {\tt ParameterR} $=0.4$ and {\tt JetPTMin} $= 20.0$).

Cut analyses to obtain the acceptance times efficiency {$A \times \varepsilon$} and the exclusion limit (using $CL_s$ procedure~\cite{Read:2002hq}) are done 
by the {\it expert mode} of {\tt MadAnalysis5}~\cite{Conte:2012fm,Conte:2014zja,Dumont:2014tja}. 
The public analysis code of {\tt MadAnalysis5} for the process (top/bottom squarks search: 0 leptons + 2 b-jets)~\cite{MA5PAD_ATLAS-SUSY-2013-05} at $8\,\text{TeV}$ 
has been written by G.~Chalons and is obtained in the {\it Public Analysis Database}~\cite{MA5PAD}.
Note that the public code {\tt MctLib} which is available in Ref.~\cite{soft:MctLib} is used for calculating $m_{\text{CT}}$~\cite{Tovey:2008ui,Polesello:2009rn}.
We use this code with minimal modification for the $14\,\text{TeV}$ case by adding different choices in $m_{\text{CT}}$ as shown in Table~\ref{tab:SRA_cuts}.  
As for (mis-)tagging rates for $b$-jets, we used the $p_T$ and $|\eta|$-dependent $b$-tagging efficiencies considered in Ref.~\cite{ATL-PHYS-PUB-2014-010}.


The production cross section $\sigma_{pp \to \tilde{b}_1 \tilde{b}_1^\ast}$, necessary to evaluate the discovery and exclusion limits, 
is reported in Ref.~\cite{LHCSUSY8TeVstop} for $8\,\text{TeV}$ and Ref.~\cite{LHCSUSY14TeVstop} for $14\,\text{TeV}$. 
The public codes {\tt Prospino2.1}~\cite{Beenakker:1996ed,Prospino2} (NLO) and {\tt NLL-fast}~\cite{NLL-fast} (NLO and NLO + NLL) can also obtain the values. 
Those values were cross-checked using {\tt Prospino2.1}.

\begin{table}[t!]
\centering
\begin{tabular}{|c||c|c|c|c|c|c|}\hline
BG type & SRA300 & SRA350 & SRA450 & SRA550 & SRA650 & SRA750 \\ \hline
$t\bar{t}$    & $32.6\pm3.0$ & $14.8\pm2.0$ & $4.3\pm1.1$ & $1.5\pm0.7$ & $0.6\pm0.4$ & $0.29\pm0.29$ \\
single top & $146\pm12$ & $83\pm8$ & $41\pm6$ & $25\pm5$ & $12.7\pm3.2$ & $8.9\pm2.5$ \\ 
$Z+\text{jets}$ & $508\pm8$ & $249\pm5$ & $70.5\pm2.7$ & $23.1\pm1.5$ & $9.1\pm1.0$ & $4.1\pm0.7$ \\
$W+\text{jets}$ & $92\pm5$ & $44\pm4$ & $9.3\pm1.7$ & $2.9\pm0.9$ & $1.6\pm0.8$ & $0.9\pm0.6$ \\
others & $5.4\pm0.5$ & $3.3\pm0.4$ & $1.59\pm0.28$ & $0.50\pm0.16$ & $0.18\pm0.09$ & $0.15\pm0.08$ \\ \hline 
\end{tabular}
\caption{
Expected numbers of events for SM backgrounds with statistical errors for an integrated luminosity of $300\,\text{fb}^{-1}$ at $14\,\text{TeV}$ from table 11 of Ref.~\cite{ATL-PHYS-PUB-2014-010}. 
The SRA regions are selected as $m_\text{CT} > 300,\ 350,\ 450,\ 550,\ 650,\ 750\,\text{GeV}$. 
}
\label{tab:14TeV_ATLAS_BGcutflow}
\end{table}

\subsubsection*{\bf Background event:}
Expected numbers of events for SM backgrounds with a $300\,\text{fb}^{-1}$ integrated luminosity at $14\,\text{TeV}$ have been already simulated in Ref.~\cite{ATL-PHYS-PUB-2014-010}. 
The relevant processes are $t\bar{t}$, single top, $Z+\text{jets}$, $W+\text{jets}$, and others. 
The expected numbers, with statistic uncertainties, are shown for SRA in every region of $m_{\text{CT}}$ in Table~\ref{tab:14TeV_ATLAS_BGcutflow}. 
We adopt the total uncertainties as used in the analysis of Ref.~\cite{ATL-PHYS-PUB-2014-010} and do not consider the pileup effect.

\subsubsection*{\bf Test analysis:}

\begin{figure}[t!]
\centering
\includegraphics[viewport=0 0 360 385, width=22em]{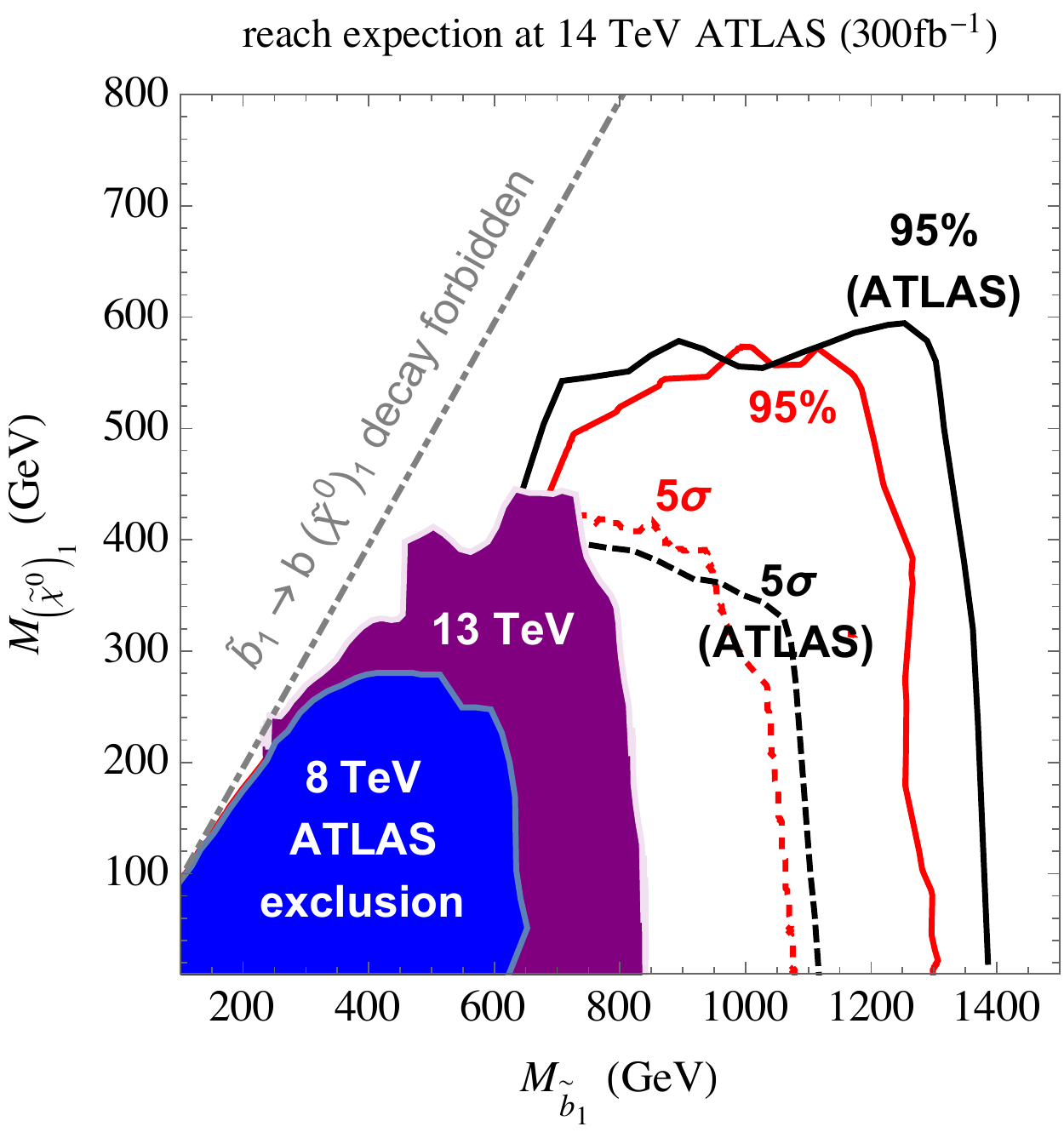}
\caption{
The expected $95\%$ CL exclusion boundary (solid lines) and the $5\sigma$ discovery reach (dashed lines) for the bottom squark pair production with $300\,\text{fb}^{-1}$ of integrated luminosity at $14\,\text{TeV}$. 
Our evaluation and the ATLAS official report~\cite{ATL-PHYS-PUB-2014-010} are shown with red and black colors, respectively. 
The ATLAS detector system was adopted in our evaluation. 
The covered region with blue (purple) color was already excluded by the $8\,\text{TeV}$ ($13\,\text{TeV}$) ATLAS analysis 
based on the data with $20.1\,\text{fb}^{-1}$ ($3.2\,\text{fb}^{-1}$) integrated luminosity~\cite{Aad:2013ija,ATLAS-CONF-2015-066}.
}
\label{fig:14TeV_SUSY_reach}
\end{figure}
Finally, we estimate the ranges of $95\%$ CL exclusion, using the $CL_s$ procedure, and of $5\sigma$ discovery in this SUSY case.
The result is shown in Fig.~\ref{fig:14TeV_SUSY_reach} along with the ATLAS official result. 
One can see that the small differences of around $50 \sim 100\,\text{GeV}$ between our result and the ATLAS official one are found in the $(M_{\tilde{b}_1},\, M_{\tilde{\chi}_1^0})$ plane.
This amount of deviations would be expected from a difference between a simplified analysis and a full calculation.
Thereby, we can conclude that our method in the analysis and simulations are reasonably good and reliable.

\subsubsection{Leptoquark case}
In the case of the $S_1$ leptoquark, the signal events from the process $p p \to S_1^* S_1 \to (b \nu_\ell)\, (\bar{b} \bar{\nu}_\ell)$ are generated by {\tt MadGraph5\_aMC@NLO} as well, 
where we have implemented the model file of the $S_1$ leptoquark with the help of {\tt FeynRules} and {\tt UFO} format~\cite{Degrande:2011ua}.
We remind that the relevant free parameter for the production process $p p \to S_1 S_1^\ast$ is only the $S_1$ mass $M_{S_1}$.  
Exclusion and discovery limits as a function of the branching ratio $\mathcal B(S_1^* \to b \bar\nu_\ell)$ and $M_{S_1}$ are subsequently derived. 
Then, we follow the same steps with the SUSY case for the parton-showering, hadronization, jet merging, and detector simulations, through {\tt pythia-pgs} and {\tt DelphesMA5tune}.
In the leptoquark case, we adopt the PDF {\tt NN23LO1}~\cite{Ball:2012cx} in parton-level event generations.
As for the cut analysis, we apply the same procedure as in Table~\ref{tab:SRA_cuts} to the signal events for the $S_1$ leptoquark, that is, an appropriate SRA region is automatically imposed by {\tt MadAnalysis5}. 
The LQ pair production cross section is evaluated by {\tt Prospino2.1}~\cite{Beenakker:1996ed,Prospino2} at NLO, (which has also been computed in Ref.~\cite{Mandal:2015lca}). 
We employ the SM background events and its total uncertainties as provided in Ref.~\cite{ATL-PHYS-PUB-2014-010} for the present case. 
The pileup effect is not considered as well.

\subsection{$S_1^\ast \to c \tau$}
To confirm that the $S_1$ leptoquark boson is the origin of the anomaly in $\bar B \to D^{(*)} \tau \bar\nu$, 
we should observe the non-zeroness of the couplings $g_{1L}^{3i}$ and $g_{1R}^{23\ast}$ ($i=1,\,2,\, \text{or}\ 3$). 
As shown in the previous subsection, we can probe the contribution of $g_{1L}^{3i} \neq 0$ through the $S_1$ search in the $(b \nu_\ell)\, (\bar{b} \bar{\nu}_\ell)$ final state.
On the other hand, we need to investigate the decay $S_1^\ast \to c \tau$ for $g_{1R}^{23\ast}\neq 0$, which is not simple due to jets originating from the charm-quark ($c$-jets) and decays of tau-lepton. 
A general feature of $S_1^\ast \to c \tau$ at the LHC is, however, similar to that of $S_1^\ast \to b \tau$. 
The process $pp \to S_1^*\, S_1 \to (b \tau)\, (\bar{b} \bar{\tau})$ has been analyzed by the CMS group based on the $8\,\text{TeV}$ data in Ref.~\cite{Khachatryan:2014ura}, 
and was applied to obtain the current bound by recasting the $(b \tau)\, (\bar{b} \bar{\tau})$ analysis to the $(c \tau)\, (\bar{c} \bar{\tau})$ case in Sec.~\ref{Sec:current_ctauctau}. 
For the 14~TeV search, we directly apply a similar method in Ref.~\cite{Khachatryan:2014ura} to the process $pp \to S_1^*\, S_1 \to (c \tau)\, (\bar{c} \bar{\tau})$.

Some optional modifications of the method (for requirements of jets and leptons) are also discussed. 
As for a (mis-)tagging efficiency of $c$-jet, a further discussion is necessary and we investigate several cases as will be shown later. 
Our analysis method based on Ref.~\cite{Khachatryan:2014ura} is summarized as follows.

\subsubsection{Procedure of our analysis}

We focus on the events where one of the two tau-leptons decays into a light lepton $\ell$ (electron or muon) such as $\tau \to \ell \bar{\nu}_\ell \nu_\tau$  
and the other one decays hadronically (denoted as $\tau_\text{h}$) as $\tau_\text{h} \to \text{hadrons} + \nu_\tau$. 
In Ref.~\cite{Khachatryan:2014ura}, the two signal regions: $e \tau_{\text{h}}$ and $\mu \tau_{\text{h}}$ are separately considered. 
In our analysis, we consider two cases for $\ell = \mu$ and $\ell = \mu, e$.

The trigger cuts are imposed so that the light lepton (jet) satisfies the conditions $p_\text{T} > 30\,\text{GeV}$, $|\eta| < 2.1\, (2.4)$, and the light leptons and jets are isolated as $\Delta R > 0.5$~\cite{Khachatryan:2014ura}.

At the first step after the trigger cut and isolation, we require a $\tau_\text{h}$-jet. 
In our analysis simulation, a candidate for $\tau_\text{h}$-jet is selected among reconstructed jets by applying the conditions $p_\text{T} > 50\,\text{GeV}$ and $|\eta| < 2.3$. 
The selected candidate, along with (without) a {\it parton-level} tau lepton 
within the range $\Delta R < 0.5$, is classified as a {\it true} ({\it fake}) $\tau_\text{h}$-jet candidate. 
Then, we identify a {\it true} ({\it fake}) candidate\footnote
{For a $\tau_\text{h}$-jet originating from the {\it true} category, the electric charge of the {\it parton-level} tau lepton corresponds to that of the charge of the initial $\tau$ of $\tau_\text{h}$.
Whereas when a {\it fake} candidate is mis-identified as a $\tau_\text{h}$-jet, the corresponding electric charge of the initial $\tau$ of $\tau_\text{h}$ is randomly determined because of the absence of the corresponding data.
}
as a real $\tau_\text{h}$-jet by taking (mis-)tagging efficiency into account. 
For the {\it true} candidate, we uniformly use a tagging rate of $0.5$, found in Refs.~\cite{CMS_tauIR,CMS-DP-2014-015} in the tight operating point for the hadron plus splits~(HPS) and the multivariate analysis~(MVA) algorithms.
The mis-tagging rate for the {\it fake} candidate is also obtained in Refs.~\cite{CMS_tauIR,CMS-DP-2014-015} as a function of $p_\text{T}$. 
For the HPS algorithm, the following form is obtained through our data fitting, 
\begin{align}
&(1.23193 \cdot 10^{-10})\,p_\text{T}^3 + (-1.28812 \cdot 10^{-7})\,p_\text{T}^2 + (4.81842 \cdot 10^{-5})\,p_\text{T} \notag \\
&\quad +(0.124279)\,\log{p_\text{T}}/p_\text{T} + (-0.00820209).  \label{eq:HPS_formula} 
\end{align} 
In our analysis, we adopt the HPS algorithm.
A major reason why we perform $\tau_{\text{h}}$-jet tagging without using the function installed in {\tt DelphesMA5tune} is to improve statistics by accepting all events and subsequently reweighting them based on the tagging rates. 
The factor for reweighting is defined as the probability that only one candidate is tagged and others (if exist) are not tagged.

For the next step after $\tau_\text{h}$-jet identification, we find $c$-jets in a similar manner to the above. 
We note that in our analysis for the $(c \tau)\, (\bar{c} \bar{\tau})$ final state, we do not tag $b$-jets since it is not necessary. 
Since the present detector simulation does not provide a $c$-jet tagging module, we need to implement it in our analysis simulation. 
Namely, {\it true} and {\it fake} candidates for $c$-jet are selected among reconstructed jets by the same condition with the $\tau_\text{h}$-jet case. 
Next, we take into account (mis-)tagging efficiencies of $c$-jet candidates. 
In our study, we consider three different choices for the efficiencies, reported in different studies~\cite{Perez:2015lra,Aad:2015gna,ATL-PHYS-PUB-2015-001}.  
The values are written as 
\begin{align}
&(\text{Case 1}) \quad  \epsilon_{c \to c} = 50\%, \quad \epsilon_{b \to c} = 20\%,\quad \epsilon_{\text{light} \to c} = 0.5\%, & \text{from Ref.~\cite{Perez:2015lra}},
\label{EQ:c-tagging_rate_1} \\
&(\text{Case 2}) \quad \epsilon_{c \to c} = 19\%, \quad \epsilon_{b \to c} = 13\%,\quad \epsilon_{\text{light} \to c} = 0.5\%, & \text{from Ref.~\cite{Aad:2015gna}},
\label{EQ:c-tagging_rate_2} \\
&(\text{Case 3}) \quad \epsilon_{c \to c} = 40\%, \quad \epsilon_{b \to c} = 25\%,\quad \epsilon_{\text{light} \to c} = 10\%,  & \text{from Ref.~\cite{ATL-PHYS-PUB-2015-001}},
\label{EQ:c-tagging_rate_3}
\end{align}
where $\epsilon_{c \to c}$ is a tagging rate and $\epsilon_{(b,\text{light}) \to c}$ indicates a mis-tagging rate of ($b$, light)-jet as $c$-jet.

We comment on the three types of ratios.
The values in Eq.~(\ref{EQ:c-tagging_rate_1}), used in the analysis of Ref.~\cite{Perez:2015lra}, are highly desirable, where a rather high tagging probability and small mis-tagging ratios are assumed. 
The second choice in Eq.~(\ref{EQ:c-tagging_rate_2}) was adopted in the analysis by ATLAS to search for a charm squark pair production at $8\,\text{TeV}$ in Ref.~\cite{Aad:2015gna}, 
where the $95\%$ CL lower bound on $M_{\tilde{c}}$ is obtained at around $560\,\text{GeV}$ assuming a massless neutralino and $\mathcal{B}(\tilde{c} \to c \tilde{\chi}^{0}_{1}) = 100\%$.
Here, the $c$-tagging rate is quite low compared with the first category in Eq.~(\ref{EQ:c-tagging_rate_1}), while the mis-tagging probabilities are still suppressed.
For identifying $c$-jets, the ATLAS group have developed the algorithm named JetFitterCharm~\cite{ATL-PHYS-PUB-2015-001}.
The values in the third category is also provided from Ref.~\cite{ATL-PHYS-PUB-2015-001} through the JetFitterCharm algorithm in a different operating point, 
where $c$-tagging rate is emphasized but the mis-tagging rates are also enhanced, especially from light jets.
Such high mis-tag rates would lead to serious deterioration in background rejection.
Later, we provide a quantitative comparison of the impact of these three choices in our simulation.

Another important aspect on $c$-jets is whether at least one or at least two $c$-jets should be required in our analysis.
The former choice is better for earning statistics, while the latter one definitely has better performance in background rejection.
We perform analyses following both of the criteria, the number of $c$-jets to be at least one or two, for a better understanding on $c$-jet identification.

\begin{table}[t!]
\centering
\begin{tabular}{cc}
\hline
\hline
Category 					& Cut and selection rule \\ 
\hline 
Leptons					& \begin{tabular}{l} (A-1) one $ \tau_\text{h}$ and one $\ell = \mu$ \\  (A-2) one $ \tau_\text{h}$ and one $\ell = \mu$ or $e$ \end{tabular}   \\ 
\hline
Electric charge 	\quad		& \quad opposite sign between $\tau_\text{h}$ and $\ell^{\pm}$ \\ 
\hline
Jet objects 				& \begin{tabular}{l} (B-1) $\geq 3$ (including $\tau_\text{h}$) \\  (B-2) $\geq 2$ (including $\tau_\text{h}$)  \end{tabular}  \\ 
\hline
$c$-tagged jet 				& \begin{tabular}{l} (B-1) at least two \\ (B-2) at least one \end{tabular} \\ 
\hline
$M(\tau_\text{h}\text{-jet},\,  \text{a chosen jet})$ 	& $> 250\,\text{GeV}$ \\ 
\hline
$S_\text{T}$ 				& \quad $>100$ -- $1000\,\text{GeV}$ for each $100\,\text{GeV}$ bin \\ 
\hline
\hline
\end{tabular}
\caption{Summary of the event selection cuts after the physics object reconstruction, which is mainly based on the choices in~\cite{Khachatryan:2014ura}.
Details of each cut are found in the main text.}
\label{tab:ctau_cuts}
\end{table}

After implementing the above procedure for the $\tau_\text{h}$-jet and $c$-jets, we perform selections and cuts to every event. 
It is summarized in Table~\ref{tab:ctau_cuts}. 
As mentioned above, we take account of two cases for the selection of a light lepton mode such as (A-1) $\ell = \mu$ and (A-2) $\ell = \mu,e$. 
We also consider the cases where the number of $c$-jets is required to be (B-1) at least two and (B-2) at least one. 
The invariant mass between $\tau_\text{h}$-jet and a chosen jet is required to be larger than $250\,\text{GeV}$.
Which jet is used for the invariant mass is determined as follows.
The two candidates $j_{1,2}$ for the jet are the leading $c$-tagged jet and the most leading jet among the other jets except  for the already picked-up leading $c$-jet and the  {$\tau_{\text{h}}$-jet}.
Finally, we adopt the selection cut as $M(\tau_\text{h}\text{-jet}, j_{1}) > 250\,\text{GeV}$ when $|M(\tau_\text{h}\text{-jet}, j_{1}) - M(\ell, j_{2})| < |M(\tau_\text{h}\text{-jet}, j_{2}) - M(\ell, j_{1})|$ is satisfied. 
When the above condition is failed, we choose $M(\tau_\text{h}\text{-jet}, j_{2})$ to the selection cut. 
This procedure is based on Ref.~\cite{Khachatryan:2014ura} for the $b$-tagged jets case.
The kinetic variable $S_\text{T}$ is defined as the scalar sum of the $p_\text{T}$ of $\ell$, $\tau_\text{h}$-jet, and the two jets $j_{1,2}$ of the two candidates for the invariant mass calculation.
The selection cut of $S_\text{T}$ is highly efficient for rejecting the irreducible $t\bar{t}$ background~\cite{Khachatryan:2014ura}. 
In our study, we prepare the cut region from $100\,\text{GeV}$ to $1000\,\text{GeV}$ every $100\,\text{GeV}$ step in advance 
and then choose an appropriate region to maximize the signal significance for each model parameter region.

\subsubsection{Event data for signal and background}
For our simulation, we generated $5 \times 10^4$ signal events for each mass of $S_1$ every $50\,\text{GeV}$ bin from $350\,\text{GeV}$ to $1600\,\text{GeV}$, 
produced by {\tt MadGraph5\_aMC@NLO} via the process $pp \to S_1^* S_1 \to (c\tau)\, (\bar{c}\bar{\tau})$ accompanying up to two additional jets (to perform jet merging). 
As for backgrounds, $10^7$ events of $t\bar{t}$ along with up to three jets and $5 \ (3) \times 10^6$ events of $W \to \ell \nu_\ell$ ($Z \to \ell \bar\ell$) along with up to four jets were generated for each $\ell = \mu$ and $e$, as well.
Note that the number of generated events is not equal to the numbers of reconstruct-level events used in our cut-based analysis since $\mathcal{O}(10)\%$ events are discarded through the jet merging procedure.
The $t\bar{t}$ events are dominant backgrounds since it includes two possible miss-tagged $c$-jets originating from $b$ quarks, one $\tau_\text{h}$-jet, and one $\tau$ decaying into $\ell$. 
The $W$+jets and $Z$+jets events give rather small contributions to the backgrounds, but might not be negligible due to their huge cross sections and possible mis-tagged $c$-jets and $\tau_\text{h}$-jet. 
The actual values of the nominal cross sections of the three background processes are summarized in Table~\ref{tab:bkg_crosssection}. 
The pure QCD background is neglected since a charged lepton is required in the final state.
The single top production is subleading in the original $(b\tau)$ case~\cite{Khachatryan:2014ura}.
Then, we ignore such two types of backgrounds in our analyses.
\begin{table}[t]
\centering
\begin{tabular}{cccc}
\hline
\hline
Channel & Cross section & Reference & PDF \\
\hline
$t\bar{t}$ & $970.5$ (pb) [NNLO+NNLL] &
\begin{tabular}{c} available in~\cite{ttbar_official_xsec}, \\ (generated by {\tt Top++v2.0}~\cite{Czakon:2013goa}) \end{tabular} &
\begin{tabular}{c} {\tt NNPDF2.3 NNLO}~\cite{Ball:2012cx} \\ (5f FFN) (Lower PDF) \end{tabular} \\
\hline
\begin{tabular}{c} $W$+jets, \\ $W \to \ell \nu_{\ell}$ \end{tabular} &
\begin{tabular}{c} $7978$ (pb) [$W^{+}$, NNLO + NLO EW] \\ $5662$ (pb) [$W^{-}$, NNLO + NLO EW] \end{tabular} &
generated by {\tt FEWZ}~\cite{Gavin:2012sy,Li:2012wna} & {\tt MSTW2008NNLO}~\cite{Martin:2009iq} \\
\hline
\begin{tabular}{c} $Z$+jets, \\ $Z \to 2\ell$ \end{tabular} &
$1207$ (pb) [NNLO + NLO EW] & generated by {\tt FEWZ}~\cite{Gavin:2010az,Li:2012wna} & {\tt MSTW2008NNLO}~\cite{Martin:2009iq} \\
\hline
\hline
\end{tabular}
\caption{Summary of the nominal cross sections of backgrounds in the $(c\tau)(\bar{c}\bar{\tau})$ channel.}
\label{tab:bkg_crosssection}
\end{table}

As well as the analysis for the process $pp \to S_1^*\, S_1 \to (b\nu_\ell)\, (\bar b \bar\nu_\ell)$, the parton-showering, hadronization, and jet merging are done via {\tt pythia-pgs}.
Also, the detector simulations are performed by {\tt DelphesMA5tune} and the reconstructed event data are stored in a {\tt root} file.
The {\tt NN23LO1} PDF is used for parton-level event generations of signals and backgrounds.

Then, the selections of candidate $c$-jets and $\tau_\text{h}$-jet, the evaluations of (mis-)tagging efficiencies for $c$-jets and $\tau_\text{h}$-jet, and the selection cuts listed in Table~\ref{tab:ctau_cuts} 
are executed in {\tt MadAnalysis5}, where we prepare the analysis code for the {\it expert mode} of {\tt MadAnalysis5}.

\section{Numerical result}
\label{Sec:Prospect}
The detailed procedures of our analysis simulations {aiming at} the two processes, 
$ pp \to S_1^*\, S_1 \to (b \nu)\, (\bar{b} \bar{\nu})$ and $pp \to S_1^*\, S_1 \to (c \tau)\, (\bar{c} \bar{\tau})$, are presented in Sec.~\ref{Sec:Analysis}. 
Based on them, we obtain prospects for the $S_1$ leptoquark model at the 14~TeV LHC explaining the $\bar B \to D^{(*)} \tau \bar\nu$ anomaly.

\subsection{Prospects of the $(b \nu)\, (\bar{b} \bar{\nu})$ channel}
\begin{figure}[t]
\centering
\includegraphics[viewport=0 0 360 367, width=18em]{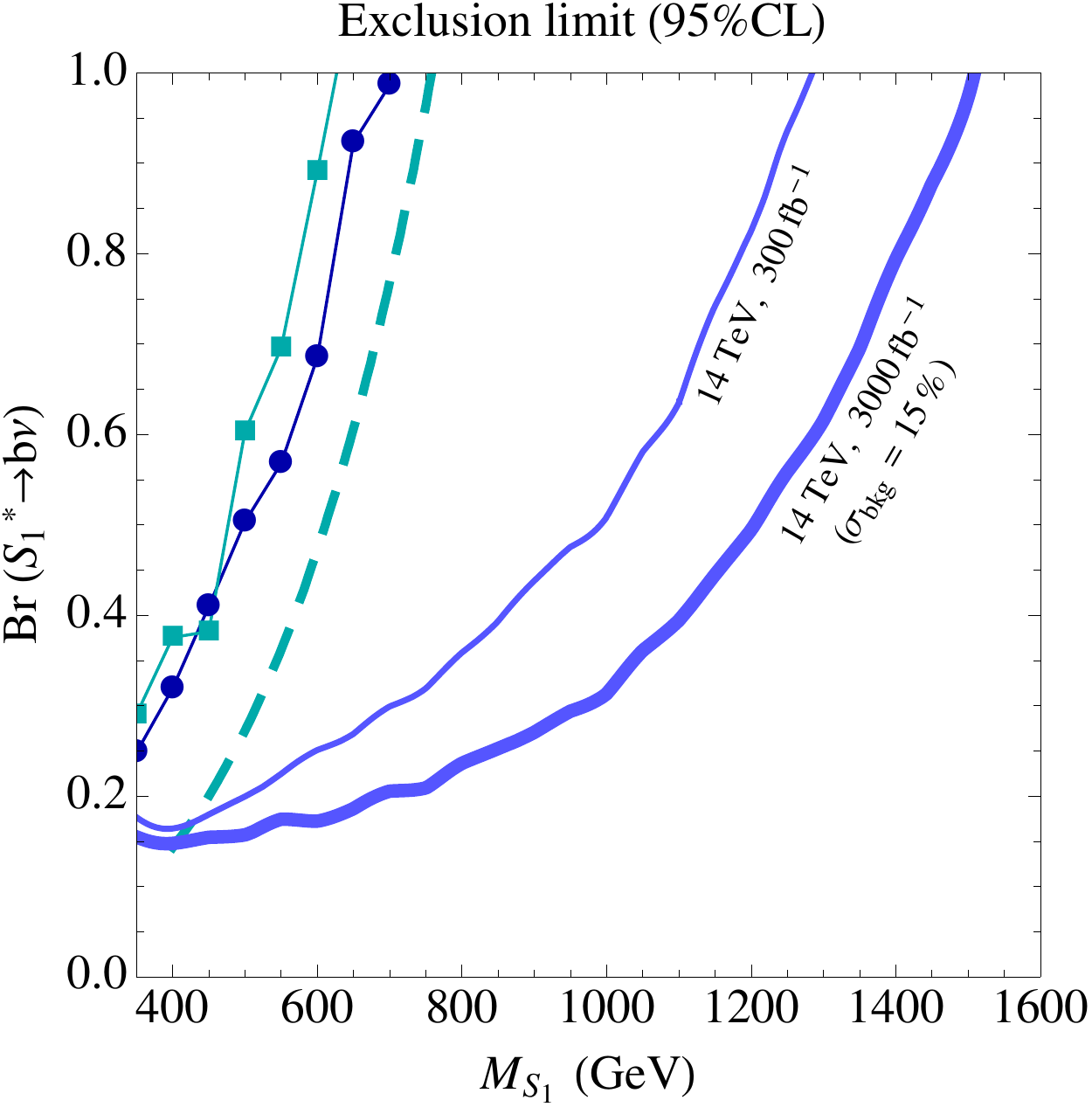}
\caption{
Prospects of the $(b \nu)\, (\bar{b} \bar{\nu})$ channel at the 14~TeV LHC {together with the constraints, given in Sec.~\ref{Sec:curentbounds}, from the 8~TeV (lines with dots) and the 13~TeV (dashed line) analyses. 
Two kinds of expectations based on different integrated luminosities ($\mathcal{L} = 300\, \text{fb}^{-1}$ and $\mathcal{L} = {3000}\, \text{fb}^{-1}$) 
with background uncertainties (the one from Ref.~\cite{ATL-PHYS-PUB-2014-010} and 15\%, respectively) are considered as indicated in the plot. }
}
\label{Fig:Br_bnu}
\end{figure}

At first, we show the prospects of $(b\nu)(\bar{b} \bar{\nu})$ channel at the 14~TeV LHC in Fig.~\ref{Fig:Br_bnu}.
The two blue solid lines indicate the exclusion limits at $95\%$ CL, 
where the first one is obtained with $\mathcal{L} = 300\,\text{fb}^{-1}$ and the total uncertainty in the backgrounds $\sigma_{\text{bkg}}$ used in Ref.~\cite{ATL-PHYS-PUB-2014-010}, 
whereas the other is obtained with $\mathcal{L} = 3000\,\text{fb}^{-1}$ and $\sigma_{\text{bkg}} = 15\%$, as presented in the figure. 
In the latter case ($\mathcal{L} = 3000\, \text{fb}^{-1}$), we expect that the background will be understood better and that $\sigma_{\text{bkg}} = 15\%$ is achievable. 
The current observed limits from the 8~TeV searches by ATLAS and CMS, as given in Sec.~\ref{Sec:curentbounds}, are also represented in the figure. 
The rough estimate from the ATLAS 13~TeV analysis is given as well. 

The result suggests that we can discard the $S_{1}$ leptoquark up to $1.3\,\text{TeV}$ ($1.5\,\text{TeV}$) with $\mathcal{L} = 300\, \text{fb}^{-1}$ ($\mathcal{L} = 3000\, \text{fb}^{-1}$), if $\mathcal{B}(S_{1}^{\ast} \to b \nu)=100\%$.
However, the 100\% branching ratio for $S_{1}^{\ast} \to b \nu$ is not obtainable 
because $g_{1L}^{3i}$ also controls the decay branch $S^{\ast}_{1} \to t \ell^{i}$ and then the possible value of $\mathcal{B}(S_{1}^{\ast} \to b \nu)$ is saturated at less than $50\%$. 
Moreover, in our setup of the model, the couplings and the $S_1$ mass are assumed to obey the condition in Eq.~(\ref{EQ:condition}) to explain the $\bar B \to D^{(*)} \tau \bar\nu$ anomaly. 
This assumption implies that $g_{1R}^{23}$ cannot be non-zero for a fixed non-zero $g_{1L}^{3i}\,(i = 3 \text{ or } 1,2)$ and $M_{S_{1}}$. 
Furthermore, $g_{1R}^{23}$ becomes sizable for a small $g_{1L}^{3i}$ and a large $M_{S_{1}}$. 
Therefore, in practice we can investigate the leptoquark through this channel up to around $1.0\,\text{TeV}$ ($1.2\,\text{TeV}$) when $\mathcal{L} = 300\, \text{fb}^{-1}$ ($\mathcal{L} = 3000\, \text{fb}^{-1}$).

\subsection{Prospects of the $(c\tau)(\bar{c}\bar{\tau})$ channel}
Next, we show the prospects of the $(c\tau)(\bar{c}\bar{\tau})$ channel as functions of $M_{S_{1}}$ and $\mathcal{B}(S_{1}^{\ast} \to c\tau)$, based on the analysis method given in Sec.~\ref{Sec:Analysis}.
As we explained, there are several possible selection criteria for the signal events in this channel,
\begin{enumerate}
\item $c$-tagging and mis-tagging ratios: (Case-1), (Case-2), (Case-3), as in Eqs.~(\ref{EQ:c-tagging_rate_1})-(\ref{EQ:c-tagging_rate_3}),
\item requirement on the number of $c$-jets: (B-1) at least two, or (B-2) at least one, 
\item requirement for the light lepton flavor:  (A-1) $\ell = \mu$, or (A-2) $\ell = \mu \text{ or } e$. 
\end{enumerate} 
These points are very important since they directly affect background rejections. 
So, we describe their effects at length in this subsection.

\begin{figure}[t]
\centering
\includegraphics[width=0.32\columnwidth]{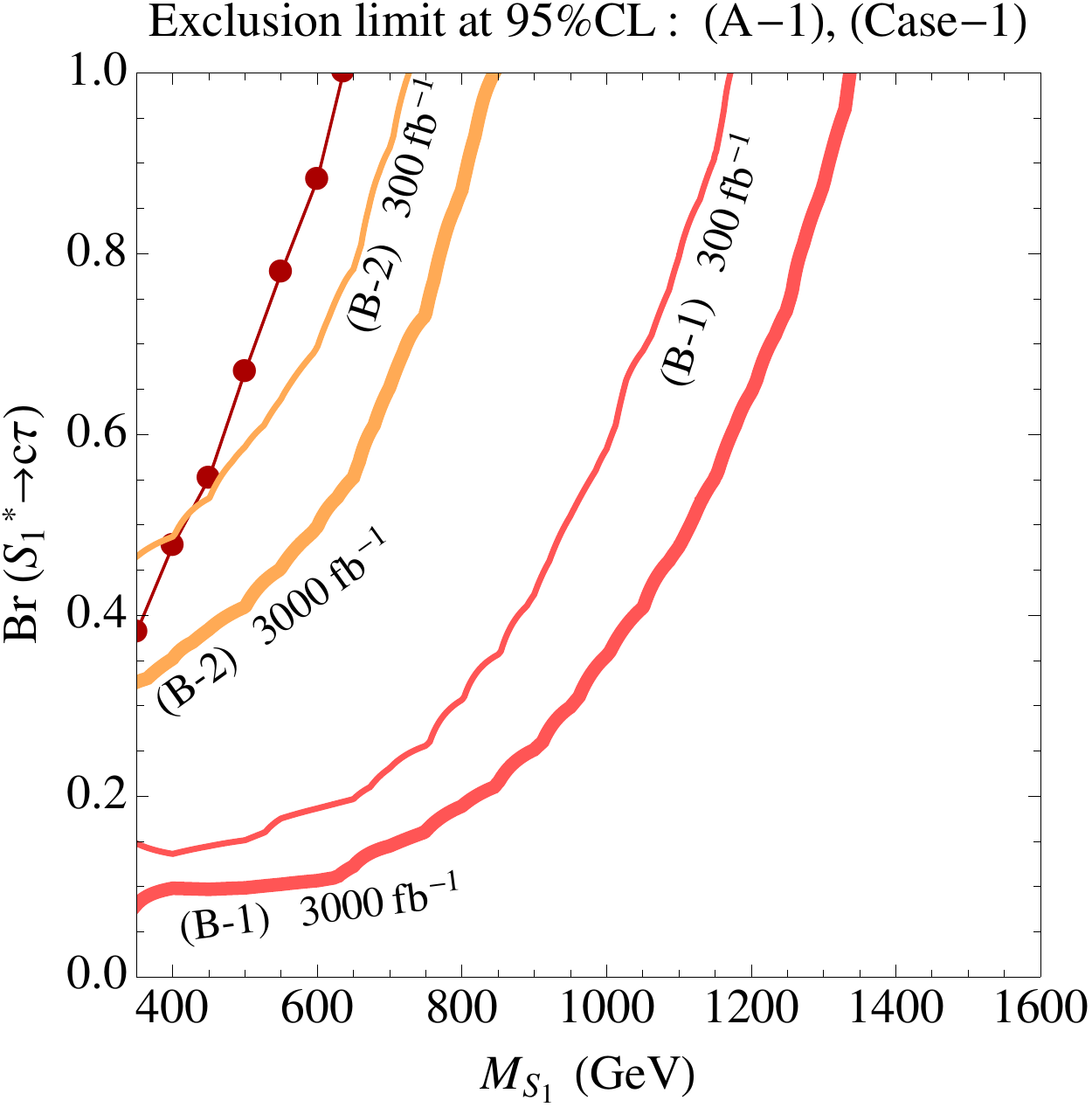}
\includegraphics[width=0.32\columnwidth]{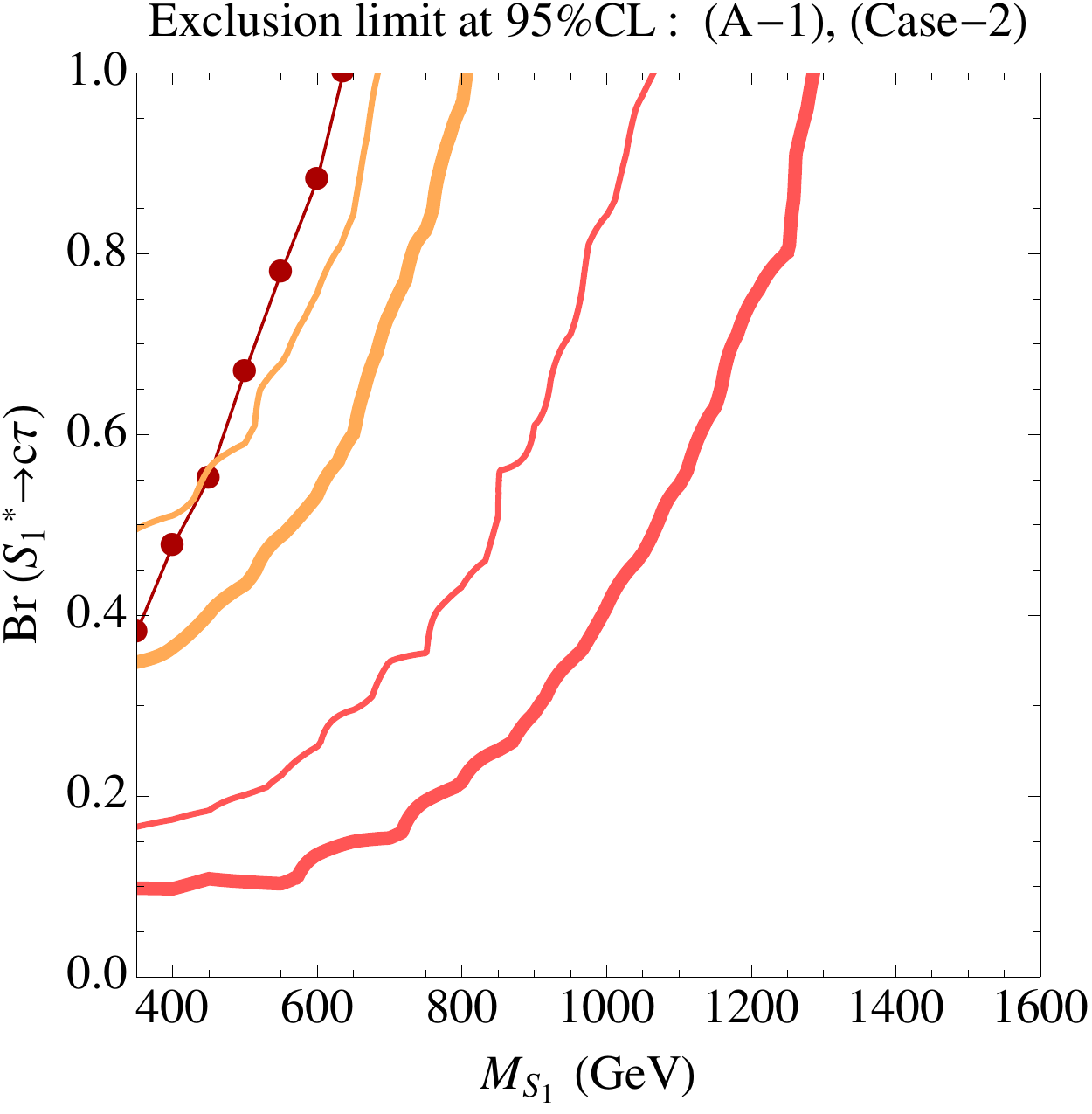}
\includegraphics[width=0.32\columnwidth]{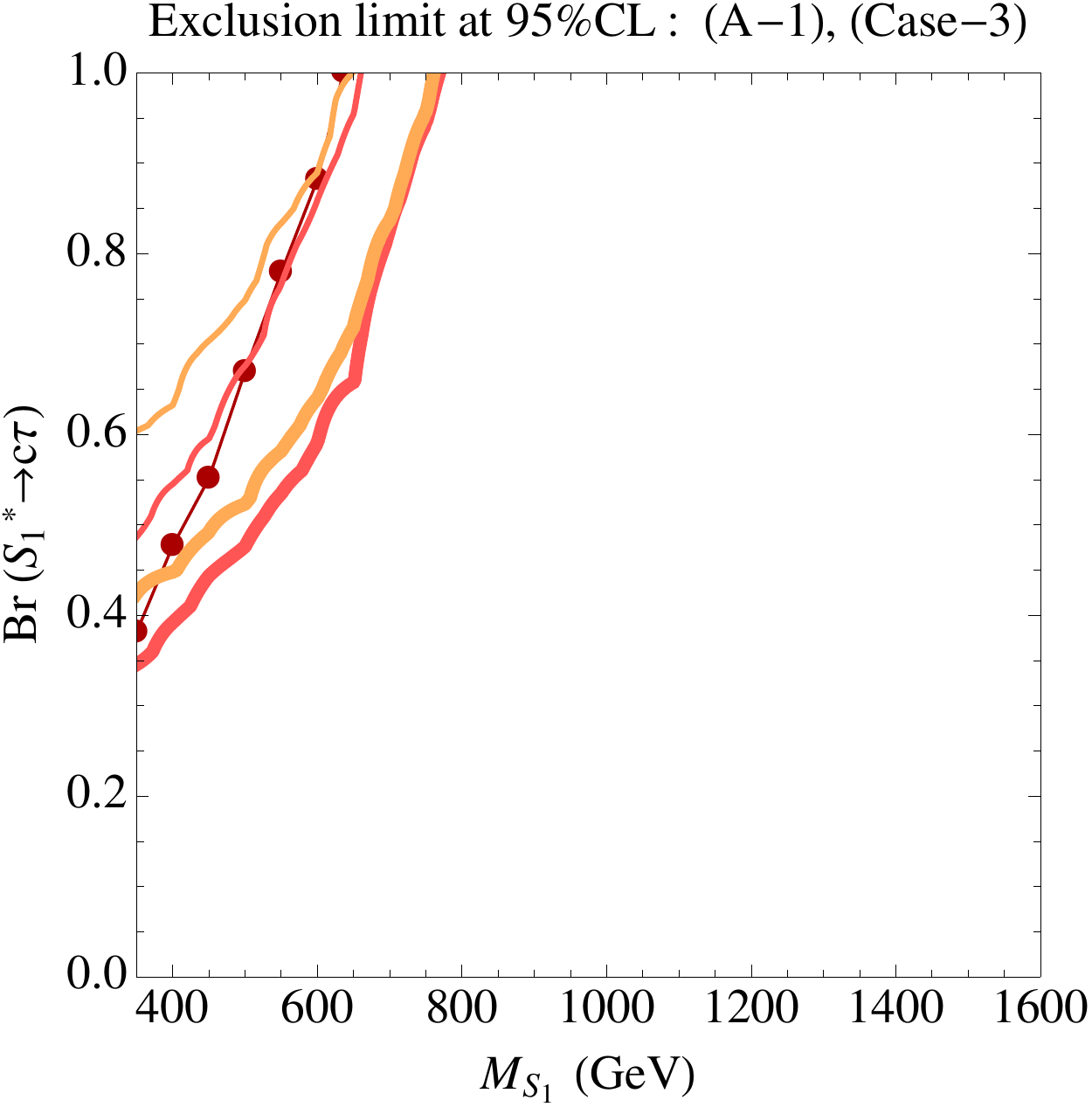} \\[1em]
\includegraphics[width=0.32\columnwidth]{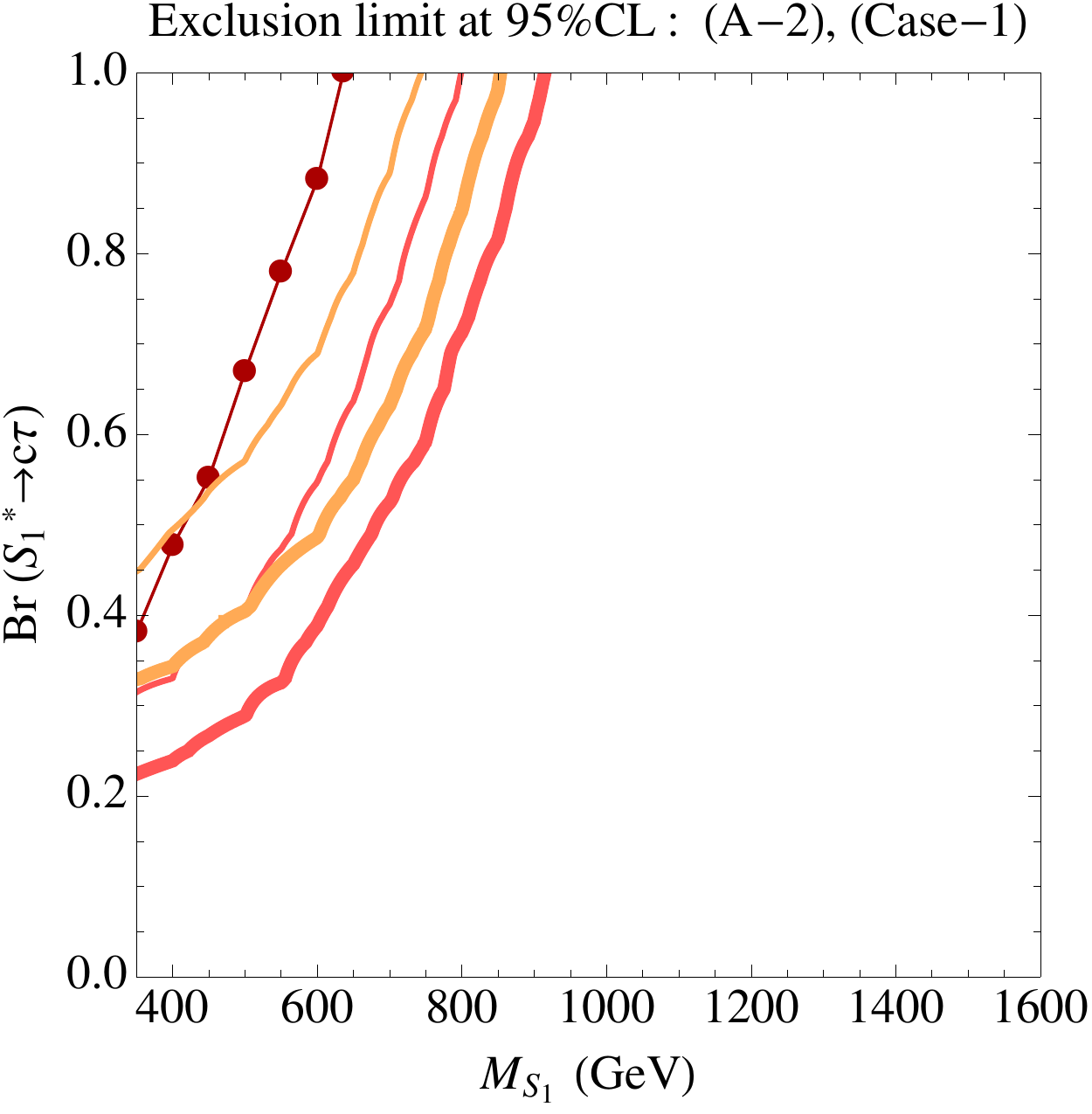}
\includegraphics[width=0.32\columnwidth]{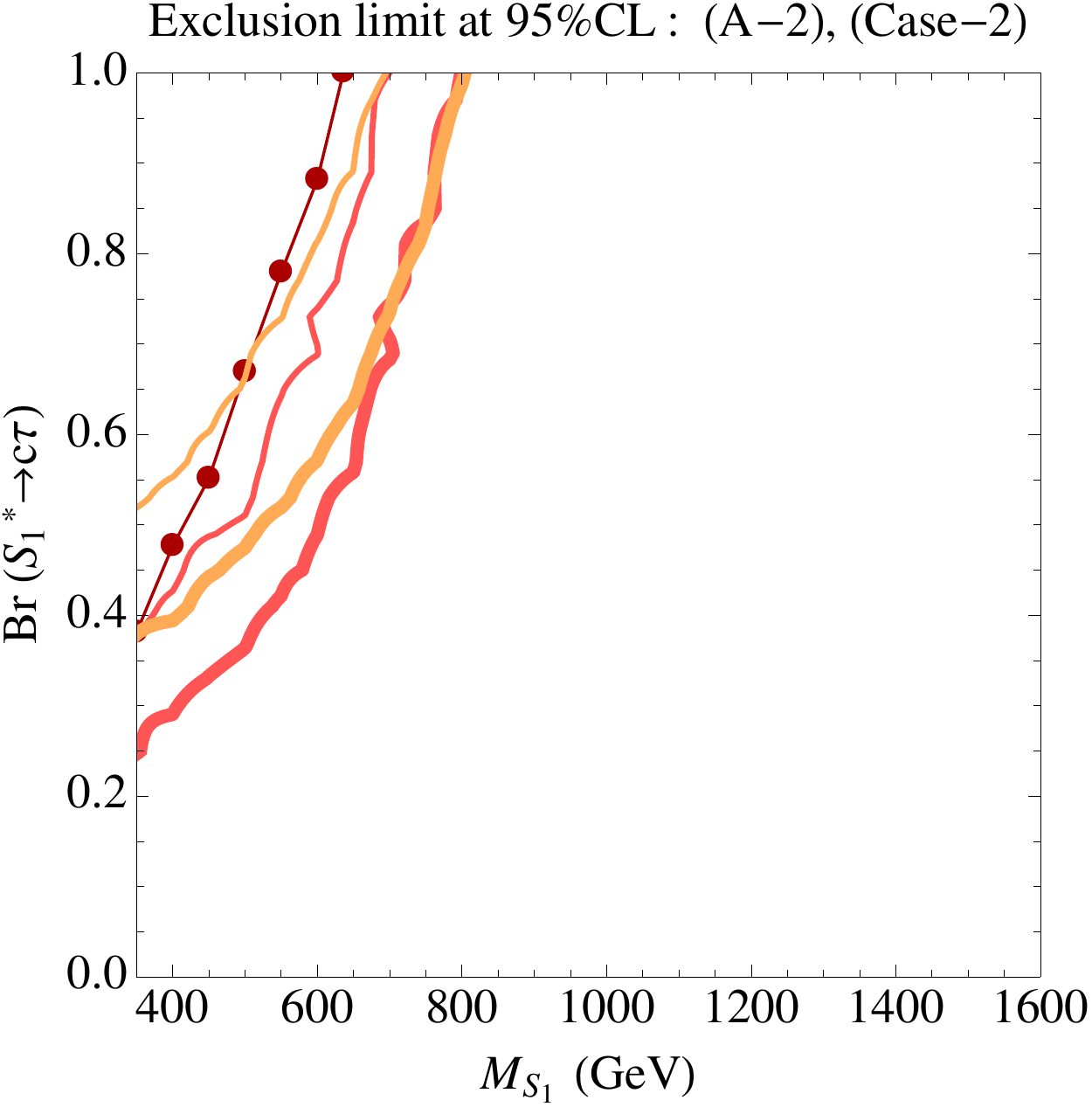}
\includegraphics[width=0.32\columnwidth]{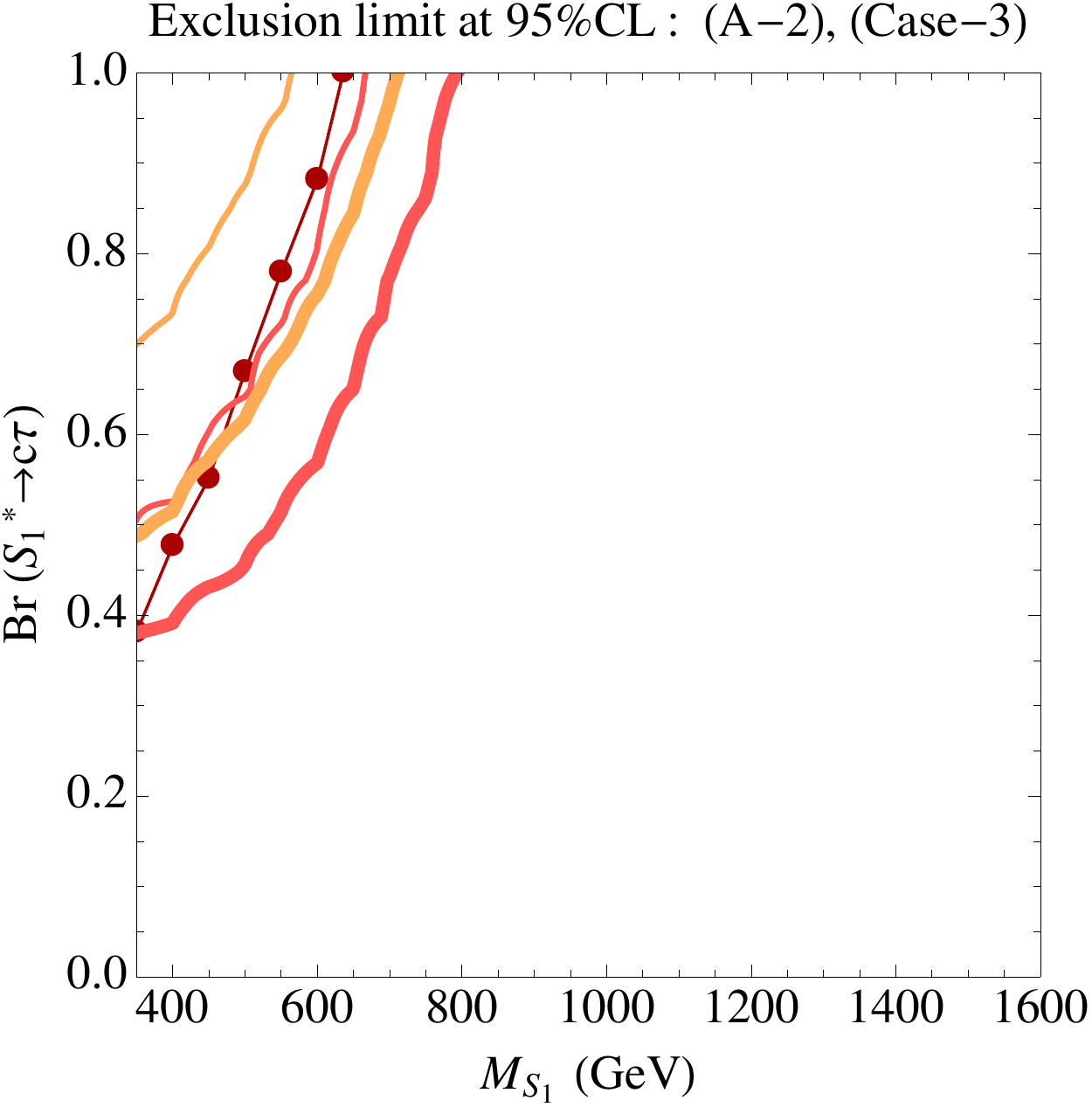}
\caption{
Prospects of the $(c\tau)(\bar{c}\bar{\tau})$ channel at the 14~TeV LHC as varying (A) the requirement on the light lepton flavor, (B) the number of required $c$-jets, and (C) the $c$-tagging/mis-tagging rates. 
The solid and thick solid lines in each panel are the expected exclusion limit at 95\% CL for the integrated luminosity with the background uncertainty, 
specified as $\mathcal{L} = 300\, \text{fb}^{-1}$ with $\sigma_{\text{bkg}} = 30\%$ and $\mathcal{L} = 3000\, \text{fb}^{-1}$ with $\sigma_{\text{bkg}} = 15\%$. 
The upper and lower panels show the results for (A-1) $\ell = \mu$ and (A-2) $\ell = \mu  {\text{ or }} e$, respectively. 
The number of required $c$-jets is chosen as at least (B-1) two and (B-2) one, which result in red and orange colored lines, respectively. 
The left, middle, and right panels indicate the result for (Case-1), (Case-2), and (Case-3), respectively, which are the three choices of $c$-tagging/mis-tagging rates adopted in our analysis. 
The red line with dots is the recast bound from the $8\,\text{TeV}$ CMS analysis for $(b \tau) (\bar{b}\bar{\tau})$. 
}
\label{Fig:Br_ctau_1}
\end{figure}
In Fig.~\ref{Fig:Br_ctau_1}, we show our numerical results for the prospects in the $(c\tau)(\bar{c}\bar{\tau})$ channel at the 14~TeV LHC. 
In this figure, we consider two cases for the integrated luminosity with the background uncertainty, 
$\mathcal{L} = 300\, \text{fb}^{-1}$ with $\sigma_{\text{bkg}} = 30\%$ and $\mathcal{L} = 3000\, \text{fb}^{-1}$ with $\sigma_{\text{bkg}} = 15\%$, denoted by solid and thick solid curves, respectively. 
The upper panels in the figure show the results for (A-1),  where the muon is required in the final state, whereas the lower panels are the results for (A-2),  where the muon or electron is required. 
The left, middle, and right panels indicate the results obtained from the different choices of $c$-tagging/mis-tagging rates (Case-1), (Case-2), and (Case-3), respectively 
as defined in Eqs.~(\ref{EQ:c-tagging_rate_1})-(\ref{EQ:c-tagging_rate_3}). 
In each panel, we show two cases for the requirement on the number of $c$-jets, (B-1) at least two and (B-2) at least one as denoted by red and orange colors, respectively. 
The red line with dots in each plot indicates our recast bound from the $8\,\text{TeV}$ CMS result on $(b \tau) (\bar{b}\bar{\tau})$ channel~\cite{Khachatryan:2014ura}.
We immediately recognize the following points:
\begin{itemize}
\item
We can rank the three choices of $c$-tagging/mis-tagging {rates} as
\begin{align}
(\text{Case }1) > (\text{Case }2) \gg (\text{Case }3).
\label{EQ:c-tag_rank}
\end{align}
The result claims that (Case-1) works the most effectively. 
This is definitely obvious since this configuration is a desired one; however such high $c$-tagging and low mis-tagging rates may be beyond the current technology. 
On the other hand, the efficiencies of (Case-2) are already realized and used in experiment. 
Although the $c$-tagging rate in (Case-2) is lower than that in (Case-1), 
we can see that good performance is obtained in (Case-2) for our model, similarly to (Case-1). 
From the upper middle panel of Fig.~\ref{Fig:Br_ctau_1}, we conclude that we can search for the $S_{1}$ leptoquark boson through the $(c\tau)(\bar{c}\bar{\tau})$ channel up to $1.05\,\text{TeV}$ and $1.3\,\text{TeV}$, 
when accumulating $\mathcal{L} = 300\,\text{fb}^{-1}$ of data at $14\,\text{TeV}$ with $\sigma_{\text{bkg}} = 30\%$ and $\mathcal{L} = 3000\,\text{fb}^{-1}$ with $\sigma_{\text{bkg}} = 15\%$, respectively. 
The last one, (Case-3), is insignificant because of the high misidentification rate, especially in $\epsilon_{\text{light} \to c}$.
\item
One can find that requiring at least two $c$-tagged jets, (B-1), results in the better expected exclusion than (B-2). 
This is simply due to the fact that the background rejection by the requirement of at least two $c$-jets is more efficient than that of at least one $c$-jet, 
since the $c$-jet tagging efficiencies are not high enough and requiring two $c$-jets helps us to improve separability. 
\item
The requirement for the light lepton to be muon (A-1), $\ell = \mu$, works well compared with (A-2), $\ell = \mu \text{ or } e$ 
(remind that the signal region (A-2) considers both $\mu$ and $e$ in the same signal region). 
This implies that an electron channel would not significantly improve exclusion.  
In our analysis, we select events with one leptonic $\tau$  (and one hadronic $\tau$). 
Hence, the primary background is $pp \to t\bar{t} \to b \bar{b} W^{+} W^{-}$ where one of the sequential decays is $W \to \tau \nu_{\tau}$. 
When we enlarge the allowed configuration from $\ell = \mu$ to $\ell = \mu \text{ or } e$, 
both of the signal and the primary background receive similar gains and the deterioration in the background overwhelms the improvement in the signal 
because the nominal cross section is much greater than that of the signal. 
\end{itemize}
As a conclusion, the best choice in {the requirements for} the number of $c$-tagged jets and the light lepton flavor from the leptonic $\tau$ is (A-1) $\ell = \mu$ and (B-1) at least two $c$-jets. 
Performances of the three types of $c$-tagging/mis-tagging rates are investigated and graded as in Eq.~(\ref{EQ:c-tag_rank}).

\subsection{Combined results}
\label{Sec:CombinedResult}
\begin{figure}[t]
\centering
\includegraphics[viewport=0 0 360 363, width=18em]{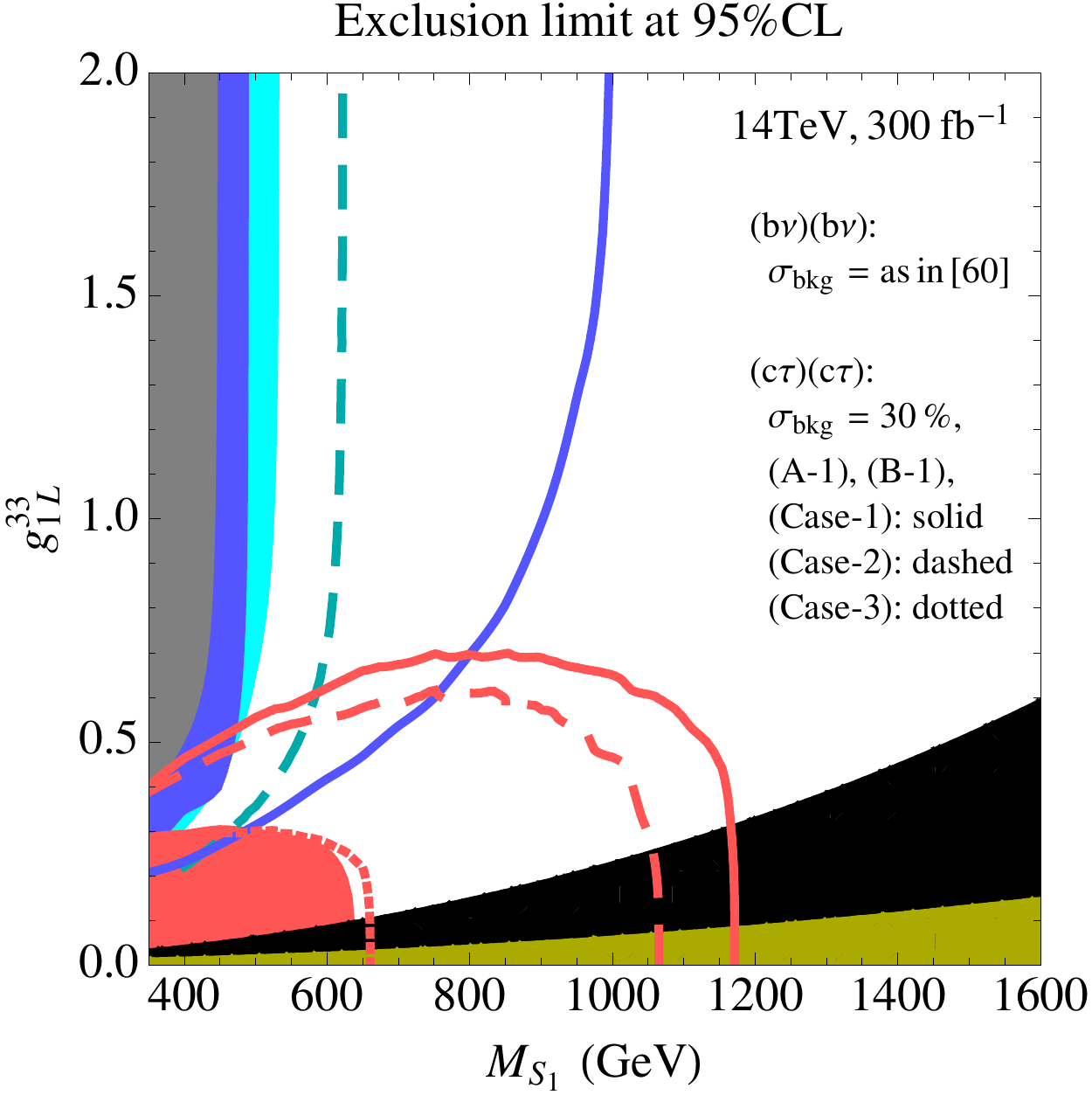}
\quad
\includegraphics[viewport=0 0 360 363, width=18em]{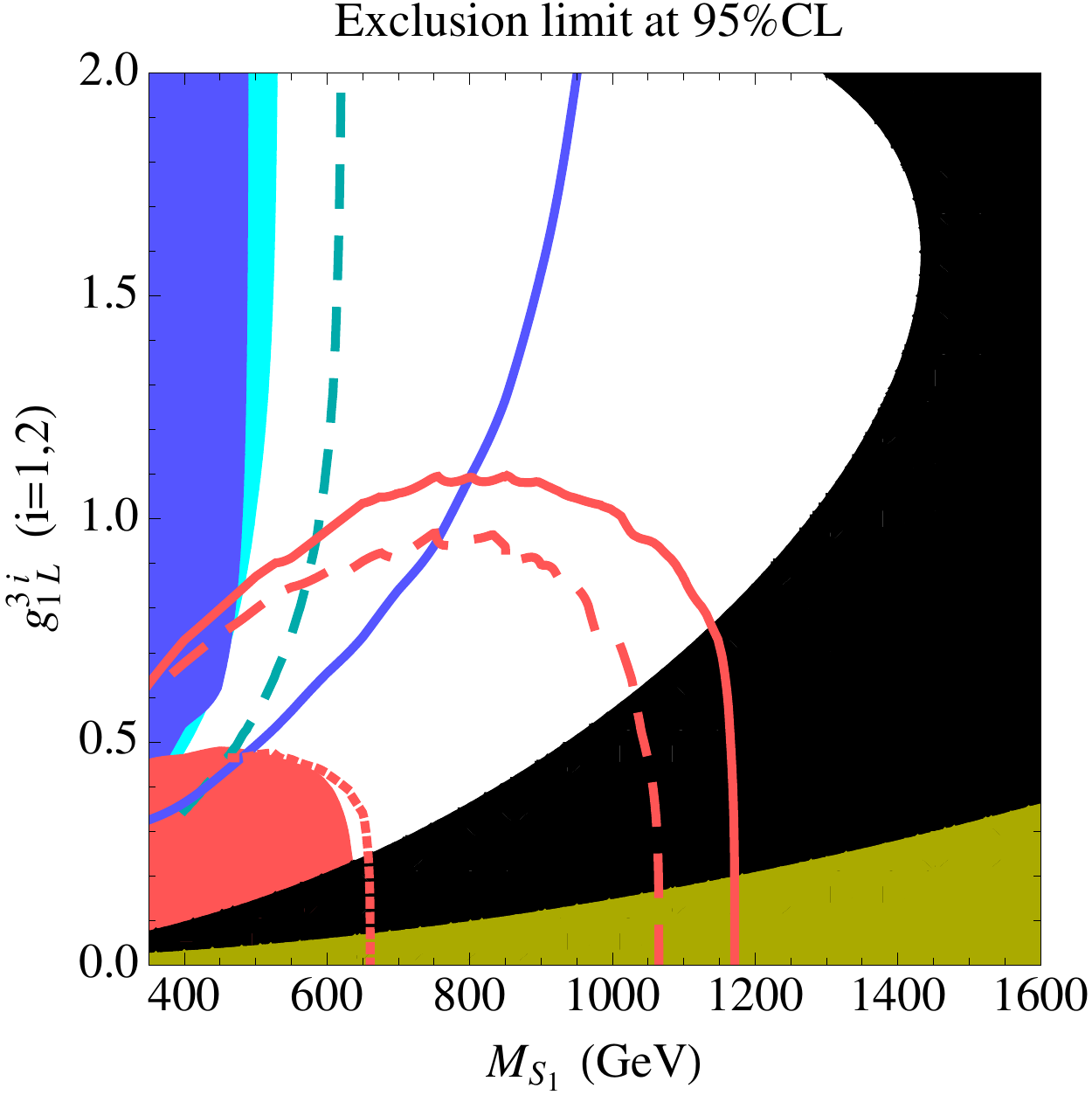}
\caption{
Prospects of exclusions for the 14~TeV LHC  when $\mathcal{L} = 300\, \text{fb}^{-1}$ data is collected. 
The plots present 95\% CL exclusions for the coupling $g_{1L}^{3i}$ and the mass $M_{S_{1}}$ from both the $(b\nu)(\bar{b}\bar{\nu})$ and $(c\tau)(\bar{c}\bar{\tau})$ channels.
The blue curve shows the $95\%$ exclusion limit from the $(b\nu)(\bar{b}\bar{\nu})$ channel, while the red curves describe the ones from the $(c\tau)(\bar{c}\bar{\tau})$ channel, 
where the three different $c$-tagging/mis-tagging probabilities defined as (Case-1), (Case-2), and (Case-3) are adopted in solid, dashed, and dotted curves, respectively. 
Here, we depict the excluded regions from the 8~TeV {and 13~TeV} results.
The black regions represent the areas with $\Gamma_{S_{1}}/M_{S_{1}} \geq 20\%$. 
The dark-yellow parts are theoretically unacceptable since $g_{1R}^{23} \geq 4\pi$. 
}
\label{Fig:coup_1}
\end{figure}

Here, we translate the results for the expected and current exclusion limits on the branching ratios shown above into those on the coupling of the $S_1$ leptoquark model, 
in order to declare future prospects for probing the $\bar B \to D^{(*)} \tau \bar\nu$ anomaly in this model. 
In Fig.~\ref{Fig:coup_1}, we summarize the results for the 14~TeV LHC at 95\% CL for $\mathcal{L} = 300\, \text{fb}^{-1}$ of accumulated data, 
which present prospects for the coupling $g_{1L}^{3i}$ and the mass $M_{S_{1}}$ from both the $(b\nu)(\bar{b}\bar{\nu})$ and $(c\tau)(\bar{c}\bar{\tau})$ channels.
The blue curve shows the $95\%$ exclusion limit from the $(b\nu)(\bar{b}\bar{\nu})$ channel, 
while the red curves describe the ones from the $(c\tau)(\bar{c}\bar{\tau})$ channel with three different $c$-tagging/mis-tagging probabilities, 
(Case-1,2,3) as defined in Eqs.~(\ref{EQ:c-tagging_rate_1})-(\ref{EQ:c-tagging_rate_3}) with solid, dashed, dotted curves, respectively.  
For the $(c\tau)(\bar{c}\bar{\tau})$ analysis, (A-1) $\ell = \mu$, (B-1) at least two $c$-jet, and $\sigma_{\text{bkg}}=30\%$ are required in this figure. 
The background uncertainty for the $(b\nu)(\bar{b}\bar{\nu})$ channel is given as in Ref.~\cite{ATL-PHYS-PUB-2014-010} ($\sim 30\%$  in high $m_{\text{CT}}$ signal regions), the same as before in this paper.
We also show the constraints from {the} $8\,\text{TeV}$ and $13\,\text{TeV}$ LHC data which we discussed before.
The black regions represent the areas with $\Gamma_{S_{1}}/M_{S_{1}} \geq 20\%$, where the narrow-width approximation is not reliable.
The dark-yellow parts should be discarded as theoretically unacceptable since perturbativity is violated for $g_{1R}^{23} \geq 4\pi$.

Remind that, in our setup, the couplings ($g_{1L}^{3i}$, $g_{1R}^{23}$) and the mass ($M_{S_1}$) are related by the condition in Eq.~(\ref{EQ:condition}) to explain the $\bar B \to D^{(*)} \tau \bar\nu$ anomaly. 
Hence, $g_{1R}^{23}$ is determined with the condition in the figure. 
From Eq.~(\ref{EQ:condition}), we recognize that the resultant $g_{1R}^{23}$ tends to be larger in the case of $i=1 \text{ or }2$ than $i=3$ 
when we compare the two cases with the common $M_{S_{1}}$ and values of $g_{1L}^{33}$ and $g_{1L}^{3i}$  $(i = 1 \text{ or } 2)$ being identical. 
Then, the following relations are expected, 
\begin{align}
\mathcal{B}(S_{1}^{\ast} \to b\nu)|_{i=3} > \mathcal{B}(S_{1}^{\ast} \to b\nu)|_{i=1\text{ or }2},\quad
\mathcal{B}(S_{1}^{\ast} \to c\tau)|_{i=3} < \mathcal{B}(S_{1}^{\ast} \to c\tau)|_{i=1\text{ or }2}. 
\end{align} 
Thus, the coverage of the $95\%$ exclusion contour from the $(c\tau)(\bar{c}\bar{\tau})$ channel tends to be broader in $i=1\text{ or }2$ compared with $i=3$, 
while the opposite trend is found in the contour from the $(b\nu)(\bar{b}\bar{\nu})$ channel. 
The efficiencies of the three $c$-tagging/mis-tagging rates in the Case-1, 2, 3 are directly reflected in the explored ranges as following the order in Eq.~(\ref{EQ:c-tag_rank}). 
Through the cooperation of the $(b\nu)(\bar{b}\bar{\nu})$ and $(c\tau)(\bar{c}\bar{\tau})$ searches with an accumulated luminosity of $\mathcal{L} = 300\, \text{fb}^{-1}$ at the 14~TeV LHC run~II,  
we can exclude the $S_{1}$ leptoquark boson explaining the $B$ physics anomaly up to at least $0.8\,\text{TeV}$ for both $i=3$ and $i=1\text{ or }2$. 
For small and large $g_{1L}^{3i}$, $M_{S_1} \lesssim 1\,\text{TeV}$ can be ruled out.

\begin{figure}[t!]
\centering
\includegraphics[viewport=0 0 360 363, width=18em]{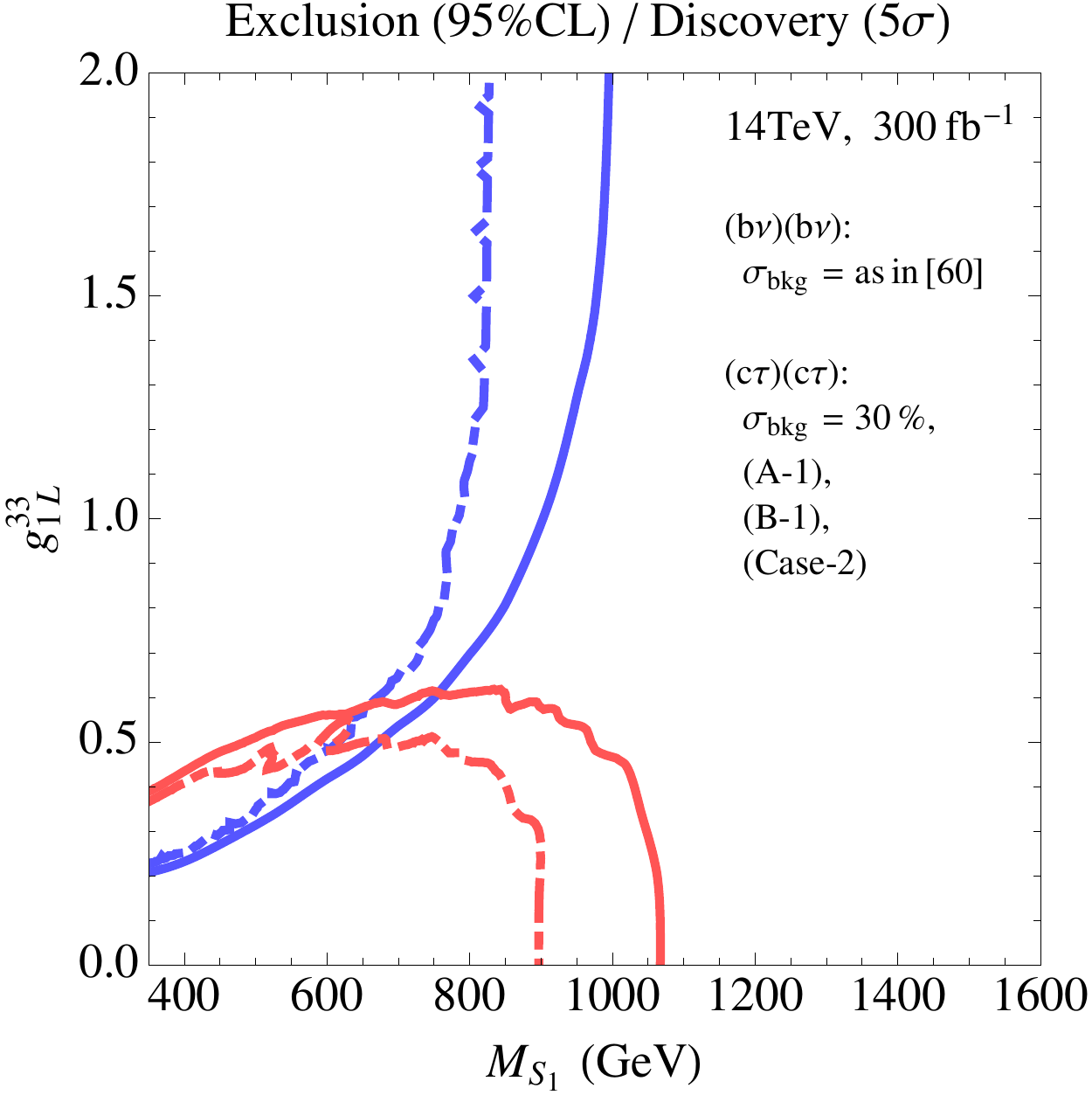}
\quad
\includegraphics[viewport=0 0 360 363, width=18em]{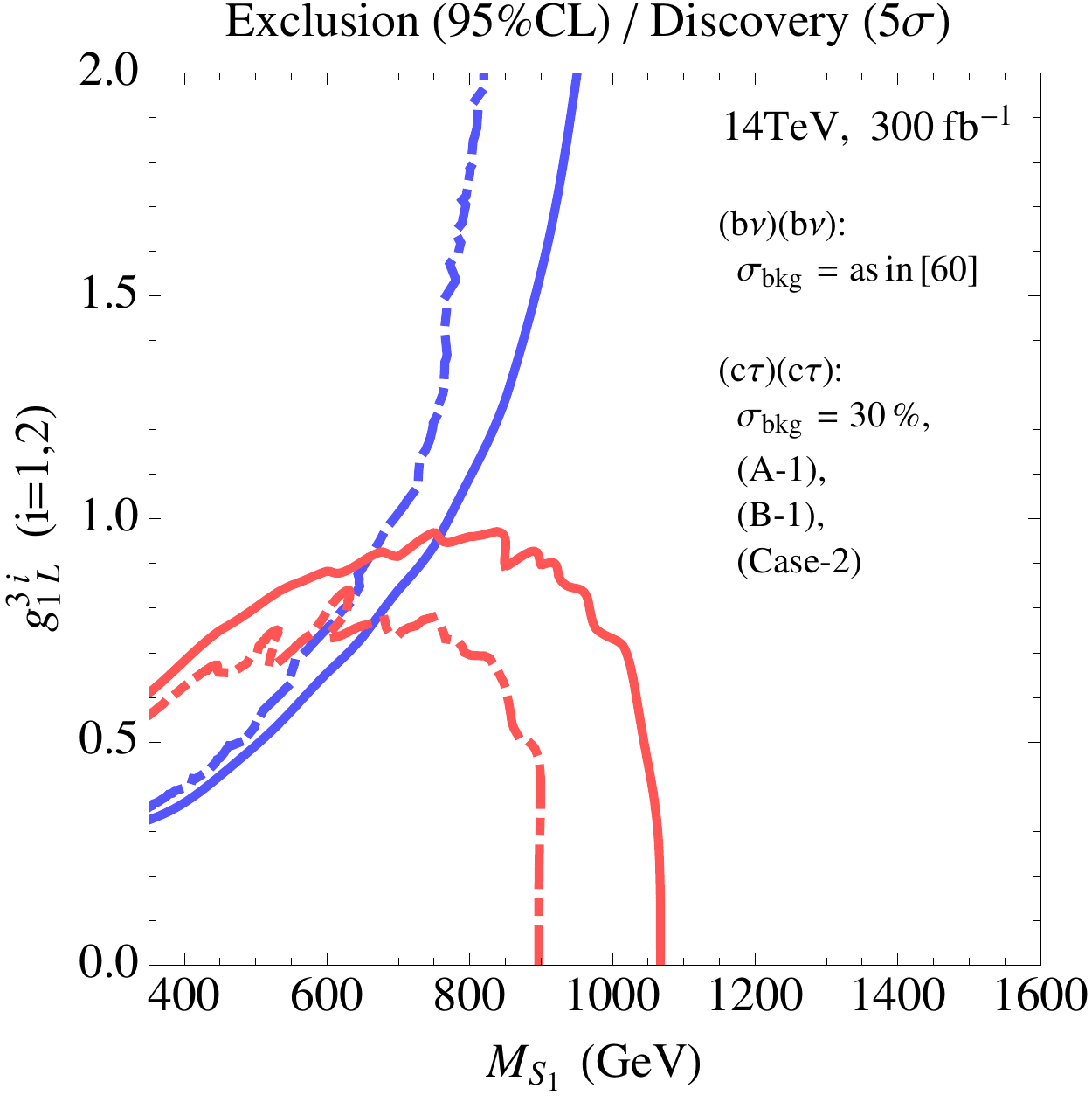} \\[1em]
\includegraphics[viewport=0 0 360 363, width=18em]{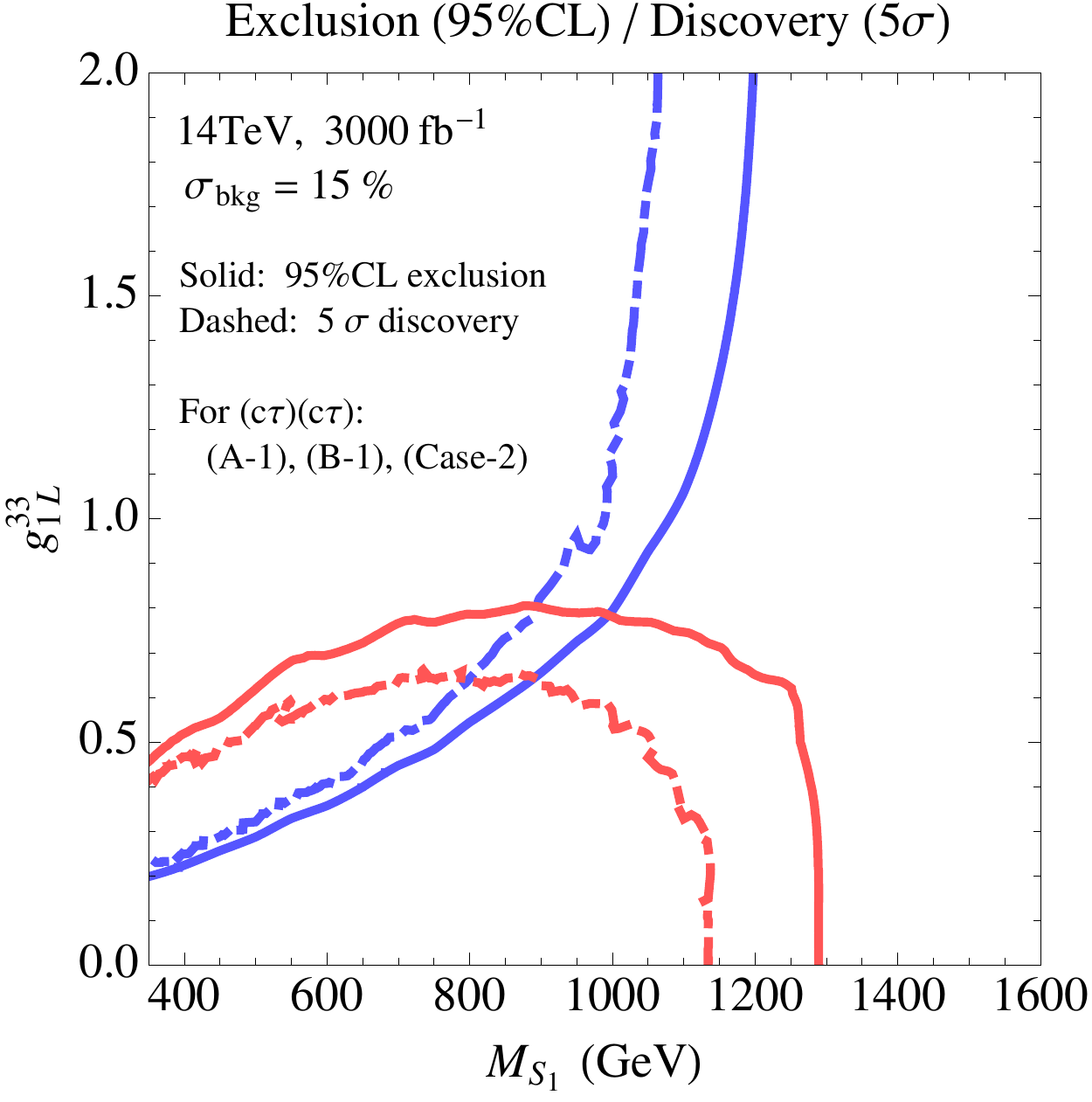}
\quad
\includegraphics[viewport=0 0 360 363, width=18em]{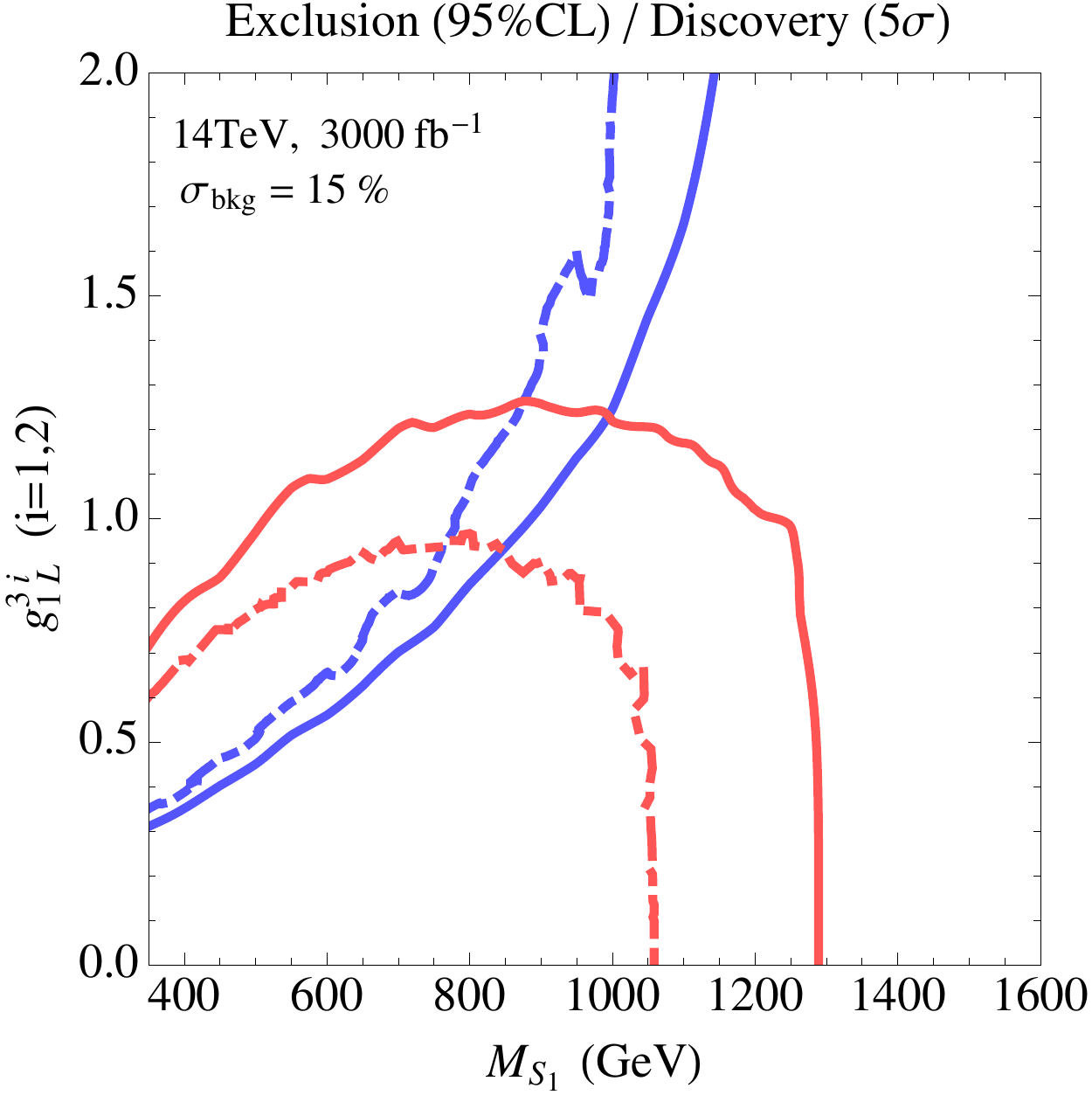} 
\caption{
Future prospects at the 14~TeV LHC with $\mathcal{L} =  {300} \text{ and } 3000\,\text{fb}^{-1}$ for 95\% exclusion and $5\sigma$ discovery potentials of the $S_1$ leptoquark boson on the plane of $(M_{S_{1}}, g_{1L}^{3i})$. 
The background uncertainty is taken as  {$\sigma_{\text{bkg}} = 30 \text{ and } 15\%$, respectively}. 
The solid and dot-dashed curves correspond to the 95\% exclusion and $5\sigma$ discovery reaches, respectively. 
The blue and red colors indicate the results from the $(b\nu)(\bar{b}\bar{\nu})$ and $(c\tau)(\bar{c}\bar{\tau})$ channels, respectively. 
For the $(c\tau)(\bar{c}\bar{\tau})$ case, the (A-1), (B-1), and (Case-2) choices are adopted in the analysis. 
}
\label{Fig:S1discovery}
\end{figure}
In Fig.~\ref{Fig:S1discovery}, 95\% CL exclusion and $5\sigma$ discovery potentials for  {$\mathcal{L} = 300 \text{ and } 3000\,\text{fb}^{-1}$} at 14~TeV are shown, 
where the total uncertainty in the backgrounds is assumed to be $\sigma_{\text{bkg}} = 30 \text{ and } 15\%$, and the $(c\tau)(\bar{c}\bar{\tau})$ analysis is done with (A-1), (B-1), and (Case-2) choices. 
The 95\% CL excluded ranges in the $(M_{S_{1}}, g_{1L}^{3i})$ parameter plane for $\mathcal{L} = 3000\,\text{fb}^{-1}$ are broaden as $1.0\,\text{TeV} \sim 1.3\,\text{TeV}$, 
compared with those for $\mathcal{L} = 300\,\text{fb}^{-1}$. 
We also find that the $S_1$ leptoquark boson, which can explain the $\bar B \to D^{(*)} \tau \bar\nu$ anomaly, can be discovered 
from both the $(b\nu)(\bar{b}\bar{\nu})$ and $(c\tau)(\bar{c}\bar{\tau})$ channels with $M_{S_1} \lesssim 600/800\,\text{GeV}$ when we accumulate data with $\mathcal{L} = 300/3000\,\text{fb}^{{-1}}$. 
There is also a possibility that the $S_1$ boson with $M_{S_1} \lesssim 1.1\,\text{TeV}$ is discovered only in either the $(c\tau)(\bar{c}\bar{\tau})$ or $(b\nu)(\bar{b}\bar{\nu})$ search.

As we have discussed, properties of jets originating from $b$ and $c$ quarks are similar and misidentification rates between them tend to be high in general. 
Due to that, it can happen that processes from the $S_1$ pair production other than $(b \nu) (\bar{b} \bar{\nu})$ and $(c\tau)\, (\bar{c}\bar{\tau})$ are detected as ``signals'' through our cut analysis. 
We call it as a misidentified signal. 
For example, the decay branches $S_1^*S_1 \to (t\tau)(\bar{t}\bar{\tau})$ and $S_1^*S_1 \to (c\tau)(\bar{t}\bar{\tau}), (t\tau)(\bar{c}\bar{\tau})$ fake $S_1^*S_1 \to (c\tau)\, (\bar{c}\bar{\tau})$ 
when one or two $b$-jets via the top decay are misidentified as $c$-jets. 
Indeed, we have seen that these two misidentified signals do not change our conclusion in this paper, but are not completely negligible. 
We have checked that other misidentified signals are completely negligible. 
We explore this issue in detail in Appendix~\ref{App:DetailedAnalysis}.

\section{Summary}
\label{Sec:Summary}
We have investigated the LHC potential to probe the $S_1$ leptoquark model that can explain the $\bar B \to D^{(*)} \tau \bar\nu$ anomaly
in light of existing LHC results at 8 and 13~TeV, and provided expected exclusion bounds and discovery reach at the 14~TeV LHC in terms of the parameters of this model.

At first, we have briefly reviewed the $\bar B \to D^{(*)} \tau \bar\nu$ anomaly, 
expressed in terms of the deviations of the observables $R(D)$ and $R(D^*)$ between the current combined experimental results and the SM predictions. 
It turns out that current results exhibit a deviation with significance of around $4\sigma$. 
The previous studies in Refs.~\cite{Tanaka:2012nw,Biancofiore:2013ki,Dorsner:2013tla,Sakaki:2013bfa} suggest that the deviations can be explained by several leptoquark models. 
Based on Ref.~\cite{Sakaki:2013bfa}, we have provided the latest allowed ranges for the couplings in the leptoquark models. 
Then we have seen that three types of leptoquark bosons, $S_1$, $R_2$, and $U_1$ can explain the anomaly while being consistent with all other flavor constraints.

Among them, we have focused on the $S_1$ leptoquark boson in order to study the LHC potential to probe the $\bar B \to D^{(*)} \tau \bar\nu$ anomaly. 
In order to explain the anomaly, the minimal setup yields $g_{1L}^{3i} \neq 0$, $g_{1R}^{23} \neq 0$, and vanishing values for all other couplings. 
The coupling $g_{1L}^{3i}$ controls the decays $S_1^* \to t \ell^i$ and $S_1^* \to b \nu_{\ell^i}$, whereas $g_{1R}^{23} \neq 0$ gives rise to $S_1^* \to c \tau$. 
Since the leptoquark boson is dominantly pair produced at the LHC through QCD interactions, there are six possible channels for the signal.

Several existing 8~TeV LHC searches can be used to constrain our model. 
We have translated the results of ATLAS and CMS searches for pair-produced bottom squarks~\cite{Aad:2013ija,Khachatryan:2015wza} decaying as $\tilde b_1 \to b \widetilde{\chi}^0_1$ into constraints for the $S_1$ boson. 
A direct bound on the scalar leptoquark boson from $(b \nu) (\bar{b} \bar{\nu})$ was also provided by ATLAS~\cite{Aad:2015caa}. 
Moreover, we have considered the constraints from the CMS search~\cite{Khachatryan:2015bsa} for third-generation scalar leptoquark bosons decaying into $(t \tau) (\bar{t}\bar{\tau})$. 
We have estimated the current bound on $(c \tau) (\bar{c}\bar{\tau})$ by recasting the leptoquark search for the $(b \tau) (\bar{b}\bar{\tau})$ channel in Ref.~\cite{Khachatryan:2014ura}. 
This recasting is based on our study for the tagging and mis-tagging efficiencies between $b$ and $c$ quarks, with the help of Refs.~\cite{CMS:2013vea,Chatrchyan:2012jua}. 
Finally, preliminary results of the search for bottom squarks at the 13~TeV LHC were also taken into account. 
In summary, the constraints from the current available LHC searches at 8~TeV imply that
$M_{S_1} < 400\,\text{GeV}$, $M_{S_1} < 530\,\text{GeV}$, and $M_{S_1} < 640\,\text{GeV}$ are ruled out for $g_{1L}^{33} \sim 0.3$, $g_{1L}^{33} \gtrsim 0.5$ and $g_{1L}^{33} \lesssim 0.2$, respectively. 
We reach a similar conclusion in the case of nonzero $g_{1L}^{3i}$ ($i=1,2$).

To extract a maximum potential at the 14~TeV LHC to search for the $S_1$ boson in our setup, we have performed detailed cut analyses that include simulation of detector effects. 
We have applied the cut analysis given for the $(b  \widetilde{\chi}^0_1) (\bar{b} \widetilde{\chi}^{0*}_1)$ channel to our $(b \nu) (\bar{b}\bar{\nu})$ channel 
and validated the expected exclusion/discovery limits on $(M_{\tilde b_1}, M_{\widetilde{\chi}^0_1})$ in the SUSY model, as was already reported by the ATLAS collaboration~\cite{ATL-PHYS-PUB-2014-010}.

As for the cut analysis in the $(c \tau) (\bar{c}\bar{\tau})$ channel, we have employed the method for $(b \tau) (\bar{b}\bar{\tau})$ given by CMS~\cite{Khachatryan:2014ura} 
and tuned it to the 14~TeV LHC study for the $(c \tau) (\bar{c}\bar{\tau})$ signal. 
The following three important topics were discussed: (A) the requirement for the light lepton flavor, (B) the requirement on the number of $c$-jets, and (C) the $c$-tagging rates. 
In the given method, one of the tau-leptons is identified by the light lepton $\ell$ through the decay. 
In our analysis, we have considered the two cases as (A-1) $\ell=\mu$ and (A-2) $\ell=\mu \text{ or } e$. 
The original method for $(b \tau) (\bar{b}\bar{\tau})$ suggests that only one of the quark flavors ($b$) is tagged in the analysis. 
Instead, we have considered the two cases such that (B-1) at least two $c$-jets and (B-2) at least one $c$-jet are tagged in our analysis for $(c \tau) (\bar{c}\bar{\tau})$. 
Finally we have studied the three possibilities for the $c$-tagging/mis-tagging rates such as (Case-1) from Ref.~\cite{Perez:2015lra}, (Case-2) from Ref.~\cite{Aad:2015gna}, and (Case-3) from Ref.~\cite{ATL-PHYS-PUB-2015-001}, 
since the efficiency of the $c$-tagging algorithms at 14 TeV is not yet known.

After implementing the above method, we have generated and analyzed the signal events in the processes $pp \to S_1^*\,S_1 \to (b \nu) (\bar{b}\bar{\nu})$ and $pp \to S_1^*\,S_1 \to (c \tau) (\bar{c}\bar{\tau})$ 
with the use of {\tt MadGraph5\_aCM@NLO}, {\tt pythia-pgs}, {\tt DelphesMA5tune}, and {\tt MadAnalsysis5} in the cluster system provided at CTPU-IBS. 
Then we have finally obtained the exclusion limits on the $S_1$ leptoquark boson, expected at the 14~TeV LHC when $\mathcal{L} = 300\,\text{fb}^{-1}$ of data is accumulated.  
Our results suggest that the $S_{1}$ leptoquark boson up to at least $0.8\,\text{TeV}$ mass can be excluded at 95\% CL for both $i=3$ and $i=1\text{ or }2$ cases of $g_{1L}^{3i}$. 
For large and small $g_{1L}^{3i}$, $M_{S_1} \lesssim 1\,\text{TeV}$ can be ruled out from the $(b\nu)(\bar{b}\bar{\nu})$ and $(c\tau)(\bar{c}\bar{\tau})$ searches, respectively. 
We have also evaluated the 95\% CL exclusion and $5\sigma$ discovery potentials at a future 14~TeV center-of-mass energy, 
assuming that $\mathcal{L} = 3000\,\text{fb}^{-1}$ of data is collected and the background uncertainty is improved as $\sigma_{\text{bkg}} = 15\%$. 
The $95\%$ CL excluded ranges of $M_{S_1}$ are changed as $1.0\,\text{TeV} \sim 1.3\,\text{TeV}$.  
It has been found that the $S_1$ leptoquark boson with mass less than $0.8\,\text{TeV}$ can be discovered from both the $(b\nu)(\bar{b}\bar{\nu})$ and $(c\tau)(\bar{c}\bar{\tau})$ channels. 
A discovery only from either the $(c\tau)(\bar{c}\bar{\tau})$ or $(b\nu)(\bar{b}\bar{\nu})$ search can be expected up to $M_{S_1} \lesssim 1.1\,\text{TeV}$. 
We emphasize that the $\bar B \to D^{(*)} \tau \bar\nu$ anomaly, explained by the $S_1$ leptoquark boson, can be probed at the LHC search 
only if both the signals from $(b\nu)(\bar{b}\bar{\nu})$ and $(c\tau)(\bar{c}\bar{\tau})$ are discovered.

We briefly comment on prospects for the $(t \ell)(\bar{t} \bar{\ell})$ final state.
Although this channel has not yet been surveyed at the LHC, it may have good prospects since there are at least two charged leptons in the final state. 
In Ref.~\cite{Davidson:2011zn}, the $95\%$ CL lower bound on the mass was evaluated as $m_{\text{LQ}} \gtrsim 160\,\text{GeV}$ for $\mathcal{B}(\text{LQ} \to t \mu) = 1$ 
via the $t\bar{t}$ production cross section $\sigma_{t\bar{t}}$ measured by the D0 experiment at the Tevatron, 
from the final state $\ell_{i}^{\pm} \ell^{\mp}_{j} + {\not\!\! E_{\text{T}}} + \geq 3\,\text{jets}$ using $4.3\,\text{fb}^{-1}$ data at $\sqrt{s} = 1.96\,\text{TeV}$~\cite{D0note_6038-CONF}.
This bound is rather weak compared with $m_{\text{LQ}} \gtrsim 300\,\text{GeV}$, 
obtained by the search for the second generation leptoquark through $\text{LQ} \to q \mu$ based on the $1.0\,\text{fb}^{-1}$ data assuming $\mathcal{B}(\text{LQ} \to q \mu)=1$~\cite{Abazov:2008np}. 
On the other hand, refinement of the analysis cuts would lead to improvements in the sensitivity to the $(t \ell)(\bar{t} \bar{\ell})$ final state (see~\cite{CMS-PAS-EXO-16-007,Aaboud:2016qeg} for the latest LHC analyses at $\sqrt{s} = 13\,\text{TeV}$ for the second generation leptoquark.).

Finally, we mention that the leptoquark study in this paper is a simplified one, 
where only two leptoquark couplings to the second and third generation fermions are nonzero, and the $SU(2)_{L}$ singlet $S_1$ leptoquark boson is chosen for simplicity. 
In this model, however, nonzero proton decay amplitudes are written down with renormalizable interactions in general, 
even though the proton decay is problematic only in the presence of nonzero couplings to the first generation fermions. 
A more realistic candidate would be the doublet leptoquark $R_{2}$, where proton decay does not occur at the renormalizable level.
An exhaustive study including detailed collider analyses on $R_{2}$ would be an interesting further direction.


\begin{acknowledgments}
We are grateful to Wonsang Cho for providing a cluster system to generate a huge number of signal and background events. 
We are also thankful to Dipan Sengupta for helping us with event generation of the background processes and also giving advice for the NNLO cross section for the background processes.
KN also thanks Shigeki Matsumoto, Satoshi Mishima, Mihoko Nojiri, Takaaki Nomura, Chan Beom Park, Kohsaku Tobioka, Tsutomu Yanagida, and Hiroshi Yokoya for fruitful discussions.
We acknowledge the CTPU-IBS cluster system for executing massive computations.
This work is supported in part by IBS-R018-D1 for RW and BD. 
\end{acknowledgments}

\appendix
\section{Experimental results of $R(D)$ and $R(D^*)$}
\label{App:DetailedExperiment}
The present experimental results from the BaBar experiment~\cite{Lees:2012xj,Lees:2013uzd} have been given by 
\begin{equation}
   R(D)_\text{BaBar}=0.440\pm 0.072 \,, \quad R(D^*)_\text{BaBar}=0.332\pm 0.030 \,,
\end{equation}
where their correlation is reported as $\rho_\text{BaBar} = -0.27$. 
The recent results reported from the Belle~\cite{Huschle:2015rga} and LHCb~\cite{Aaij:2015yra} collaborations are shown as 
\begin{align}
   &R(D)_\text{Belle}=0.375\pm 0.069 \,, \quad R(D^*)_\text{Belle}=0.293\pm 0.041 \,, \quad \rho_\text{Belle} = -0.36 \,,\\
   &R(D^*)_\text{LHCb}=0.336\pm 0.040 \,. 
\end{align}
Then we obtained the combined results as 
\begin{align}
   R(D)_\text{exp}=0.393\pm 0.048 \,, \quad R(D^*)_\text{exp}=0.321\pm 0.021 \,, \quad \rho_\text{exp} = -0.31 \,.
\end{align}
With using this, we have evaluated the deviations as in Eqs.~(\ref{Eq:deviationRD}) and (\ref{Eq:deviationRDst}) and plotted the contour as in Fig.~\ref{Fig:Comparison}.

We now briefly explain the way in which the observables $R(D)$ and $R(D^*)$, defined in Eq.~(\ref{Eq:RDdefinition}), are measured. 
The BaBar collaboration~\cite{Lees:2012xj} reconstructed only the purely leptonic decays of the tau lepton such as $\tau^- \to e^- \bar\nu_e\nu_\tau$ and $\tau^- \to \mu^- \bar\nu_\mu\nu_\tau$, 
so that the signal ($\bar B \to D^{(*)} \tau^- \bar\nu_\tau$) and the normalization ($\bar B \to D^{(*)} \ell^- \bar\nu_\ell$ for $\ell = e$ and $\mu$) events can be identified using the same particles in the detector. 
Then signal and normalization events are extracted after several parameter fits to distributions are performed. 
This method can reduce various sources of uncertainty in $R(D)$ and $R(D^*)$. 
The recent Belle result in Ref.~\cite{Huschle:2015rga} was also improved in a similar way. 
The analysis for the LHCb is totally different~\cite{Aaij:2015yra} since the $B$ mesons are produced from the proton-proton collision.  
The muonic tau decay mode is utilized at LHCb.

As for the normalization modes $\bar B \to D^{(*)} \ell^- \bar\nu_\ell$, the averaged decay rates for $\ell = e$ and $\mu$ are used for the theoretical predictions on $R(D^{(*)})$. 
These decay processes have been observed to measure $|V_{cb}|$ in Refs.~\cite{Aubert:2007rs,Aubert:2009ac,Dungel:2010uk,Glattauer:2015teq}. 
We note that differences between the results from $\ell = e$ and $\mu$ decay modes are not seen in the determination of $|V_{cb}|$, 
which implies that the lepton flavor universality between $\bar B \to D^{(*)} e^- \bar\nu_e$ and $\bar B \to D^{(*)} \mu^- \bar\nu_\mu$ holds within uncertainties.

\section{Misidentified signals}
\label{App:DetailedAnalysis}
In our main study, we focused on the $(b \nu) (\bar{b} \bar{\nu})$ and $(c\tau)\, (\bar{c}\bar{\tau})$ channels as signal events in the search. 
As introduced in Sec.~\ref{Sec:CombinedResult}, misidentified signals, arising from other leptoquark processes than the ones primarily considered, may arise and should be discussed. 
In particular, the processes $S_1^*S_1 \to (t\tau)(\bar{t}\bar{\tau})$ and $S_1^*S_1 \to (c\tau)(\bar{t}\bar{\tau}), (t\tau)(\bar{c}\bar{\tau})$ are dominant misidentified signals in our model. 
They can contribute to the signal in the search for $S_1^* \to c\tau$. 
We have investigated such misidentified signals and evaluated their exclusion potential in the $(M_{S_1}, g_{1L}^{33})$ plane of the $S_1$ leptoquark model.

\begin{figure}[t!]
\centering
\includegraphics[viewport=0 0 360 363, width=18em]{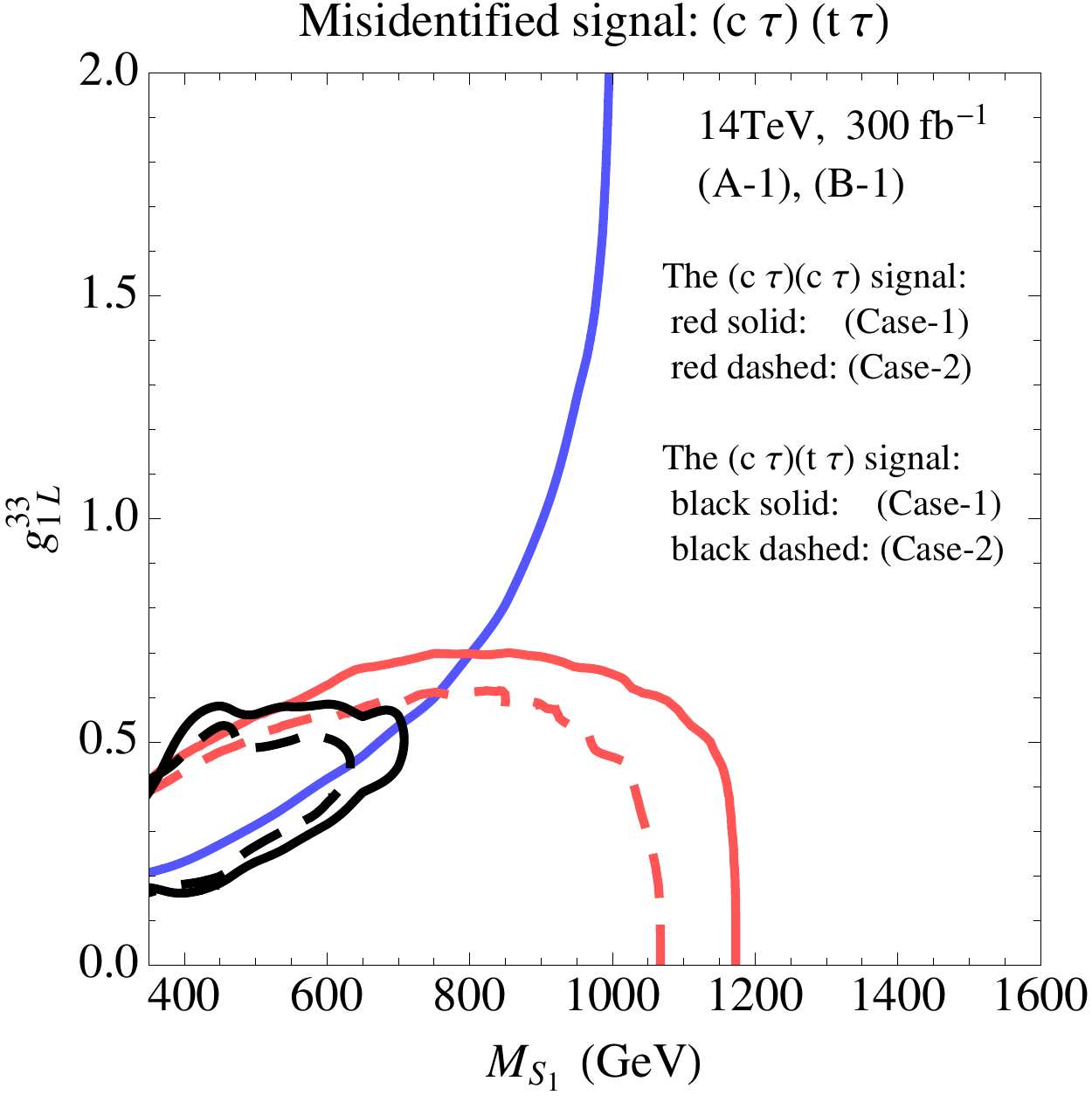} \quad
\includegraphics[viewport=0 0 360 363, width=18em]{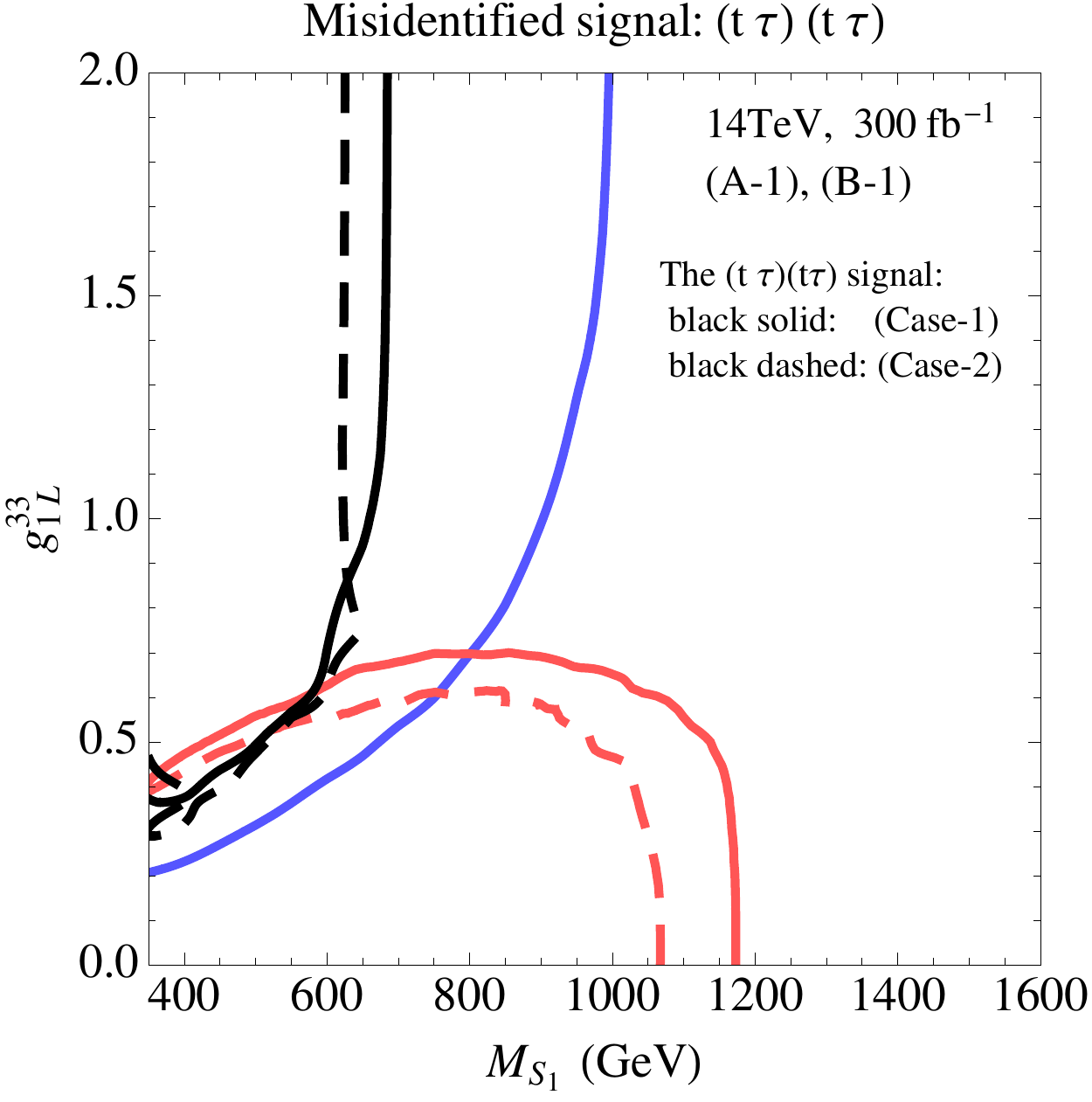}
\caption{
The $95\%$ exclusion limits from the misidentified signals of $S_1^*S_1 \to (t\tau)(\bar{t}\bar{\tau})$ and $S_1^*S_1 \to (c\tau)(\bar{t}\bar{\tau}), (t\tau)(\bar{c}\bar{\tau})$ 
for the 14~TeV LHC with $\mathcal{L} = 300\, \text{fb}^{-1}$ and $\sigma_{\text{bkg}}=30\%$, along with the results from the normal signals as given in Sec.~\ref{Sec:CombinedResult}. 
The black curves show the results of the misidentified signals, whereas the blue and red curves are from $(b \nu) (\bar{b} \bar{\nu})$ and $(c\tau)\, (\bar{c}\bar{\tau})$. 
The $c$-tagging/mis-tagging rates are chosen as indicated in the figure. 
}
\label{Fig:MisidSignal}
\end{figure}
In Fig.~\ref{Fig:MisidSignal}, we show the $95\%$ exclusion limits from the signal through the misidentification of $(c\tau)(\bar{t}\bar{\tau}), (t\tau)(\bar{c}\bar{\tau})$ and $(t\tau)(\bar{t}\bar{\tau})$, 
where we set $\mathcal{L} = 300\, \text{fb}^{-1}$, $\sigma_{\text{bkg}}=30\%$, (A-1), and (B-1). 
The black curves indicate the $95\%$ exclusion limits from the misidentified signals of $[(c\tau)(\bar{t}\bar{\tau}), (t\tau)(\bar{c}\bar{\tau})]$ and $[(t\tau)(\bar{t}\bar{\tau})]$ presented in the left and right panels, respectively. 
The solid and dashed curves are obtained for Case-1 and Case-2, respectively. 
The blue and red curves are the results from the normal signals $(b \nu) (\bar{b} \bar{\nu})$ and $(c\tau)\, (\bar{c}\bar{\tau})$  (for Case-1 and Case-2), as shown in Sec.~\ref{Sec:Prospect}.

Although the misidentification of the $(c\tau)(\bar{t}\bar{\tau}), (t\tau)(\bar{c}\bar{\tau})$, and $(t\tau)(\bar{t}\bar{\tau})$ channels affect the evaluation of expected exclusion limits, 
it turns out that our conclusion obtained from the $(b \nu) (\bar{b} \bar{\nu})$ and $(c\tau)\, (\bar{c}\bar{\tau})$ analyses is not improved significantly when the misidentifications are taken into account. 
This is because that the excluded regions from these misidentified signals are fully covered by those from the original signals. 
The other possible misidentified signals such as $(c\tau)(\bar{b}\bar{\nu})$ are vetoed in the cut analysis.

Misidentifications for the signal $(b \nu) (\bar{b} \bar{\nu})$ can also occur. 
The processes $(t \tau) (\bar{b} \bar{\nu}), (b \nu) (\bar{t} \bar{\tau})$, $(c \tau) (\bar{b} \bar{\nu}), (b \nu) (\bar{c} \bar{\tau})$ are candidates for the misidentified signals. 
We have also studied these signals and found that they are completely negligible since the exclusion potentials do not exceed $60\%$ CL in all regions of the parameter space.


\bibliographystyle{utphys}
\bibliography{reference,reference_collider,reference_add}

\end{document}